\documentclass[aps]{revtex4}
\usepackage{eurosym}
\usepackage{amsfonts}
\usepackage{amsmath}
\usepackage{amssymb,epsf}
\usepackage{color}
\usepackage{graphicx}
\usepackage{float}
\usepackage{caption}
\usepackage{subfig}
\usepackage{epstopdf}
\usepackage{hyperref}

\begin{document}

\title{$AdS_{4}$ dyonic black holes in gravity's
rainbow}
\author{S. Panahiyan$^{1,2,3}$\footnote{
email address: shahram.panahiyan@uni-jena.de}, S. H. Hendi$^{4,5}$\footnote{
email address: hendi@shirazu.ac.ir} and N. Riazi$^{6}$\footnote{
email address: n$_{-}$riazi@sbu.ac.ir}}
\affiliation{$^1$Helmholtz-Institut Jena, Fr\"{o}belstieg 3, D-07743 Jena, Germany  \\
	$^2$GSI Helmholtzzentrum f\"{u}r Schwerionenforschung, D-64291 Darmstadt, Germany \\
	$^3$Theoretisch-Physikalisches Institut, Friedrich-Schiller-University Jena, D-07743 Jena, Germany\\
$^4$ Physics Department and Biruni Observatory, College of Sciences, Shiraz
University, Shiraz 71454, Iran\\
$^5$ Research Institute for Astronomy and Astrophysics of Maragha (RIAAM),
P.O. Box 55134-441, Maragha, Iran\\
$^6$ Physics Department, Shahid Beheshti University, Tehran 19839, Iran}

\begin{abstract}
In this paper, we
investigate thermodynamical structure of dyonic black holes
in the presence of gravity's rainbow. We confirm that for super magnetized
and highly pressurized scenarios, the number of black holes' phases is reduced
to a single phase. In addition, due to specific coupling of rainbow
functions, it is possible to track the effects of temporal and spatial parts
of our setup on thermodynamical quantities/behaviors including equilibrium
point, existence of multiple phases, possible phase transitions and
conditions for having a uniform stable structure. 
\end{abstract}

\maketitle

\section{Introduction}

General relativity and quantum mechanics are two celebrated theories of physics that have been used to describe different phenomena on widely different scales. Despite the validity of these two theories, there has been an ongoing investigation to unify them and obtain a quantum theory of gravity. One of the proposals in this regard is gravity's rainbow by Magueijo and Smolin \cite{Magueijo1}.

Gravity's rainbow is a generalization of the doubly special relativity (DSR) to general relativity background. Strictly speaking, DSR is an extension of special relativity where an additional limit is imposed on the properties of a particle \cite{DSR1,DSR2,DSR3}. This additional limit puts an upper bound on energies that a particle can have (Planck energy). This results into deformation of the usual energy-momentum dispersion relation, nonlinearity of the laws of energy and momentum conservation, and clarification of a threshold between a quantum and a classical description \cite{DSR1,DSR2,DSR3}. Generalization of DSR to general relativity results into non-trivial modification of the principles and equations of general relativity. For gravity's rainbow, the geometry of spacetime is energy dependent and for each quantum of energy, there is a different classical geometry. Therefore, there is a spectrum of energies building up our spacetime and explaining the gravitational interaction. The fundamental origin of the rainbow metric is identified to be in the principle of relative locality \cite{Amelino1} and their phenomenologies were addressed in Ref. \cite{Magueijo2,Amelino2}. In addition, based on quantum field theory, a general formalism for emergence of a rainbow metric from a quantum cosmological model was introduced in Ref. \cite{Assanioussi}. 

The gravity's rainbow has been extensively investigated in literature. It was shown that breaking Lorentz symmetry in high-energy regime (predicted by other theories of quantum gravity) could be realized within the framework of gravity's rainbow \cite{Mattingly,Lobo,Nilsson}. The Starobinsky model of inflation was investigated in Ref. \cite{Chatrabhuti} and also it was shown that using gravity's rainbow, a period of cosmological inflation may arise without introducing an inflaton field \cite{Barrow,Garattini1}. In addition, the possibility of removing big bang singularity in gravity's rainbow was pointed out in Refs. \cite{Awad,Santos,HMEP}. Existence of remnants for black holes after evaporation \cite{Ali1,Ali2} and providing solutions for information paradox \cite{Ali3,Gim1}  are other end results of employing gravity's rainbow. It should be also noted that gravity's rainbow admits uncertainty principle \cite{Galan,Kim,Gim2}, it is the UV completion
of general relativity \cite{Magueijo1} and it is connected to Ho\v{r}ava-Lifshitz gravity \cite{Garattini2}. Wormholes \cite{Garattini3}, neutron stars \cite{HBEP}, cosmic strings \cite{Momeni,Bakke} and white dwarfs \cite{WD} have also been investigated in the presence of gravity's rainbow. Recently, Vaidya-rainbow spacetime \cite{Heydarzade} and possible conformal transformations to gravity's rainbow \cite{He} were also investigated. The effects of gravity's rainbow on black hole solutions \cite{Hendi1,Hendi9,Hendi4,Hendi6,Bezerra}, their geometrical properties \cite{Leiva,Ali4} and thermodynamical behaviors \cite{Galan1,Ling1,Hendi2,Hendi3,Kim1,Hendi5,Alsaleh1,Alsaleh2,Feng1,Hendi7,Feng2,Dehghani1,Dehghani2,Hendi8,PL} are other subjects that have attracted a lot of attentions. So far, the dyonic black holes in the presence of gravity's rainbow have not yet been investigated. 

The dyonic black holes are a family of the magnetically charged static black hole solutions \cite{DS1,DS2,DS3,DS4,DS5,DS6,DS7,DS8,DS9}. In fact, it was shown that large dyonic black holes in $AdS$ spacetime correspond to stationary solutions of the equations of relativistic magnetohydrodynamics on the conformal boundary \cite{Caldarelli}. The dyonic black holes have been widely investigated to address different issues/phenomena in $AdS/CFT$ studies such as: the Hall conductivity and zero momentum hydrodynamic response function \cite{Hartnoll}, magnetic dependency of superconductors' properties \cite{Albash}, transport coefficients and DC longitudinal conductivity \cite{Goldstein}, and the paramagnetism/ferromagnetism phase transitions \cite{HD1,HD2,Dutta,HRP}. Considering these applications, in this paper, we intent to construct $AdS_{4}$ dyonic black holes in the context of gravity's rainbow and investigate their geometrical and thermodynamical properties.

The main results of this paper show that: I) The existence of black hole solutions is physically limited by the contributions of gravity's rainbow. II) The electric-magnetic duality that previously was reported for dyonic black holes is eliminated here. This is due to specific coupling of rainbow functions. This results into distinct effects of electric and magnetic charges on the properties of solutions. III) The number of thermodynamical phases, phase transitions between them and the type of phase transition is bounded by upper and lower limits imposed by energy functions and magnetic charge. 

The structure of the present paper is as follows: First, action and metric are given and exact solutions are obtained. The conditions for having black hole solutions or naked singularity are extracted. Using obtained black hole solutions, thermodynamical quantities are calculated and by studying their behaviors, we show that new constraints should be imposed on the values of different parameters to have reasonable physical behavior. The possibility of thermal phase transition and its type are also investigated. The paper is finally concluded by some closing remarks.

\section{Black hole solutions} \label{Black hole solutions}

\subsection{Metric and Lagrangian ansatz} \label{Metric and lagrangian ansatz}

The core stone of the gravity's rainbow lays within the structure of
spacetime and its geometry. In other words, particles with different energy
experience different gravitational fields. This property is introduced in
the so-called doubly general relativity to incorporate the effects of
quantum gravity. To do so, instead of modifying the action of system, the
metric is modified. In other words, in this approach, the effects of quantum
gravity becomes directly apparent in the metric of spacetime. However, the
metric is not arbitrarily modified. The guideline of modification of the
metric comes from the modified version of the energy-momentum dispersion
relation. In doubly special relativity, the energy-momentum dispersion
relation is given by 
\begin{equation}
E^{2}f^{2}(\varepsilon )-p^{2}g^{2}(\varepsilon )=m^{2},  \label{MDR}
\end{equation}%
in which the dimensionless energy ratio is $\varepsilon =E/E_{P}$ where $E$
and $E_{P}$ are, respectively, the energy of test particle and the Planck
energy. The energy of test particle can not exceed the Planck energy,
therefore we have the limit of $0<\varepsilon \leq 1$. In addition, $%
f(\varepsilon )$ and $g(\varepsilon )$ are energy functions restricted by
the following condition in the infrared limit 
\begin{equation}
\lim\limits_{\varepsilon \rightarrow 0}f(\varepsilon )=1,\qquad
\lim\limits_{\varepsilon \rightarrow 0}g(\varepsilon )=1.
\end{equation}

In next step, using the analogy between the energy-momentum four vector $(E,%
\vec{p})$ and time-space $(t,\vec{x})$, one can employ the energy
functions to build an energy dependent spacetime with following recipe 
\begin{equation}
\hat{g}(\varepsilon )=\eta ^{ab}e_{a}(\varepsilon )\otimes e_{b}(\varepsilon
),  \label{rainmetric}
\end{equation}%
in which 
\begin{equation}
e_{0}(\varepsilon )=\frac{1}{f(\varepsilon )}\tilde{e}_{0},\qquad
e_{i}(\varepsilon )=\frac{1}{g(\varepsilon )}\tilde{e}_{i},
\end{equation}%
where $\tilde{e}_{0}$ and $\tilde{e}_{i}$ are the energy independent frame
fields. Now, since we are interested in topological solutions, the general
form of the metric will be 
\begin{equation}
ds^{2}=-\frac{\psi (r)}{f^{2}(\varepsilon)}dt^{2}+\frac{1}{
g^{2}(\varepsilon)}\left[ \frac{dr^{2}}{\psi (r)}+r^{2}d\Omega _{k}^{2}%
\right] ,  \label{metric}
\end{equation}
with 
\begin{equation}
d\Omega _{k}^{2}=\left\{ 
\begin{array}{cc}
d\theta ^{2}+\sin ^{2}\theta d\varphi ^{2}, & k=1 \\ 
d\theta ^{2}+d\varphi ^{2}, & k=0 \\ 
d\theta ^{2}+\sinh ^{2}\theta d\varphi ^{2}, & k=-1%
\end{array}%
\right. .  \label{dOmega}
\end{equation}

It should be noted that $k$ determines the type of horizon topology that
solutions would have. $k=-1, 0 \text{ and } 1$ correspond to 
\textit{hyperbolic}, \textit{flat} and \textit{spherical} horizons, respectively.

In general, since different components of the action and corresponding field
equations are calculated by using the metric, employing the rainbow function
will provide us with non-trivial solutions which have the trace of quantum
gravity corrections.

In this paper, we are interested in $4$-dimensional dyonic solutions. The
action of such system is given by 
\begin{equation}
\mathcal{I}=-\frac{1}{16\pi G}\int d^{4}x\sqrt{-g}\left[ \mathcal{R}%
-2\Lambda -F^{\mu \nu }F_{\mu \nu }\right] ,  \label{Action}
\end{equation}%
where $\mathcal{R}$ is the scalar curvature, $\Lambda =-\frac{3}{l^{2}}$ is
the negative cosmological constant and $F_{\mu \nu }=\partial _{\mu }A_{\nu
}-\partial _{\nu }A_{\mu }$ is the electromagnetic field tensor in which $%
A_{\mu }$ is the gauge potential. Variation of the action with respect to
metric and gauge potential results into the following field equations 
\begin{eqnarray}
e_{\mu \nu }\equiv G_{\mu \nu }+\Lambda g_{\mu \nu }-\left[ 2F_{\mu \lambda
}F_{\nu }^{\lambda }-\frac{1}{2}g_{\mu \nu }F^{\sigma \rho }F_{\sigma \rho }%
\right] &=&0,  \label{Field equation} \\
&&  \notag \\
\partial _{\mu }\left( \sqrt{-g}F^{\mu \nu }\right) &=&0,
\label{Maxwell equation}
\end{eqnarray}%
where in the above equation, $G_{\mu \nu }$ is the Einstein tensor. The term
"\textit{dyonic solutions}" correspond to the presence of a magnetic charge in
the solutions. There are several methods (gauge potential) to construct
magnetic solutions. Here, we use the following gauge one-form potential to
obtain magnetic solutions 
\begin{equation}
A=-\frac{q_{E}}{r} dt+q_{M}d\varphi\left\{ 
\begin{array}{cc}
\cos \theta & k=1 \\ 
\theta & k=0 \\ 
\cosh \theta & k=-1%
\end{array}%
\right. ,  \label{gaugeP}
\end{equation}%
in which $q_{E}$ and $q_{M}$ are two constants related to total electric and
magnetic charges, respectively. Using the Maxwell equation \eqref{Maxwell
equation} with the metric \eqref{metric}, one can show that the nonzero
components of the electromagnetic tensor are 
\begin{equation}
F_{tr}=-F_{rt}=\frac{q_{E}}{r^{2}}\text{ \ \ \ \ \ \ \& \ \ \ \ \ \ }F_{\varphi
\theta }=-F_{\theta \varphi }=q_{M}\left\{ 
\begin{array}{cc}
\sin \theta & k=1 \\ 
\theta & k=0 \\ 
\sinh \theta & k=-1%
\end{array}%
\right. .  \label{Ftr}
\end{equation}

\subsection{Solutions and their properties} \label{Solutions and their properties}

In this section and the following ones, we extract field equations and use them to obtain solutions. In addition, we determine the conditions for having black hole solutions and
investigate the effects of different parameters on the properties of
solutions.

\subsubsection{Solutions}

Using the metric \eqref{metric} and field equations \eqref{Field equation},
it is possible to obtain different components of the field equations as 
\begin{eqnarray}
e_{tt} &=&e_{rr}=g^2(\varepsilon)r^{2}[k-f(r)]-\Lambda
r^{4}-g^2(\varepsilon) (\frac{df(r) }{dr})r^{3}-g^2(\varepsilon)[
f^2(\varepsilon)q_{E}^{2}+g^2(\varepsilon)q_{M}^{2}]=0,  \notag \\
e_{\theta \theta} &=&e_{\varphi \varphi}=g^2(\varepsilon) r^{3} [2(\frac{%
df(r) }{dr})+(\frac{d^2f(r) }{dr^2}) r]+2 \Lambda r^{4}-2g^2(\varepsilon)[
f^2(\varepsilon)q_{E}^{2}+g^2(\varepsilon)q_{M}^{2}]=0,  \label{fields}
\end{eqnarray}
which by solving them, the following metric function is obtained
\begin{eqnarray}
\psi(r)=k+\frac{f^2(\varepsilon) q_{E}^2}{r^2}+\frac{g^2(\varepsilon) q_{M}^2%
}{r^2}-\frac{\Lambda r^2}{3 g^2(\varepsilon)}-\frac{m}{r},
\label{metric function}
\end{eqnarray}
where $m$ is an integral constant called geometrical mass and it is related
to total mass of the black hole.

Evidently, the topological signature of the horizon and geometrical mass are
the only parts of the metric function which are not coupled with rainbow
functions. One important consequence of this issue is the fact that geometrical horizon
of the solutions is not affected/determined by energy functions. Therefore,
one can speculate that the horizon structure of solutions with gravity's
rainbow is not distinguishable from those in the absence of gravity's
rainbow. This is similar to what is observed in Gauss-Bonnet gravity
generalization \cite{GB} but in contrast to dilaton gravity where the
topological factor is coupled and affected by dilaton parameters \cite%
{Dilaton}.

Another important effect of gravity's rainbow on the solutions is the fact
that magnetic and electric charges are coupled with different rainbow
functions. This leads to a better distinguishability between magnetic and
electric charges. Also, this indicates that our gravitational system treats
these charges differently. The energy functions are motivated and determined
by the specific details of the system. There are cases where these two
energy functions have opposite behavior. In this regard, it is
possible that in specific regimes, while the effects of magnetic charge
become significant, the effects of electric charge become negligible or
vice versa. Such differences in coupling is rooted in the gauge potential %
\eqref{gaugeP}. The electric part of the solutions comes from the temporal
coordinate of the gauge potential while the magnetic part comes from spatial
coordinates. Due to specific coupling of energy functions with temporal and
spatial coordinates in the metric \eqref{metric}, this difference in
coupling for magnetic and electric charges has emerged in the metric
function. Considering that some of the thermodynamical quantities are
obtained based on metric function, one expects to see such difference in
coupling in some of the thermodynamical quantities. This will be explored later.

One last issue is related to the electric-magnetic duality. In the absence of gravity's rainbow, the electric and magnetic charges have total symmetry for swapping subscribes "\textit{E}" and "\textit{M}". This is due to electric-magnetic duality nature of the solutions. In contrast, generalization to gravity's rainbow results into omitting this property. Therefore, there is no total symmetry for swapping subscripts "\textit{E}" and "\textit{M}" for solutions in the presence of gravity's rainbow. This again is due to different coupling of rainbow function in temporal and spatial coordinates. This is one of the main features of generalization to gravity's rainbow. 

\subsubsection{Curvature scalar and asymptotic behavior} \label{Curvature scalar and asymptotic behavior} 

Strictly speaking, the obtained solutions could be interpreted as black holes if two conditions are
met: I) Existence of a non-removable singularity. II) Presence of at least
one event horizon covering this singularity. In some papers, it was argued
that the first condition could be relaxed while the second one stays intact. Despite these arguments, geometrically and physically, it is important
to see if our spacetime with metric function \eqref{metric function}
has a singularity. To address this issue, one can study curvature scalars.
It is sufficient that one of these scalars admit an irremovable singularity
to conclude presence of curvature singularity in our spacetime.

The scalar curvature that we employ is Kretschmann scalar. The reason for
this is two folds: investigating possible singularity and studying the
asymptotic behavior of the solutions. The Kretschmann scalar is given by 
\begin{equation*}
K=R_{\alpha \beta \gamma \delta }R^{\alpha \beta \gamma \delta
}=g^{4}(\varepsilon )\left( \frac{d^{2}\psi\left( r\right) }{dr^{2}}\right)^2 +%
\frac{4g^{4}(\varepsilon )}{r^{2}}\left( \frac{d\psi\left( r\right) }{dr}%
\right) ^{2}+\frac{4g^{4}(\varepsilon )}{r^{4}}\left( \psi\left( r\right)
-k\right) ^{2},
\end{equation*}%
where by using the obtained metric function \eqref{metric function}, we will
have 
\begin{equation}
K=\frac{56g^{4}(\varepsilon )\left( f^{2}(\varepsilon
)q_{E}^{2}+g^{2}(\varepsilon )q_{M}^{2}\right)^2 }{r^{8}}-\frac{%
48mg^{4}(\varepsilon )\left( f^{2}(\varepsilon )q_{E}^{2}+g^{2}(\varepsilon
)q_{M}^{2}\right) }{r^{7}}+\frac{8\Lambda ^{2}}{3}+\frac{12 g^{4}(\varepsilon
)m^{2}}{r^{6}}.  \label{Kretschmann}
\end{equation}

\begin{figure}[!htb]
\centering
\subfloat[$q_{M}=0.1$, $k=1$, $g(\varepsilon)=f(\varepsilon)=1.1$ and $\Lambda=-1$.]{
        \includegraphics[width=0.25\textwidth]{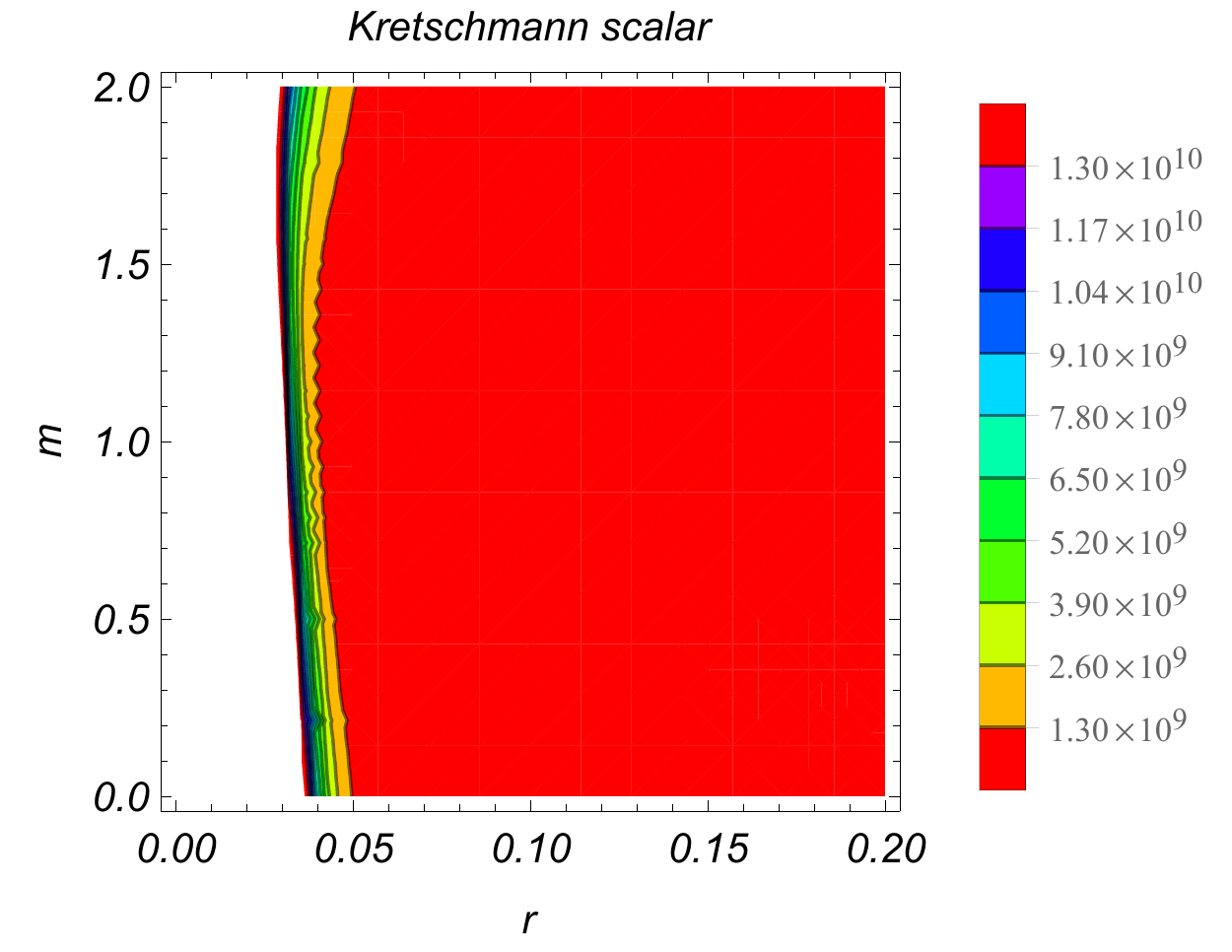}}  
\subfloat[$q_{M}=0.1$, $k=1$, $g(\varepsilon)=1.1$, $m=3$ and $\Lambda=-1$.]{
        \includegraphics[width=0.25\textwidth]{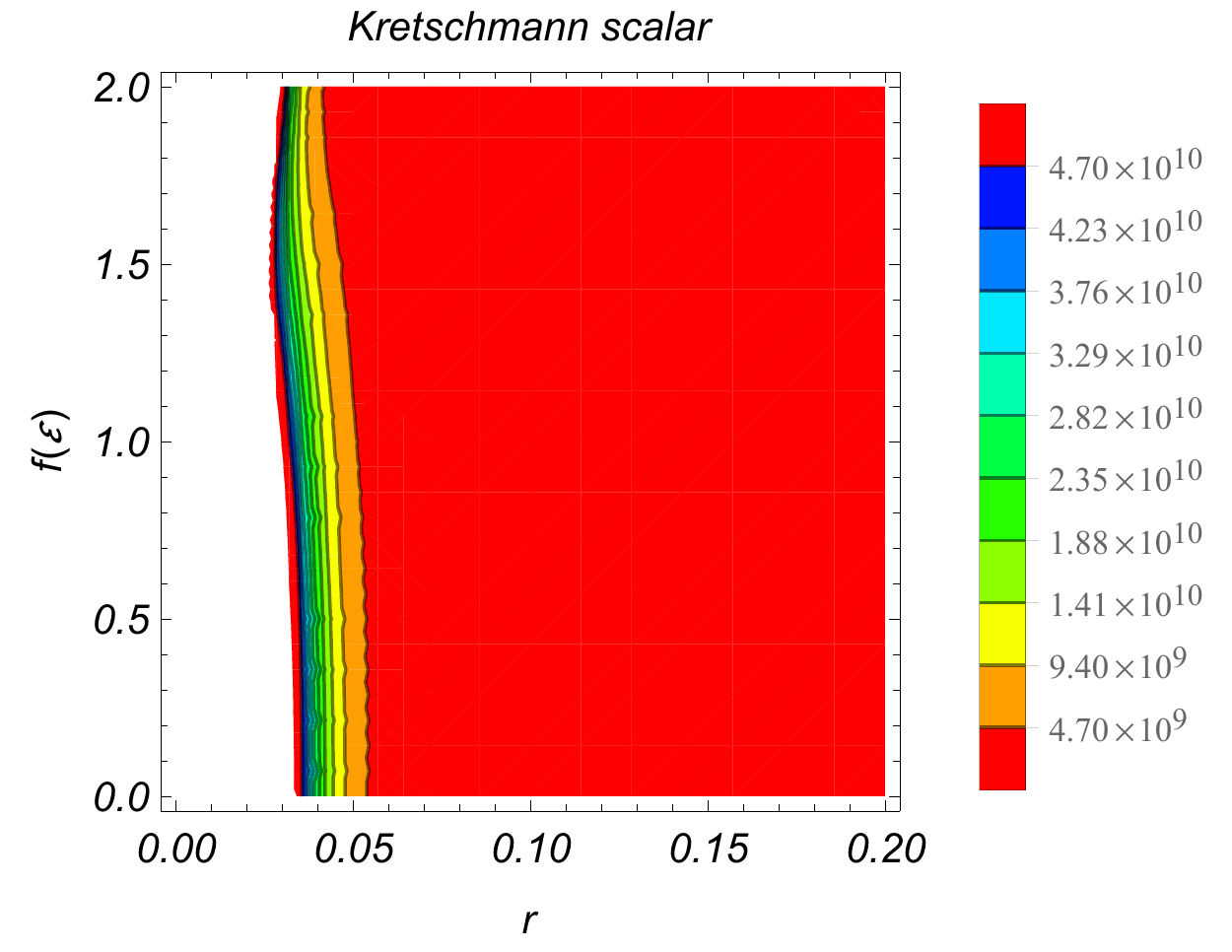}}  
\subfloat[$q_{M}=0.1$, $k=1$, $f(\varepsilon)=1.1$, $m=3$ and $\Lambda=-1$.]{
        \includegraphics[width=0.25\textwidth]{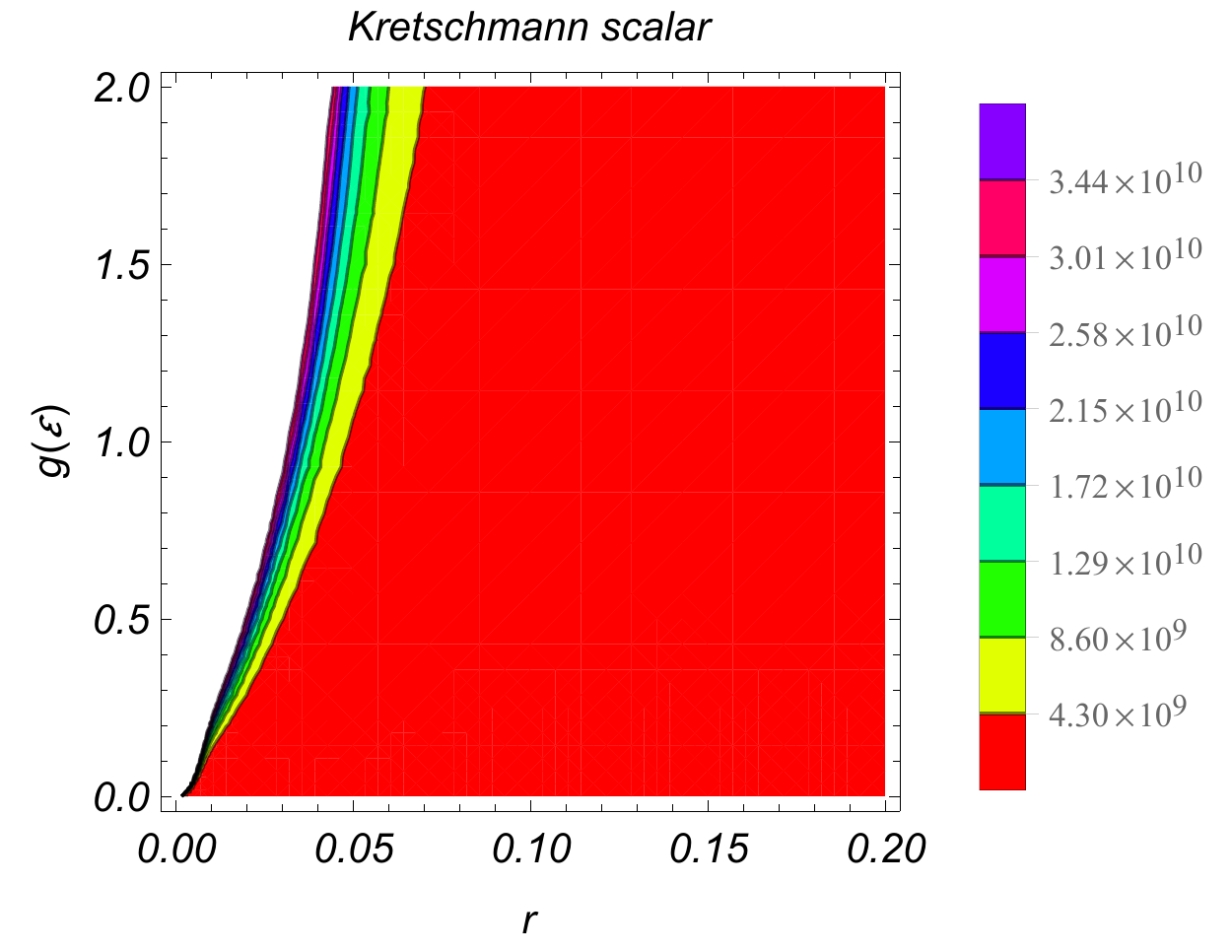}}  
\subfloat[$k=1$, $g(\varepsilon)=f(\varepsilon)=1.1$, $m=3$ and $\Lambda=-1$.]{
        \includegraphics[width=0.25\textwidth]{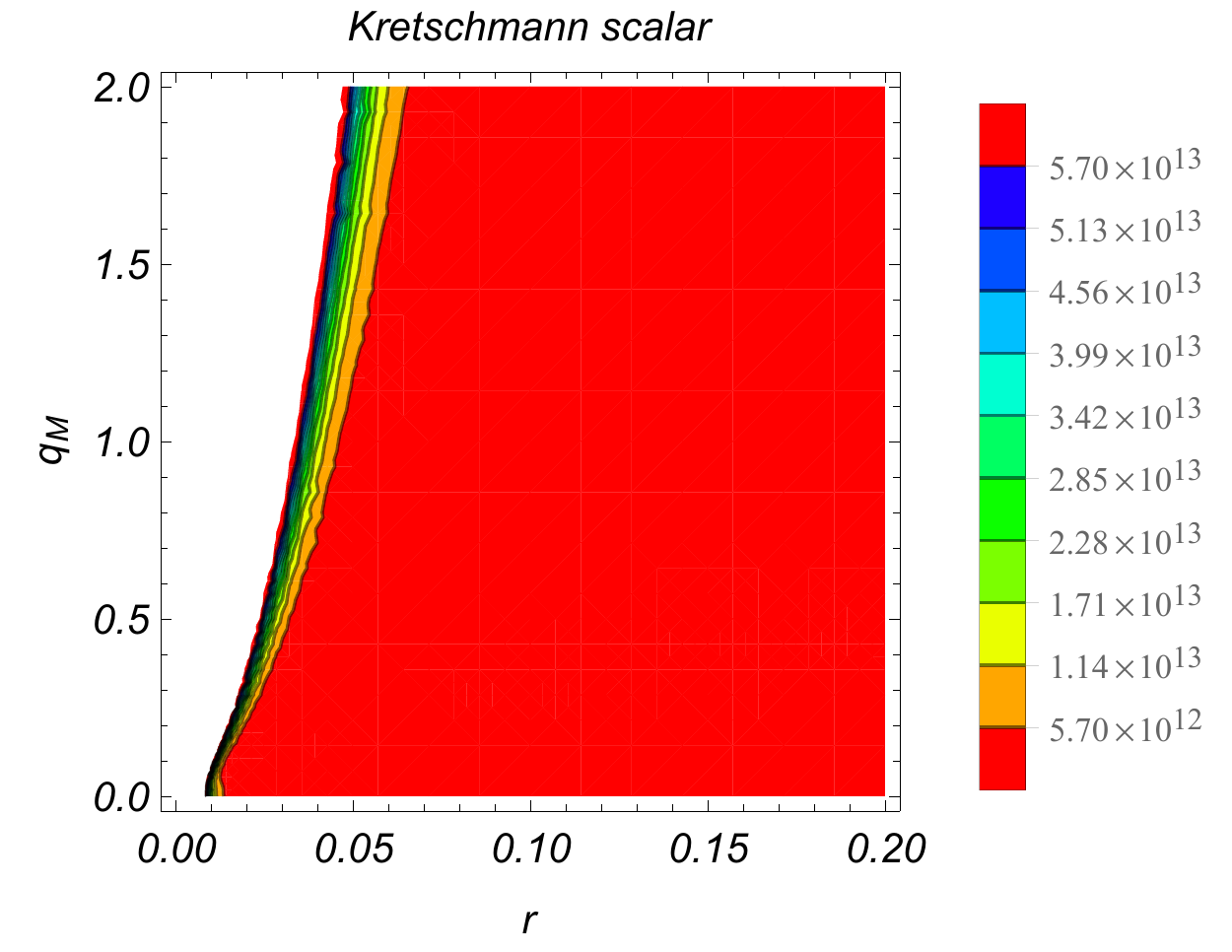}} \newline
\caption{Variation of the $K$ as a function of different parameters for $%
q_{E}=0.1$.}
\label{Fig1}
\end{figure}

For the limit $r\longrightarrow 0$, the dominant term of Eq. \eqref{Kretschmann} is
\begin{equation*}
\lim_{r\longrightarrow 0}K=\frac{56g^{4}(\varepsilon )\left(
f^{2}(\varepsilon )q_{E}^{2}+g^{2}(\varepsilon )q_{M}^{2}\right)^2 }{r^{8}}+O(%
\frac{1}{r^{7}}),
\end{equation*}%
which confirms two important points: I) Kretschmann scalar diverges at the
origin which shows that there is a singularity at $r=0$. II) The behavior of solutions near singular point is
determined by the matter field (electric and magnetic charges) and gravity's
rainbow. In order to understand the effects of different parameters on the
Kretschmann, we have plotted Fig. \ref{Fig1}.

If we keep the radial component ($r$) fixed, we find that: There is a specific geometrical mass $m_{s}$ where for 
$m<m_{s}$ the Kretschmann scalar is a decreasing function of this parameter,
whereas for $m>m_{s}$, it becomes an increasing function of $m$. The same could be stated for energy functions and magnetic charge. 

The asymptotic behavior of the Kretschmann scalar is obtained as 
\begin{equation*}
\lim_{r\longrightarrow \infty }K=-\frac{8\Lambda ^{2}}{3}+O(\frac{1}{r^{6}}),
\end{equation*}%
which shows that the asymptotic behavior of the solutions is  $AdS/dS$.
Also, we see that magnetic and electric charges have not affected the asymptotic behavior.

\subsubsection{Roots of metric function} \label{Roots of metric function}

The existence of black hole is restricted to the presence of at least one
event horizon for the solutions. The event horizons are where metric
function vanishes. It is necessary, but not sufficient. $\frac{d\psi(r)}{dr}$
should be positive at the event horizon, too. Otherwise, it is Cauchy or
cosmological horizon. Considering the parameters present in the metric
function, the following cases could be pointed out:

\textbf{I)} $\Lambda=0$: In this case, the roots of metric function are
obtained in the following form 
\begin{equation}
\left. r(\psi(r)=0)\right\vert _{\Lambda=0}=\frac{m\pm\sqrt{m^2-4
f^2(\varepsilon) k q_{E}^2-4 g^2(\varepsilon) k q_{m}^2}}{2 k}.
\end{equation}

The first condition for existence of real positive valued roots for this
case is given by 
\begin{equation}
m^2\geq 4 f^2(\varepsilon) k q_{E}^2+4 g^2(\varepsilon) k q_{m}^2,
\end{equation}%
which puts a lower limit on geometrical mass. If the mentioned condition is
not satisfied, solutions suffer the absence of root and are interpreted as
naked singularity. The positivity of the root is another condition that must
be met. Considering this, one can confirm that for hyperbolic case, only
one root could be observed. For spherical black holes, if $m^2= 4
f^2(\varepsilon) k q_{E}^2+4 g^2(\varepsilon) k q_{m}^2$, solutions will
have only one root, otherwise, there exist two roots for the metric
function.

\textbf{II)} $m=0$: The roots of the metric function for this case are
extracted as 
\begin{equation}
\left. r(\psi(r)=0)\right\vert _{m=0}=\sqrt{\frac{3 g^2(\varepsilon) k}{2
\Lambda }\pm\frac{ \sqrt{12 f^2(\varepsilon)g^2(\varepsilon) \Lambda
q_{E}^2+9 g^4(\varepsilon) k^2+12 g^4(\varepsilon) \Lambda q_{M}^2}}{2
\Lambda }}.
\end{equation}

Here, two conditions are required to be satisfied for having real valued
root. The first one is 
\begin{equation}
\Lambda\geq \frac{3 g^2(\varepsilon) k}{4 f^2(\varepsilon) q_{E}^2+4
g^2(\varepsilon) q_{M}^2 }.
\end{equation}

Such condition is automatically satisfied for $dS$ solutions while for $AdS$
black holes, a restriction is imposed on the values of different parameters.
It is possible to trace out the effects of electric and magnetic charges,
and one of the energy functions on root of the metric function if $\Lambda = 
\frac{3 g^2(\varepsilon) k}{4 f^2(\varepsilon) \Lambda q_{E}^2+4
g^2(\varepsilon) q_{M}^2 }$. In this case, only one root will be available
for the metric function and it belongs to black holes with hyperbolic
horizon. The second condition for the existence of real valued root for the
metric function is given by 
\begin{equation}
\frac{3 g^2(\varepsilon) k}{2 \Lambda }\pm\frac{ \sqrt{12
f^2(\varepsilon)g^2(\varepsilon) \Lambda q_{E}^2+9 g^4(\varepsilon) k^2+12
g^4(\varepsilon) \Lambda q_{M}^2}}{2 \Lambda }\geq 0.
\end{equation}

\textbf{III)} $\Lambda=m=0$: Interestingly, in the absence of geometrical
mass and cosmological constant, it is possible to obtain root for the metric
function as 
\begin{equation}
\left. r(\psi(r)=0)\right\vert _{\Lambda=m=0}=\sqrt{-\frac{f^2(\varepsilon)
q_{E}^2+ g^2(\varepsilon) q_{M}^2}{k}},
\end{equation}
which shows that only solutions with hyperbolic horizon enjoy the existence
of event horizon in their structures. It should be noted that root in this
case is an increasing function of the electric and magnetic charges, and
rainbow functions.

\textbf{IV)} General case: It is possible to obtain the root of metric
function analytically for this case as 
\begin{equation}
r(\psi(r)=0)=
\sqrt{\frac{12 a_{1}^{1/3} g(\varepsilon)^2 k-(2 a_{1})^{2/3}+18 \sqrt[3]{2} g(\varepsilon)^2 \left(4 f(\varepsilon)^2 \Lambda  q_{E}^2-g(\varepsilon)^2 \left(k^2-4 \Lambda  q_{M}^2\right)\right)}{24 a_{1}^{1/3} \Lambda }}\pm\frac{\sqrt{a_{2}}}{2}
\end{equation}
where
\begin{eqnarray*}
a_{1} &=& 27 g(\varepsilon)^4 \left(24 f(\varepsilon)^2 k \Lambda  q_{E}^2+2 g(\varepsilon)^2 \left(k^3+12 k \Lambda  q_{M}^2\right)-9 \Lambda  m^2\right)+
 \\
 && \sqrt{729 g(\varepsilon)^8 \left(24 f(\varepsilon)^2 k \Lambda  q_{E}^2+2 g(\varepsilon)^2 \left(k^3+12 k \Lambda 
		q_{M}^2\right)-9 \Lambda  m^2\right)^2-2916 g(\varepsilon)^6 \left(g(\varepsilon)^2 \left(k^2-4 \Lambda  q_{M}^2\right)-4 f^2 \Lambda  q_{E}^2\right)^3}
\end{eqnarray*}
\begin{eqnarray*}
a_{2} &=& \frac{24 a_{1} g(\varepsilon)^2 k+2^{2/3} a_{1}^2+18 \sqrt[3]{2} g(\varepsilon)^2 \left(g(\varepsilon)^2 \left(k^2-4 \Lambda  q_{M}^2\right)-4 f(\varepsilon)^2 \Lambda  q_{E}^2\right)}{6 a_{1} \Lambda }-
 \\
&&\frac{6 \sqrt{6} g(\varepsilon)^2 m}{\Lambda 
		\sqrt{\frac{12 a_{1} g(\varepsilon)^2 k-2^{2/3} a_{1}^2+18 \sqrt[3]{2} g(\varepsilon)^2 \left(4 f(\varepsilon)^2 \Lambda  q_{E}^2-g(\varepsilon)^2 \left(k^2-4 \Lambda  q_{M}^2\right)\right)}{a_{1} \Lambda }}}
\end{eqnarray*}

\begin{figure}[!htb]
\centering
\subfloat[$q_{M}=0.5$, $k=1$, $g(\varepsilon)=f(\varepsilon)=1.1$ and $\Lambda=-1$.]{
        \includegraphics[width=0.25\textwidth]{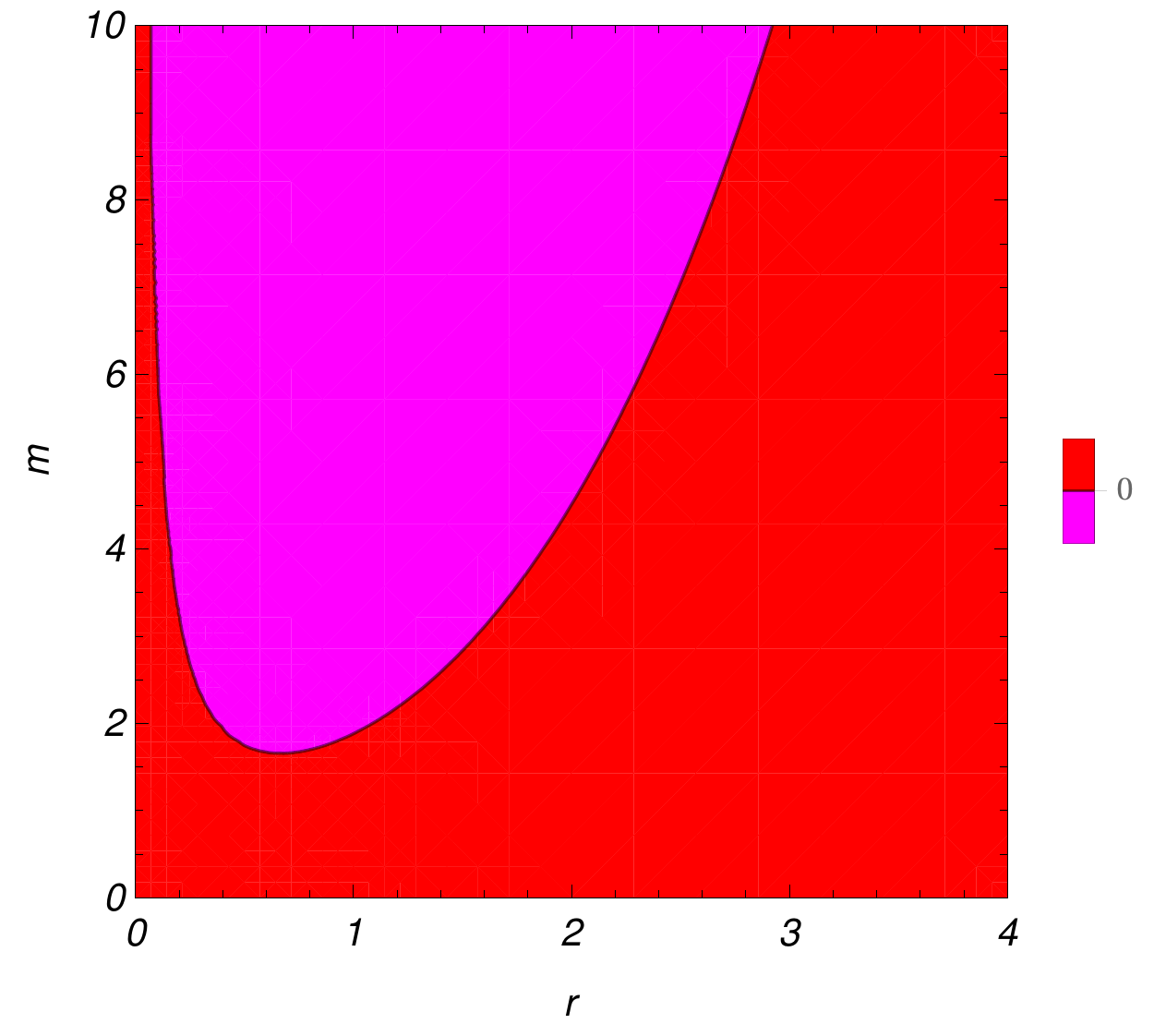}}  
\subfloat[$q_{M}=0.5$, $k=1$, $g(\varepsilon)=1.1$, $m=3$ and $\Lambda=-1$.]{
        \includegraphics[width=0.25\textwidth]{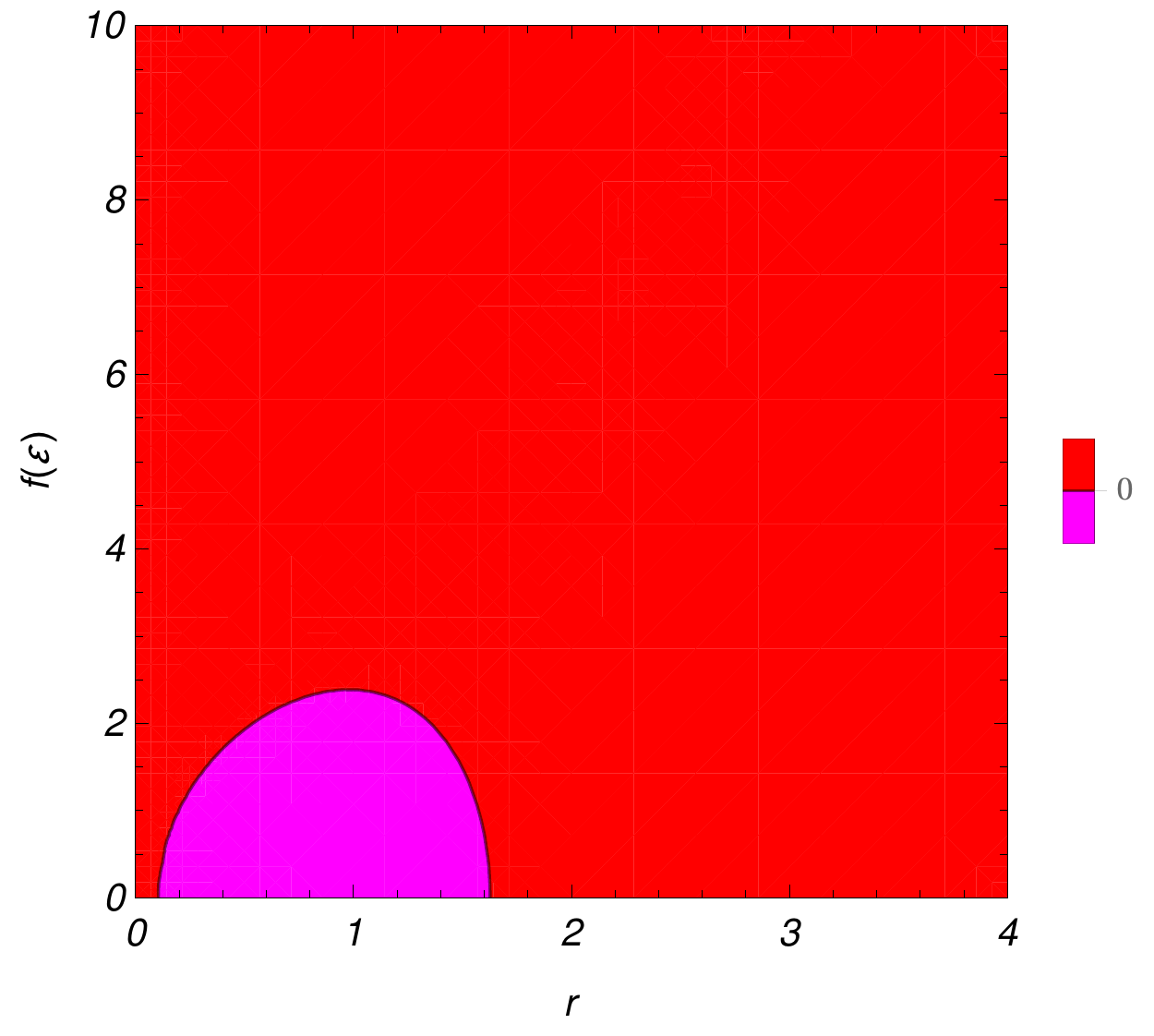}}  
\subfloat[$q_{M}=0.5$, $k=1$, $f(\varepsilon)=1.1$, $m=3$ and $\Lambda=-1$.]{
        \includegraphics[width=0.25\textwidth]{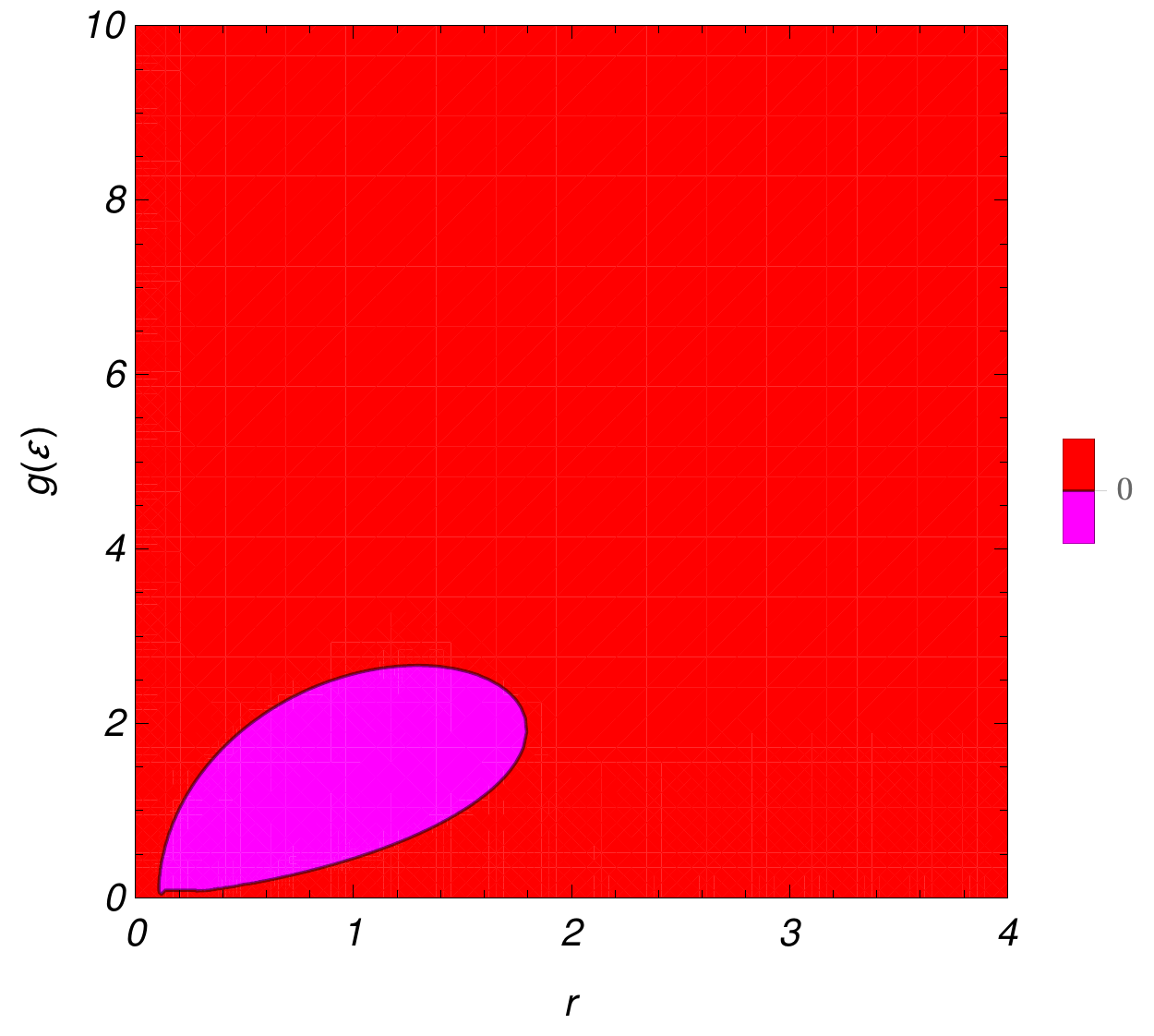}}  
\subfloat[$k=1$, $g(\varepsilon)=f(\varepsilon)=1.1$, $m=3$ and $\Lambda=-1$.]{
        \includegraphics[width=0.25\textwidth]{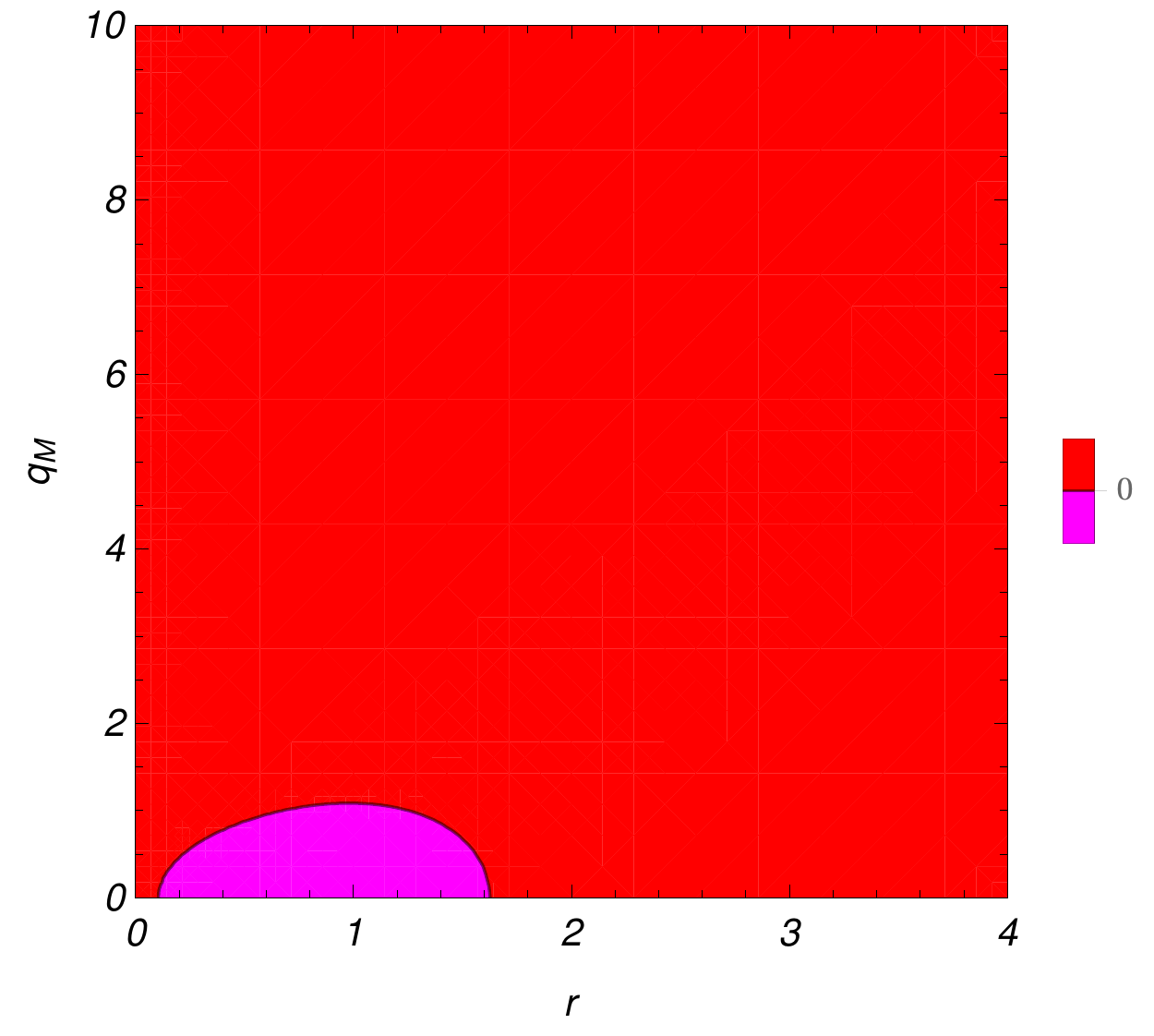}} \newline
\caption{Variation of $\protect\psi(r)$ as a function of different
parameters for $q_{E}=0.5$}
\label{Fig2}
\end{figure}

Apparently, under certain conditions, system could enjoy the existence of
one of the following cases: two roots, one root and absence of root. The
absence of root (naked singularity) takes place when the argument under
square root functions is not positive valued. In order to understand the
effects of different parameters on the number of the roots, we have plotted
Fig. \ref{Fig2}.

Evidently, for geometrical mass, there is a lower bound, $m_{bound}$ where
for $m<m_{bound}$, metric function does not have a root and the singularity
is naked. For $m=m_{bound}$ and $m>m_{bound}$ system will have one and two
roots, respectively. The smaller (larger) root is a decreasing (increasing)
function of geometrical mass. The effects of $f(\varepsilon)$ and $q_{M}$ on
number of roots is similar. There is an upper bound on these parameters
below which the system has roots. The larger (smaller)
root is a decreasing (increasing) function of $f(\varepsilon)$ and $q_{M}
$.

For $g(\varepsilon)$, we have also an upper bound. But interestingly, while
the smaller root is always an increasing function of $g(\varepsilon)$, the
larger root is first an increasing function of this parameter and then it
becomes a decreasing function of it. The reason for this behavior is due to
special coupling of $g(\varepsilon)$ with cosmological constant and magnetic
charge (see Eq. \ref{metric function}). For specific range of $g(\varepsilon)$%
, the $\Lambda$ term is dominant in behavior of the larger root, while for
the other range, the effects of magnetic charge term kicks in and becomes
dominant. It should be noted that in the absence of the magnetic charge,
such behavior would not be observed for variation of $g(\varepsilon)$.
Therefore, this specific behavior for this energy function is due to the
magnetic nature of the solutions.

So far, we established the fact that our solutions have an inherently
irremovable singularity which could be covered by at least one horizon. This
confirms that our solutions could be interpreted as black holes.

\section{Thermodynamics} \label{Thermodynamics} 

In this section, we investigate the thermodynamical
properties of our solutions. The main issue is to see how the presence of
rainbow functions and their specific coupling with other parameters would
modify the thermodynamical behavior of the solutions. Before we go on, we
introduce a new notion for our solutions. We consider the negative
cosmological constant to be a thermodynamical quantity known as pressure.
Such a proposal was used by Kubiznak and Mann in Ref. \cite%
{Kubiznak} to show the presence of van der Waals like behavior for black holes and since then, it has been widely used in literature. Therefore, from now
on, we replace the cosmological constant with pressure using the following
relation 
\begin{equation*}
\Lambda=-8\pi P.
\end{equation*}

It should be noted that replacing the cosmological constant with pressure
results into modification of the first law of the black hole thermodynamics
for these black holes as \cite{HRP,Dutta} 
\begin{equation}
dM=TdS+VdP+\Phi_{E}dQ_{E}+\Phi_{M}dQ_{M},  \label{first law}
\end{equation}
where $Q_{E}$ and $Q_{M}$ are total electric and magnetic charges, $\Phi_{E}$
and $\Phi_{M}$ are electric and magnetic potentials, and $V$ is the thermodynamic volume
of the black hole.

In what follows, first, we extract thermodynamical quantities and study
their properties. Then, we investigate the possibility of van der Waals like
phase transition for these black holes. Later on, we obtain conditions for
thermal stability/instability. It should be noted that all of the
thermodynamical quantities are calculated on the outer horizon of the black
hole, $r_{+}$.

\subsection{Thermodynamical quantities} \label{Thermodynamical quantities}

\subsubsection{Temperature} \label{Temperature}

The first thermodynamical quantity of interest is temperature. According to the pioneering work of Hawking, the temperature of black holes is related to surface gravity by the following relation \cite{HawkingT}
\begin{equation*}
T=\frac{\kappa }{2\pi }.
\end{equation*}

Therefore, the task of obtaining the temperature is limited to calculation
of the surface gravity. The surface gravity is given by 
\begin{equation*}
\kappa =\sqrt{-\frac{1}{2}\left( \nabla _{\mu }\chi _{\nu }\right) \left(
\nabla ^{\mu }\chi ^{\nu }\right) },
\end{equation*}
where $\chi ^{\nu }$ is the time-like Killing vector. The metric employed in this paper %
\eqref{metric} admits a Killing vector for temporal coordinate in the form
of $\chi =\partial _{t}$. Using this vector, temperature will be calculated
as 
\begin{equation}
T =\frac{g(\varepsilon) }{4 \pi f(\varepsilon)} \left.\frac{df\left( r\right) }{%
dr} \right\vert _{r=r_{+}}=\frac{8 \pi P r_{+}^4-g^2(\varepsilon)
\left(f^2(\varepsilon) q_{E}^2+g^2(\varepsilon) q_{M}^2-k r_{+}^2\right)}{4
\pi f(\varepsilon) g(\varepsilon) r_{+}^3},  \label{temperature}
\end{equation}

\begin{figure}[!htb]
\centering
\subfloat[$g(\varepsilon)=f(\varepsilon)=1.1$.]{
        \includegraphics[width=0.25\textwidth]{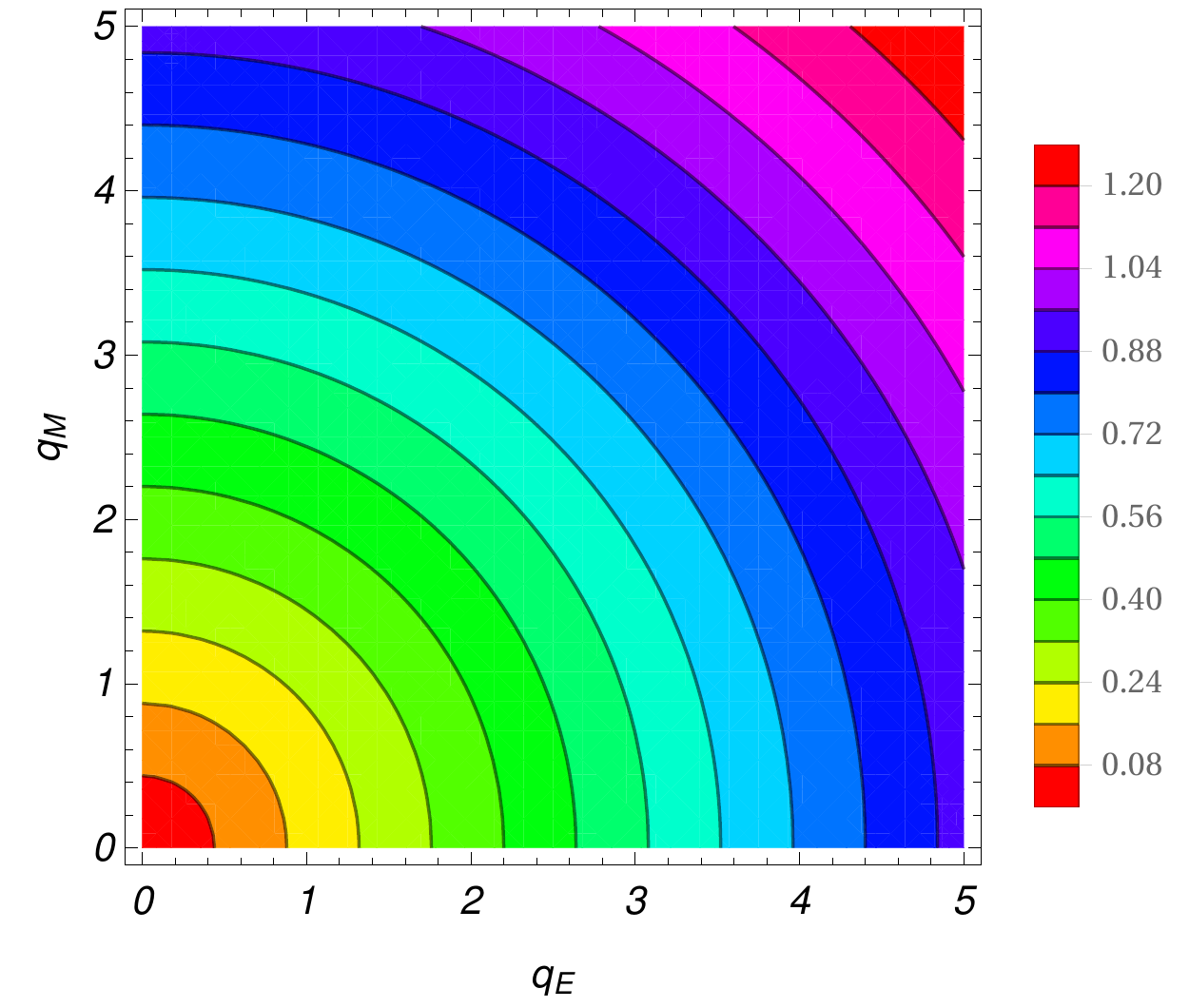}}  
\subfloat[$q_{M}=q_{E}=0.1$.]{
        \includegraphics[width=0.25\textwidth]{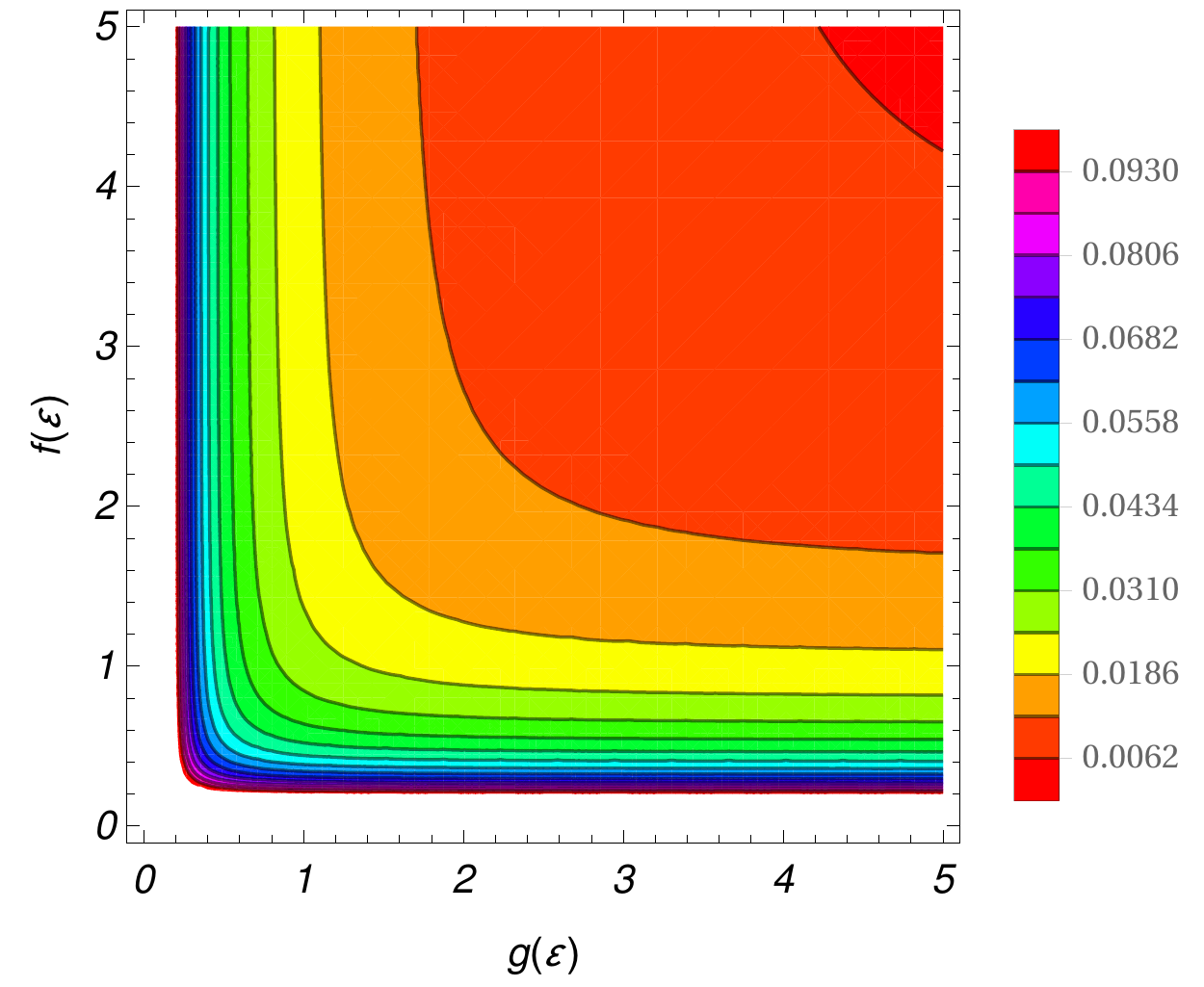}} \newline
\caption{Variation of the $T_{S}$ as a function of different parameters for $%
k=1$ and $P=0.1.$}
\label{Fig3}
\end{figure}

Evidently, the specific coupling between magnetic and electric charges with
energy functions that was observed in metric function is also present in the
temperature. Therefore, previous arguments regarding the effects of
gravity's rainbow in the metric function are also valid for the temperature.
Interestingly, if the event horizon satisfies the following condition 
\begin{equation}
f^2(\varepsilon) q_{E}^2+g^2(\varepsilon) q_{M}^2-k r_{+}^2=0,  \label{SP}
\end{equation}
the temperature reduces to 
\begin{equation}
T_{S} =\frac{2P}{f(\varepsilon) g(\varepsilon)}\sqrt{\frac{f^2(\varepsilon)
q_{E}^2+ g^2(\varepsilon) q_{m}^2}{k}},  \label{TSP}
\end{equation}
indicating a fixed temperature independent of horizon radius of the black
holes. Such temperature could only be real positive valued for black holes
with spherical horizon and it is linearly related to the pressure. In
addition, this temperature is a decreasing function of rainbow functions and
an increasing function of magnetic and electric charges. For more
clarification, we have plotted following diagrams (Fig. \ref{Fig3}) for this
specific temperature.

The value of horizon radius determines in which regime we are working. The
high energy limit of the temperature is 
\begin{equation*}
\lim_{r_{+}\longrightarrow 0 }T=-\frac{g^3(\varepsilon)
q_{M}^2+g(\varepsilon) f^2(\varepsilon) q_{E}^2}{4 \pi f(\varepsilon) r_{+}^3%
}+\frac{g(\varepsilon) k}{4 \pi f(\varepsilon) r_{+}} +O(r_{+}),
\end{equation*}
whereas the asymptotic behavior is described by the following relation 
\begin{equation*}
\lim_{r_{+}\longrightarrow \infty }T=\frac{2 P r_{+}}{ f(\varepsilon)
g(\varepsilon)}+O(\frac{1}{r_{+}}).
\end{equation*}

The high energy limit is governed by rainbow function, electric and magnetic
charges whereas the asymptotic behavior is determined by both the pressure
and rainbow functions. The difference in signature of these two limits
indicates that temperature has at least one root. Calculations show that
temperature could have up to two roots given by 
\begin{equation}
\left. r_{+}\right\vert _{T=0}=\sqrt{-\frac{g^2(\varepsilon) k}{16 \pi P }%
\pm \frac{g(\varepsilon) \sqrt{32 \pi P \left(f^2(\varepsilon)
q_{E}^2+g^2(\varepsilon) q_{M}^2\right)+g^2(\varepsilon) k^2}}{16 \pi P}}.
\label{rootT}
\end{equation}

For spherical and flat horizon black holes, only positive branch results
into real valued root for temperature. Therefore, for these two horizons,
temperature has only one root. For black holes with hyperbolic horizon,
although it seems that they could have two roots, their temperature has only
one root. This is because the second term in the root of the temperature has
higher value than the first term. Therefore, only the positive branch
results into real valued root for the temperature. Considering the number of
roots, high energy limit and asymptotic behavior, one can conclude that: the
temperature has a root. At this root, temperature changes sign from negative
to positive. In classical thermodynamic of black holes, positivity of
temperature is considered as one of the criteria for having physical
solutions. Therefore, before the temperature's root, solutions are
non-physical and the root of temperature is providing us with a lower bound.
Later (in section \ref{Stability of the solutions}), we will show that roots
of the temperature and heat capacity are identical and will show how these
roots will be modified as a function of different parameters.

The gravity's rainbow has noticeable effects on the temperature. In high
energy regime, rainbow functions affect the thermodynamical behavior of the
temperature differently compared to asymptotic behavior. In asymptotic
limit, the rainbow functions have identical effects on
temperature. In high energy limit, the situation is different. This
difference in behavior is due to the presence of the magnetic charge and its
coupling with rainbow functions which is different from the electric charge.
In section \ref{van der Waals like Behavior}, we will point out that the
temperature could exhibit a van der Waals like behavior and phase
transition. It should be noted that obtained temperature coincides with $(%
\frac{dM}{dS})_{Q_{E},Q_{M},P}$ taken from the first law of black hole
thermodynamics \eqref{first law}.

\subsubsection{Entropy} \label{Entropy} 

The method for obtaining the entropy of black holes depends
on gravities under consideration and topological structure of the black
holes. For topological black holes in the presence of Einstein gravity, the
area law proposed by Hawking and Bekenstein is a valid method for
calculating entropy \cite{Beckenstein,Hawking}. Therefore, considering the
metric \eqref{metric}, the entropy can be calculated by 
\begin{equation}
S=\frac{1}{4}\int d^2x\sqrt{\gamma}=\frac{1}{4g^2(\varepsilon)} r_{+}^2,
\label{entropy}
\end{equation}
in which $\gamma_{ab}$ is the induced metric on the $2$-dimensional
boundary. The entropy is a smooth increasing function of the horizon radius
and a decreasing function of only one of the rainbow functions. The
dependency of entropy on only one of the rainbow functions is due to
coupling of it with spatial coordinates (see Eq. (\ref{metric})). If one
considers the condition (\ref{SP}) (which resulted into a fixed temperature (%
\ref{TSP})), a fixed entropy will be obtained as 
\begin{equation}
S_{S} =\frac{f^2(\varepsilon) q_{E}^2+ g^2(\varepsilon) q_{m}^2}{4 k
g^2(\varepsilon)},  \label{SSP}
\end{equation}
which has the following interesting properties. First of all, the
non-negative entropy could only be obtained for black holes with spherical
horizon ($k>0$). In addition, this entropy is expressed in term of the matter field
and rainbow functions. It is an increasing function of the $f(\varepsilon)$,
electric and magnetic charges and a decreasing function of $g(\varepsilon)$.

\subsubsection{Mass} \label{Mass} 

The mass of black holes, without replacing cosmological
constant with pressure in it, is depicted as internal energy. By inserting
the cosmological constant as pressure, the role of the mass is
changed to enthalpy. There are several
methods for calculating the mass. One of them is employing Hamiltonian
approach. The other method is using counterterm method. Both of these
methods result into a general form for mass expressed as 
\begin{equation*}
M = \frac{f(\varepsilon)}{8 \pi g(\varepsilon) } m.
\end{equation*}

Evaluating the metric function on horizon ($f(r=r_{+})=0$), solving it with
respect to geometrical mass and replacing cosmological constant with
pressure result into the following relation for total mass of these black holes 
\begin{equation}
M =\frac{8 \pi P r_{+}^4+3 g^2 \left(f^2(\varepsilon)
q_{E}^2+g^2(\varepsilon) q_{M}^2+k r_{+}^2\right)}{24 \pi f(\varepsilon)
g^3(\varepsilon) r_{+}}.  \label{mass}
\end{equation}

Evidently, the mass is an increasing function of the pressure, rainbow
functions, electric and magnetic charges. The only term that could have
negative effect on the value of the mass is topological term. To understand
the effects of different parameters on the mass, we study the high energy and
asymptotic limits. The high energy limit of the mass is 
\begin{equation*}
\lim_{r_{+}\longrightarrow 0 }M=\frac{g^2(\varepsilon) q_{M}^2+ f^2(\varepsilon)
q_{E}^2}{8 \pi f(\varepsilon) g(\varepsilon) r_{+}}+\frac{k r_{+}}{8\pi
f(\varepsilon) g(\varepsilon)} +O(r_{+}),
\end{equation*}
whereas the asymptotic behavior is described by the following relation 
\begin{equation*}
\lim_{r_{+}\longrightarrow \infty }M=\frac{ P r_{+}^3}{3 f(\varepsilon)
g^3(\varepsilon)}+\frac{k r_{+}}{8\pi f(\varepsilon) g(\varepsilon)}+O(\frac{%
1}{r_{+}}).
\end{equation*}

For small black holes, this is the matter field (electric and magnetic
charges) that has dominant effect on the behavior of mass. Contrary to what
was observed for temperature, the high energy limit of mass is positive valued
and an increasing function of the rainbow functions, electric and magnetic
charges. On the other hand, the asymptotic behavior of mass, similar to
temperature, is positive valued, dominated by pressure and it is a
decreasing function of rainbow functions. For medium black holes, the
dominant term in the behavior of mass is the topological term. This term
could be negative for hyperbolic horizon while it is positive for spherical
and horizon flat black holes. Therefore, the mass is positive valued without
any root for arbitrary horizon radius in case of spherical and horizon flat
black holes. Whereas, for hyperbolic black holes, mass could be negative
valued and acquires roots. The roots of mass are given by 
\begin{equation*}
\left. r_{+}\right\vert _{M=0}=\sqrt{-\frac{3 g^2(\varepsilon) k}{16 \pi P }%
\pm \frac{g(\varepsilon) \sqrt{9 g^2(\varepsilon) k^2-96 \pi P
\left(f^2(\varepsilon) q_{E}^2+g^2(\varepsilon) q_{M}^2\right)}}{16 \pi P}},
\end{equation*}
which are pretty similar to the roots obtained for temperature.
Therefore, the same arguments stated for having real valued roots for
temperature is also valid here. It should be noted that according to
the first law of the black hole thermodynamics (\ref{first law}), the roots
of temperature are actually where mass acquires extremum. Remembering that
temperature could acquire only one root, mass could have up to one extremum.
Considering the high energy limit and asymptotic behavior of the mass, one
can safely conclude that extremum is a minimum. This indicates that mass has a 
lower limit on values that it can take. In addition,
depending on the place of this minimum, mass could have up to two roots with
a region of negativity. To show this, we have plotted Fig. \ref{Fig3}.

\begin{figure}[!htb]
\centering
\subfloat[$q_{M}=0.1$ and $g(\varepsilon)=f(\varepsilon)=1.1$.]{
        \includegraphics[width=0.25\textwidth]{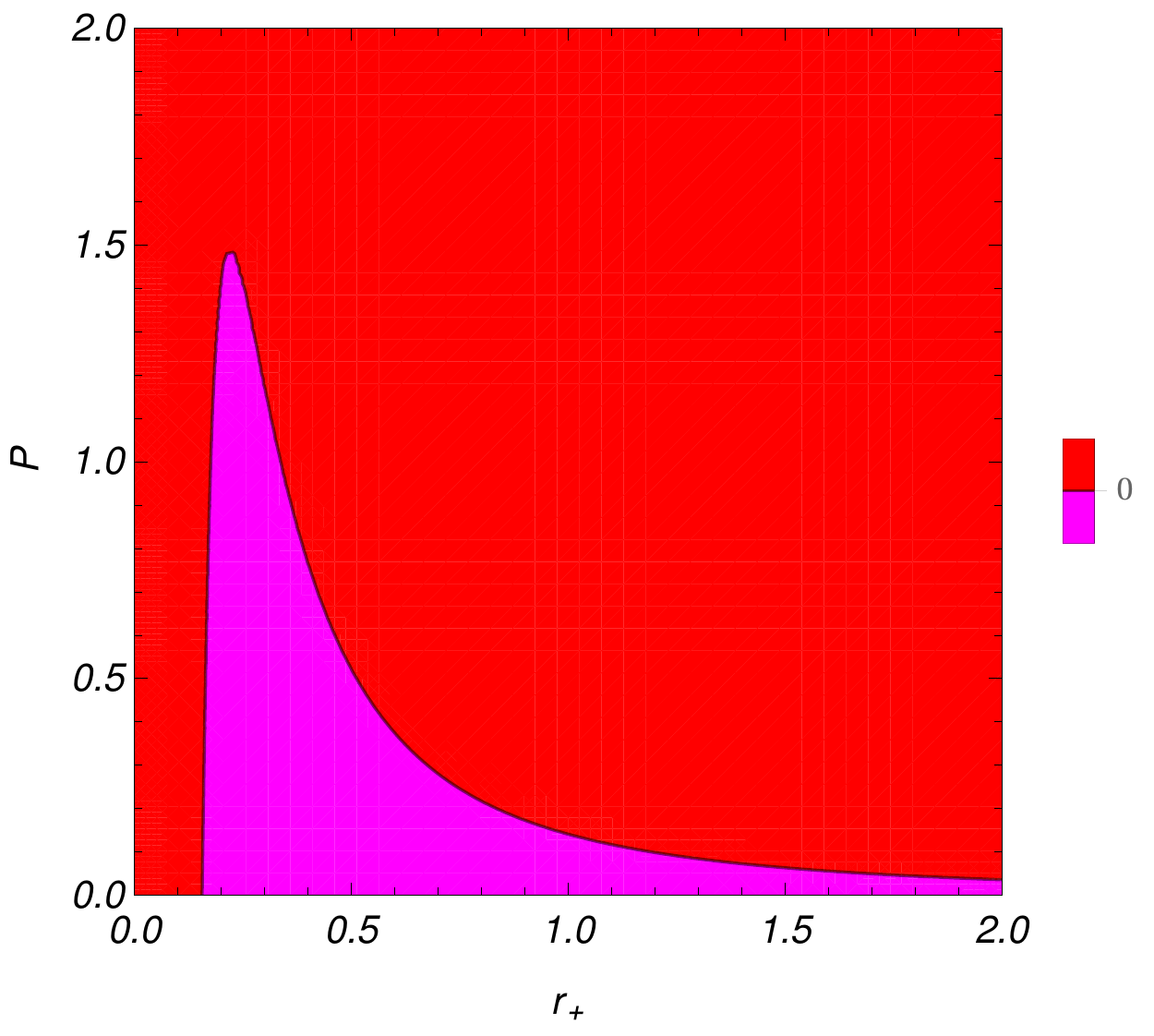}}  
\subfloat[$q_{M}=P=0.1$ and $g(\varepsilon)=1.1$.]{
        \includegraphics[width=0.25\textwidth]{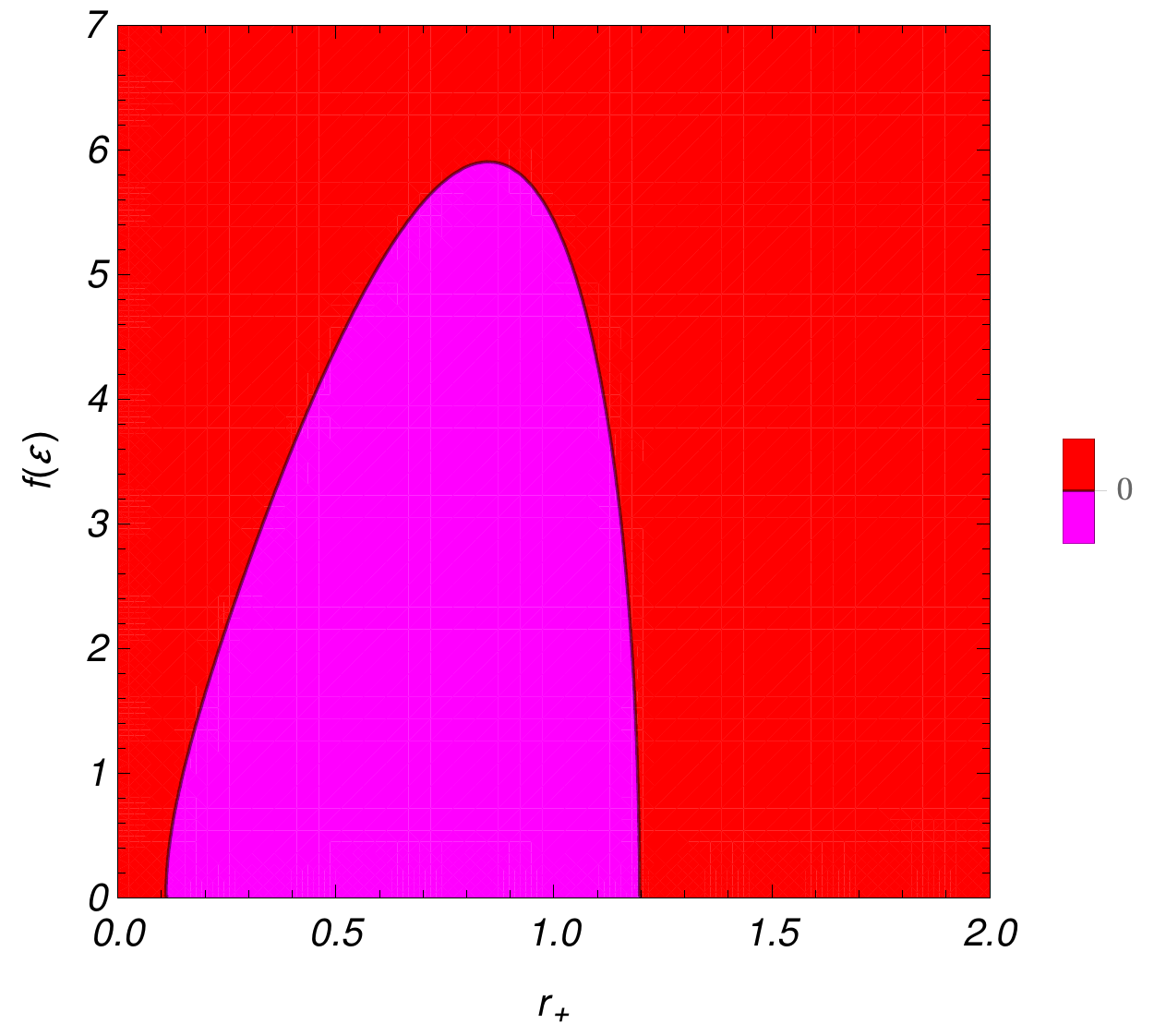}}  
\subfloat[$q_{M}=P=0.1$ and $f(\varepsilon)=1.1$.]{
        \includegraphics[width=0.25\textwidth]{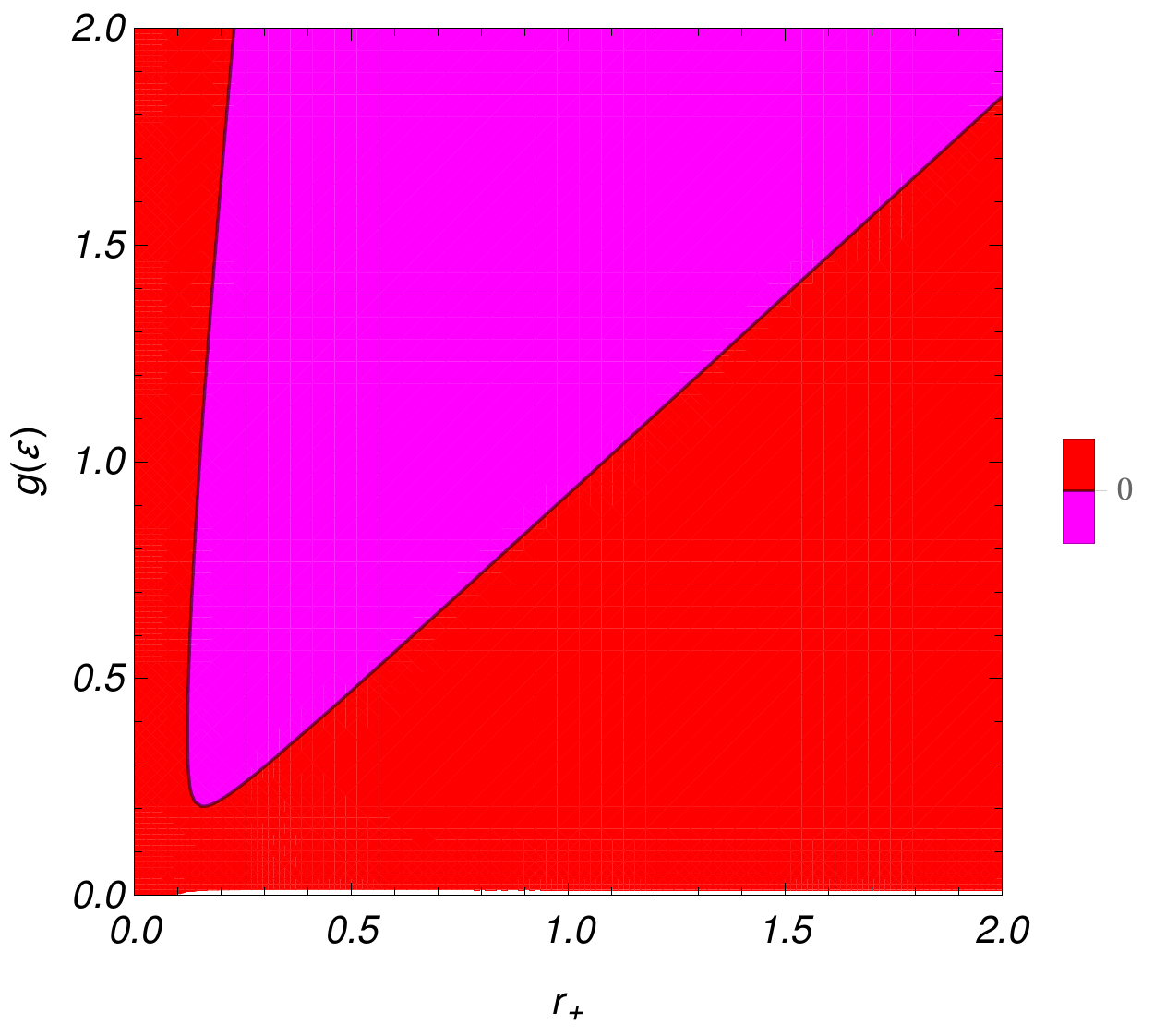}}  
\subfloat[$P=0.1$ and $g(\varepsilon)=f(\varepsilon)=1.1$.]{
        \includegraphics[width=0.25\textwidth]{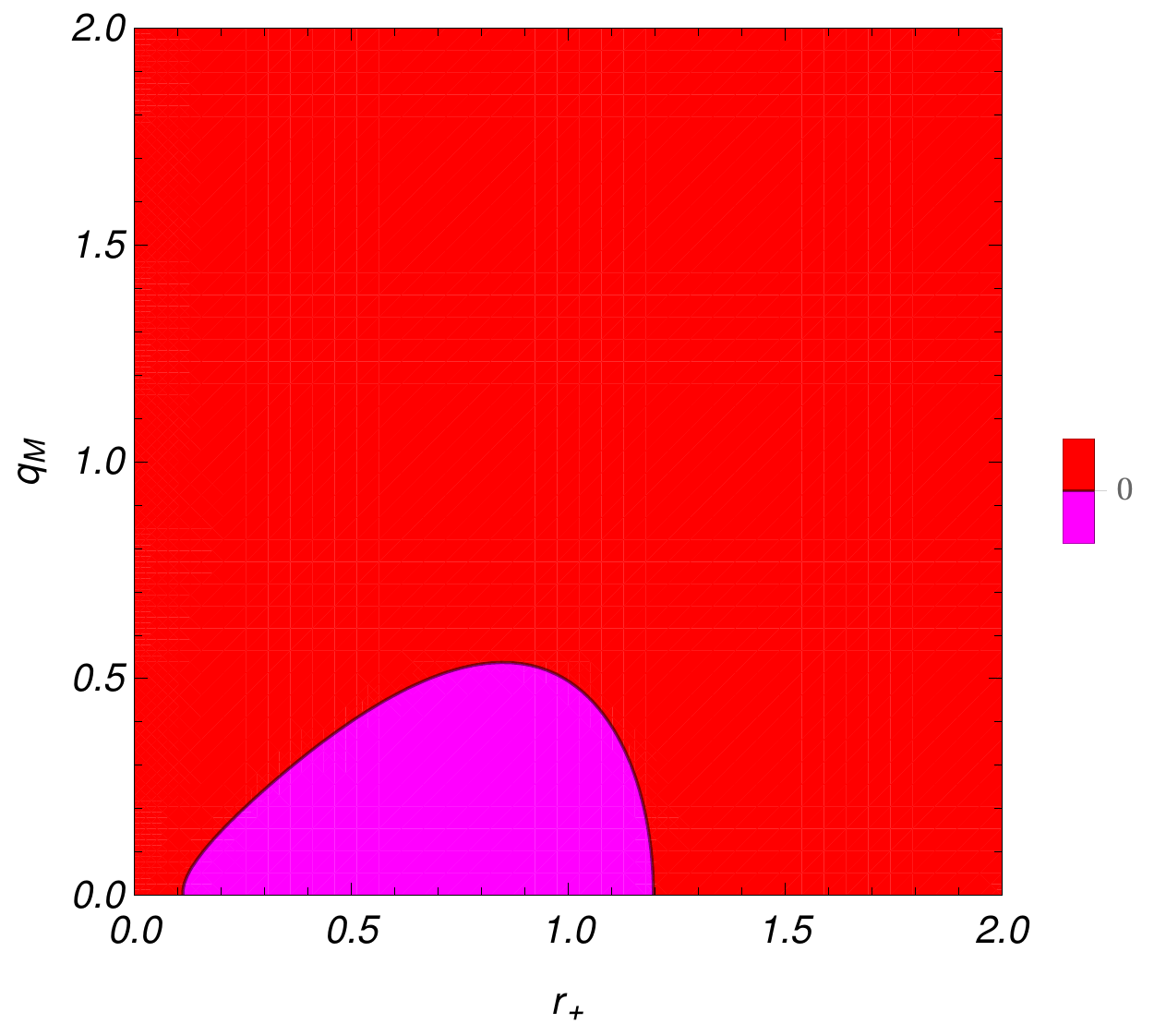}} \newline
\caption{Variation of the $M$ as a function of different parameters for $%
q_{E}=0.1$ and $k=-1$}
\label{Fig4}
\end{figure}

If we consider the positivity of mass as a condition for having physical
solutions, one can conclude that for black holes with hyperbolic horizon,
the mass is bounded. In other words, for specific regions, no physical
solutions exist for black holes due to negativity of the mass. According to
Fig. \ref{Fig4}, the region of negativity for mass is an increasing function
of the $g(\varepsilon)$ while it is a decreasing function of the $%
f(\varepsilon)$, pressure and magnetic charge. As one can see, both smaller
and larger roots of mass are increasing functions of $g(\varepsilon)$.
Whereas, larger (smaller) root is a decreasing (increasing) function of $%
f(\varepsilon)$, pressure and magnetic charge. It should be noted that
energy functions have opposite effective behavior on mass but their
effectiveness is different. One can find that the effect of $g(\varepsilon)$ on mass is more significant than $f(\varepsilon)$. Once more, we
remind the reader that $g(\varepsilon)$ is coupled with spatial coordinates
while $f(\varepsilon)$ is coupled with temporal coordinate. Considering
this, one can also state that mass is more affected by spatial coordinate
comparing to temporal coordinate of the metric.

\subsubsection{Pressure} \label{Pressure} 

Our next thermodynamical quantity of interest is pressure.
Replacing the cosmological constant with pressure changes the type of the
relation which was obtained for pressure. After replacement, this relation
becomes actually the equation of state. Therefore, using the temperature %
\eqref{temperature}, one can obtain the pressure in the following form 
\begin{equation}
P =\frac{4 \pi g(\varepsilon) f(\varepsilon) r_{+}^3 T+g^2(\varepsilon)
\left(f^2(\varepsilon) q_{E}^2+g^2(\varepsilon) q_{M}^2-k r_{+}^2\right)}{8
\pi r_{+}^4}.  \label{pressure}
\end{equation}

The first noticeable issue is the fact that there is no energy function in
the denominator of this relation for pressure. This is in contrast of what was observed for previously calculated thermodynamical quantities
(see Eqs. \eqref{temperature}, \eqref{entropy} and \eqref{mass}). Therefore,
if black holes have hyperbolic horizon, the pressure will be an increasing
function of the rainbow functions, temperature, electric and magnetic
charges. For spherical black holes, the topological term in pressure
negatively affects it. To have a better understanding of the pressure's
properties, we investigate the high energy limit given by 
\begin{equation*}
\lim_{r_{+}\longrightarrow 0 }P=\frac{g^4(\varepsilon) q_{M}^2+ g^2(\varepsilon)
f^2(\varepsilon) q_{E}^2}{8 \pi r_{+}^4}-\frac{k g^2(\varepsilon)}{8\pi
r_{+}^2} +O(r_{+}),
\end{equation*}
and asymptotic behavior obtained by 
\begin{equation*}
\lim_{r_{+}\longrightarrow \infty }P=\frac{ g(\varepsilon) f(\varepsilon) T}{2
r_{+}}-\frac{k g^2(\varepsilon)}{8\pi r_{+}^2}+O(\frac{1}{r_{+}}).
\end{equation*}

Evidently, the dominant factors for small, medium and large black holes are
matter field (electric and magnetic charges), topological term and
temperature, respectively. The topological term for hyperbolic and horizon
flat black holes is positive. Therefore, in these two cases, pressure is
positive everywhere and it has no root. For spherical black holes,
since the topological term is negative, the pressure could acquire root
given by 
\begin{equation*}
\left. r_{+}\right\vert _{P=0}=\left\{ 
\begin{array}{c}
\frac{g(\varepsilon)^2 k^2+a_{3}^{2/3}+g(\varepsilon) k \sqrt[3]{a_{3}}}{12 \sqrt[3]{a_{3}} \pi  f(\varepsilon) T}, \\ 
\\ 
\frac{4 g(\varepsilon) k \sqrt[3]{a_{3}} \pm 2 i \left(\sqrt{3} \pm i\right) g(\varepsilon)^2 k^2-2 \left(1 \pm i \sqrt{3}\right) a_{3}^{2/3}}{48 \sqrt[3]{a_{3}} \pi  f(\varepsilon) T},%
\end{array}%
\right.
\end{equation*}
where
\begin{equation*}
a_{3}=g(\varepsilon)^3 k^3-216 \pi ^2 f(\varepsilon)^2 g(\varepsilon) T^2 \left(f(\varepsilon)^2 q_{E}^2+g(\varepsilon)^2 q_{M}^2\right)+\frac{1}{2} \sqrt{\left(2 g(\varepsilon)^3 k^3-432 \pi ^2 f(\varepsilon)^2 g(\varepsilon) T^2 \left(f(\varepsilon)^2 q_{E}^2+g^2 q_{M}^2\right)\right)^2-4 g(\varepsilon)^6 k^6}.
\end{equation*}

\begin{figure}[!htb]
\centering
\subfloat[$q_{M}=0.1$ and $g(\varepsilon)=f(\varepsilon)=1.1$.]{
        \includegraphics[width=0.25\textwidth]{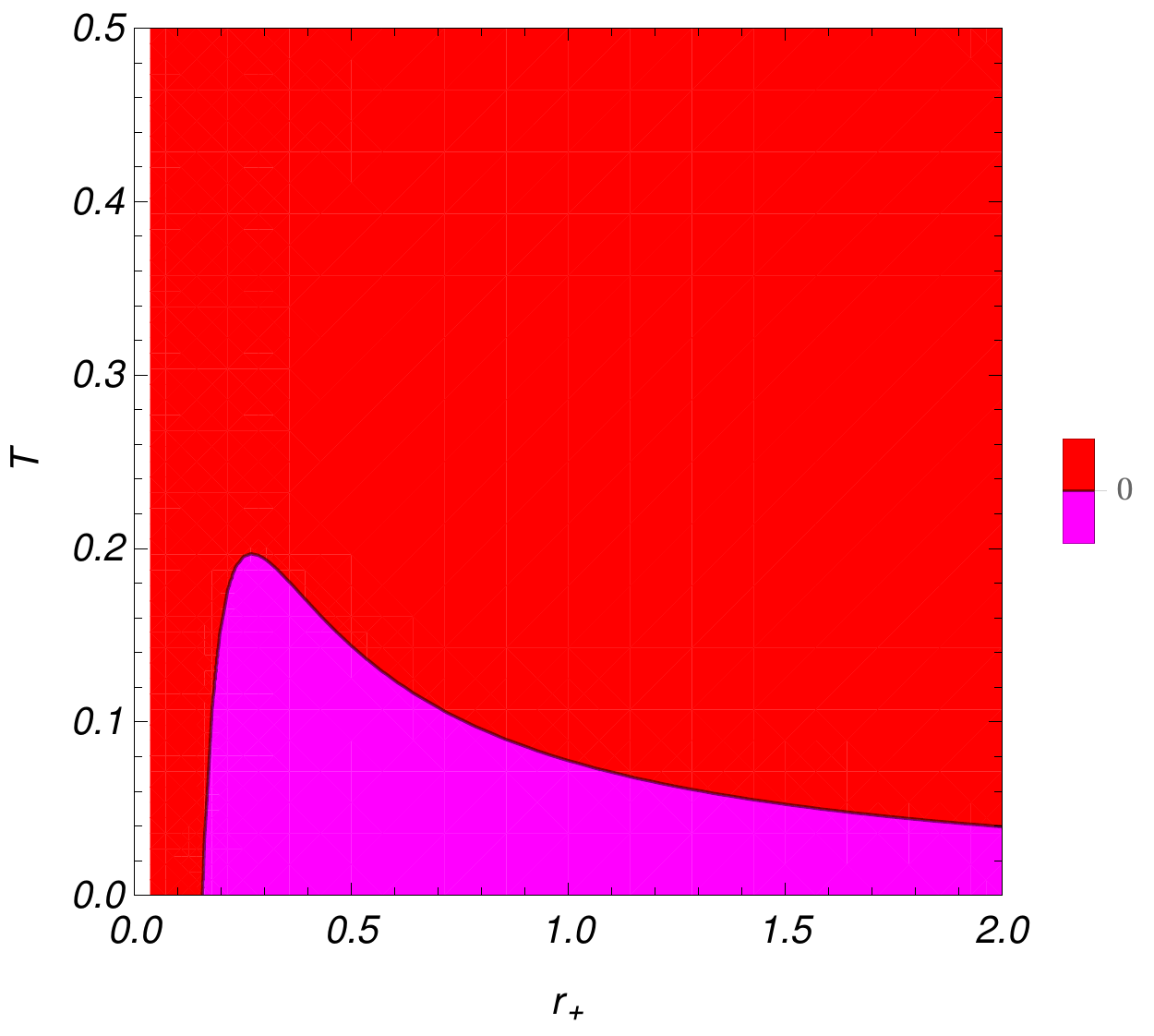}}  
\subfloat[$q_{M}=T=0.1$ and $g(\varepsilon)=1.1$.]{
        \includegraphics[width=0.25\textwidth]{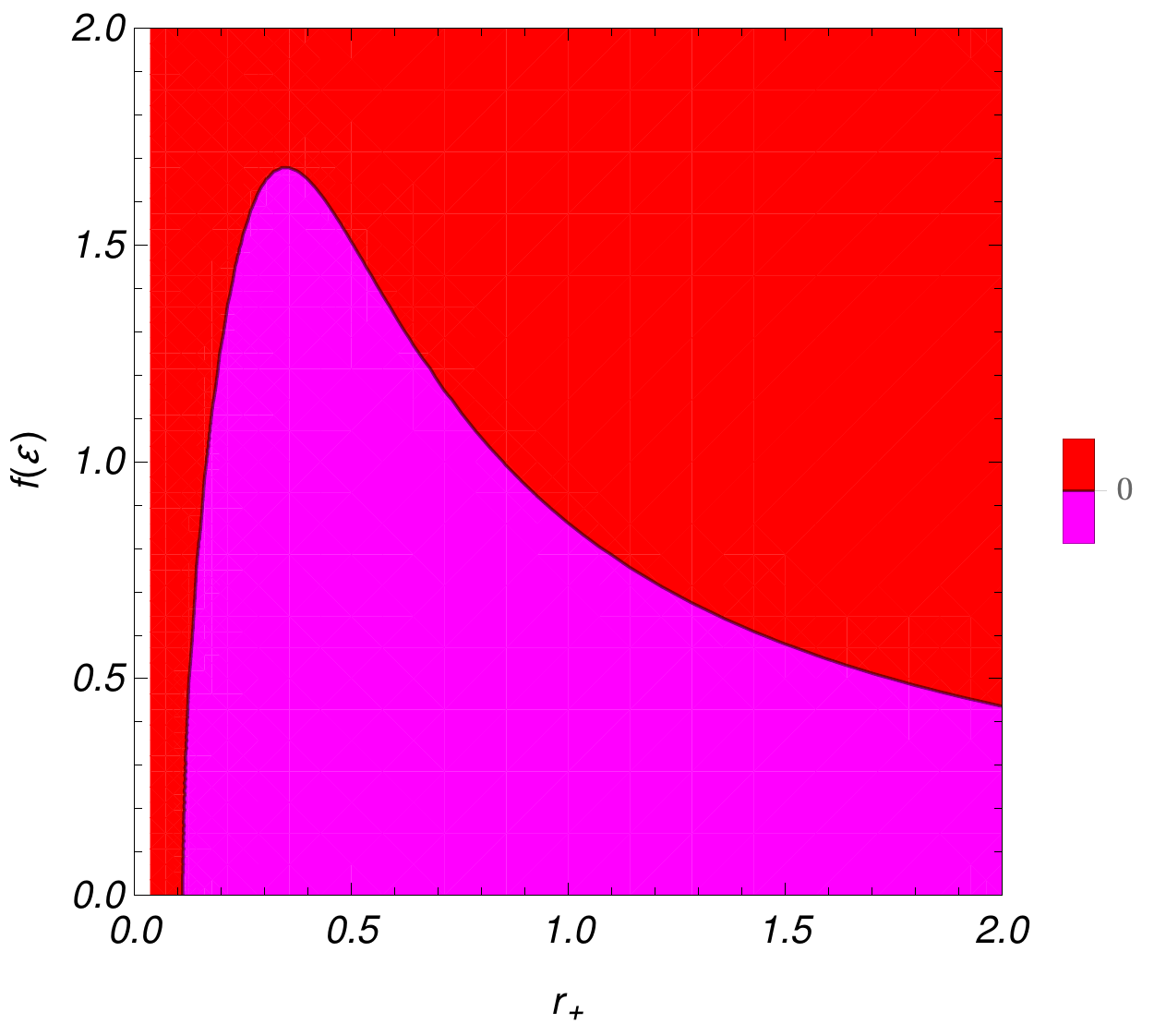}}  
\subfloat[$q_{M}=T=0.1$ and $f(\varepsilon)=1.1$.]{
        \includegraphics[width=0.25\textwidth]{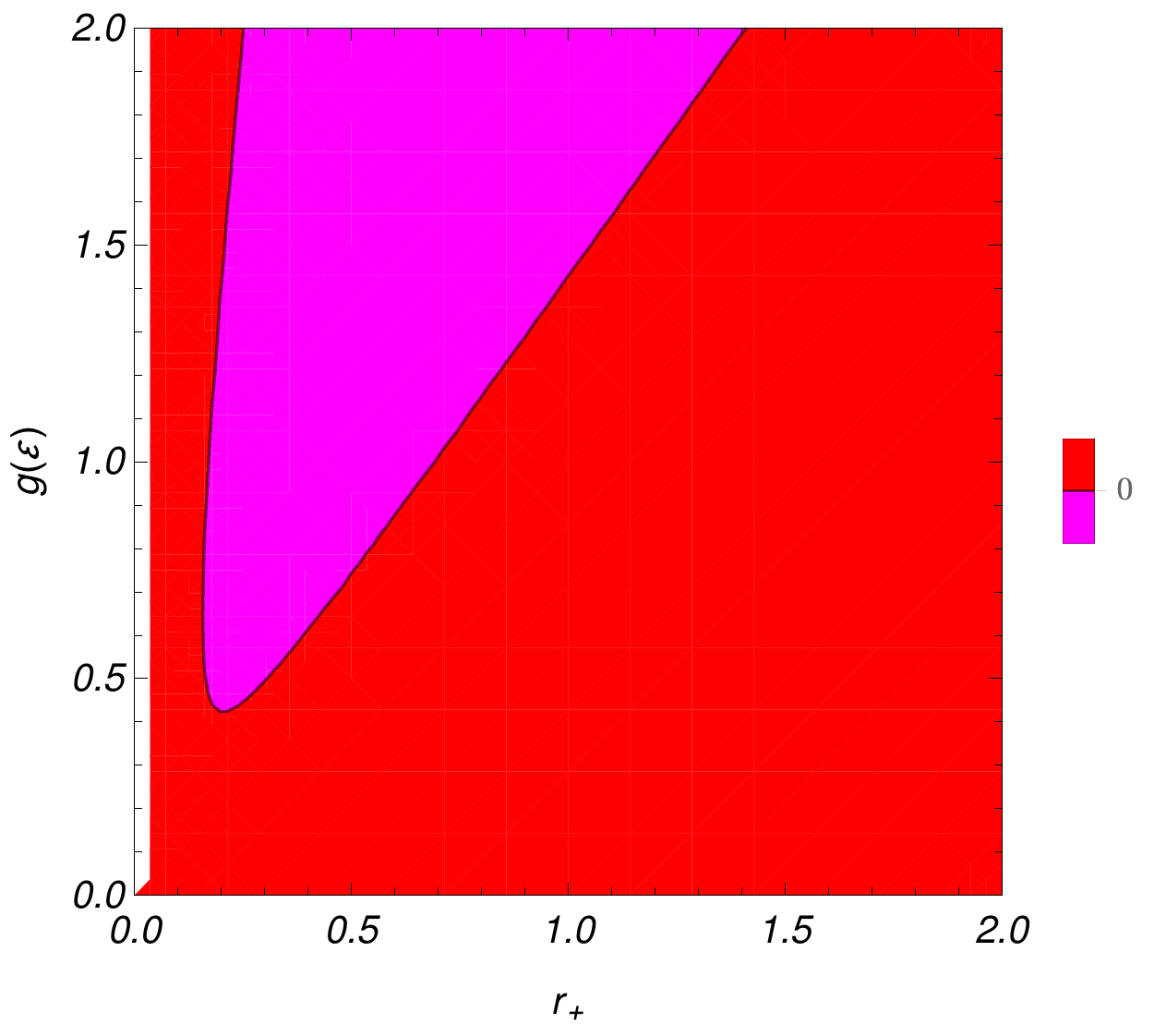}}  
\subfloat[$T=0.1$ and $g(\varepsilon)=f(\varepsilon)=1.1$.]{
        \includegraphics[width=0.25\textwidth]{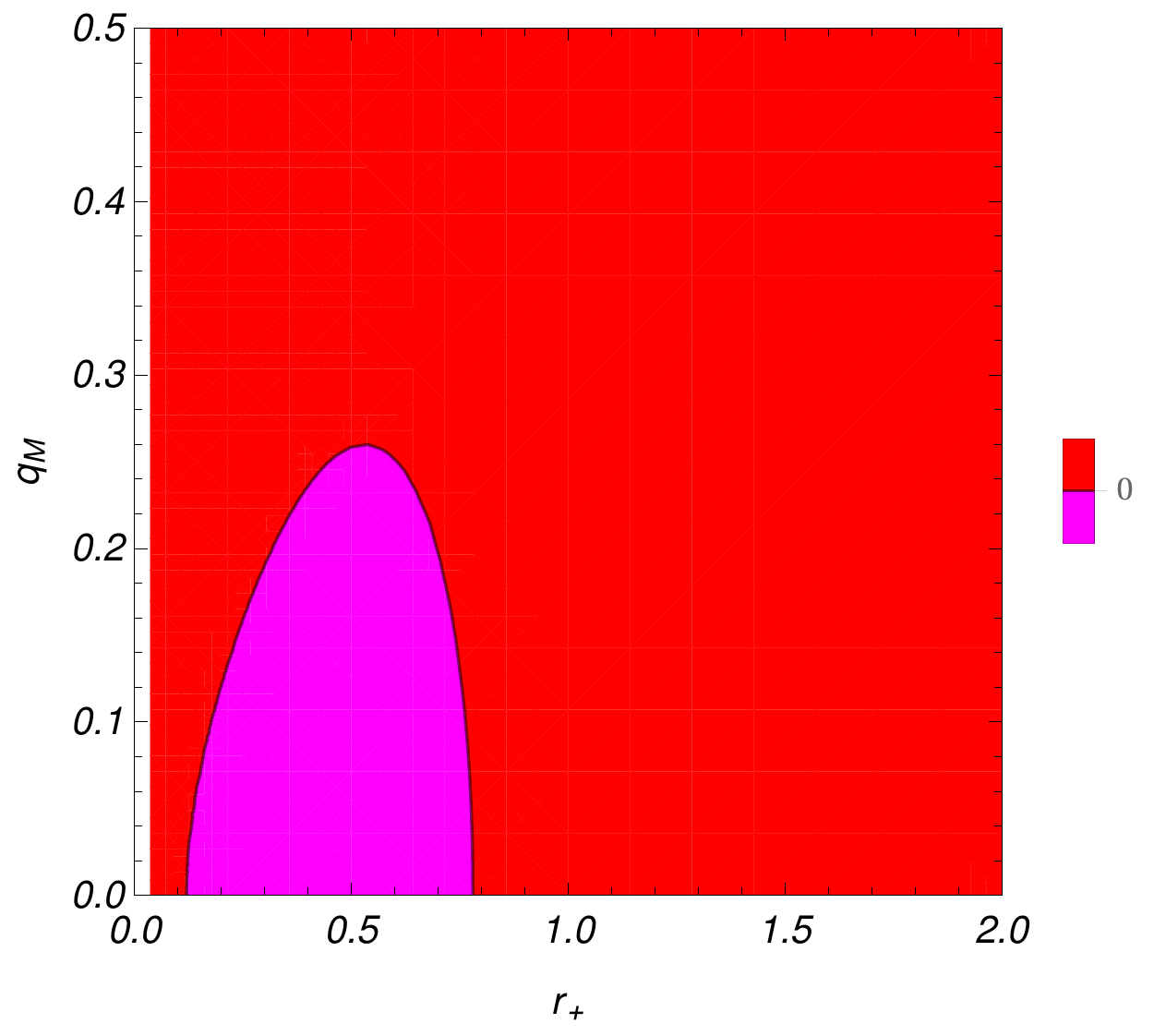}} \newline
\caption{Variation of the $P$ as a function of different parameters for $%
q_{E}=0.1$ and $k=1$.}
\label{Fig5}
\end{figure}

The effects of different parameters on the number of roots and
negativity/positivity of the pressure is investigated through Fig. \ref{Fig5}. 

The negative pressure is not physically acceptable in the context of black
hole thermodynamics. In addition, the plotted diagrams for pressure are
similar to those plotted for the mass (compare Fig. \ref{Fig5} with Fig. \ref%
{Fig4}). Therefore, the effects of different parameters on the behavior of the
pressure is similar to those discussed for the mass. It should be noted that
pressure enjoys up to two roots. Using the concept of first law of black
hole thermodynamics \eqref{first law} with the obtained mass \eqref{mass}, one
can obtain the volume of these black holes as 
\begin{equation}
V=(\frac{dM}{dP})_{q_{M},q_{E}}=\frac{r_{+}^3}{3 g^3(\varepsilon)
f(\varepsilon)},  \label{volume}
\end{equation}
which shows that total volume is linearly related to the horizon radius.
Therefore, one can use horizon radius instead of volume for describing
different phenomena and calculations. The volume is a decreasing function
of both rainbow functions. If we compare the obtained volume \eqref{volume}
with entropy \eqref{entropy}, we see that they are related with the
following relation 
\begin{equation*}
S=\frac{3 g^3(\varepsilon) f(\varepsilon)}{4 r_{+}}V.
\end{equation*}

In fact, this is expected. The reason is that entropy is calculated by using
the area of the black holes. Therefore, the obtained volume divide by $r_{+}$ has sound
relation with area of the black holes. The main issue is the dependency of
the volume on both of the rainbow functions. Both of them are in denominator
of the volume but the effect of rainbow function coupled with spatial
coordinate is more significant compared to the one coupled with temporal
coordinate.

According to conventional thermodynamic, the pressure is a decreasing
function of the volume. In any place that such principle is violated, that
region is not physically accessible for the thermodynamical system and phase
transition could take place. If we apply the thermodynamical principle to black holes, we expect the pressure to be a decreasing function
of the horizon radius (volume). The high energy limit (which diverges for $%
r_{+}=0$) and asymptotic behavior (which goes to zero for $r\longrightarrow
\infty$) show that the mentioned principle is satisfied for small and large
black holes. The remaining is the medium black holes. In order to see
whether pressure is a decreasing/increasing function of the horizon radius,
we calculate the first order derivation of the pressure with respect to
horizon radius 
\begin{equation*}
(\frac{dP}{dr_{+}})_{q_{M},q_{E},T}= -\frac{2 \pi g(\varepsilon)
f(\varepsilon) r_{+}^3 T+g^2(\varepsilon) \left(f^2(\varepsilon)
q_{E}^2+g^2(\varepsilon) q_{M}^2-k r_{+}^2\right)}{4 \pi r_{+}^5}.
\end{equation*}

\begin{figure}[!htb]
\centering
\subfloat[$q_{M}=0.1$ and $g(\varepsilon)=f(\varepsilon)=1.1$.]{
        \includegraphics[width=0.25\textwidth]{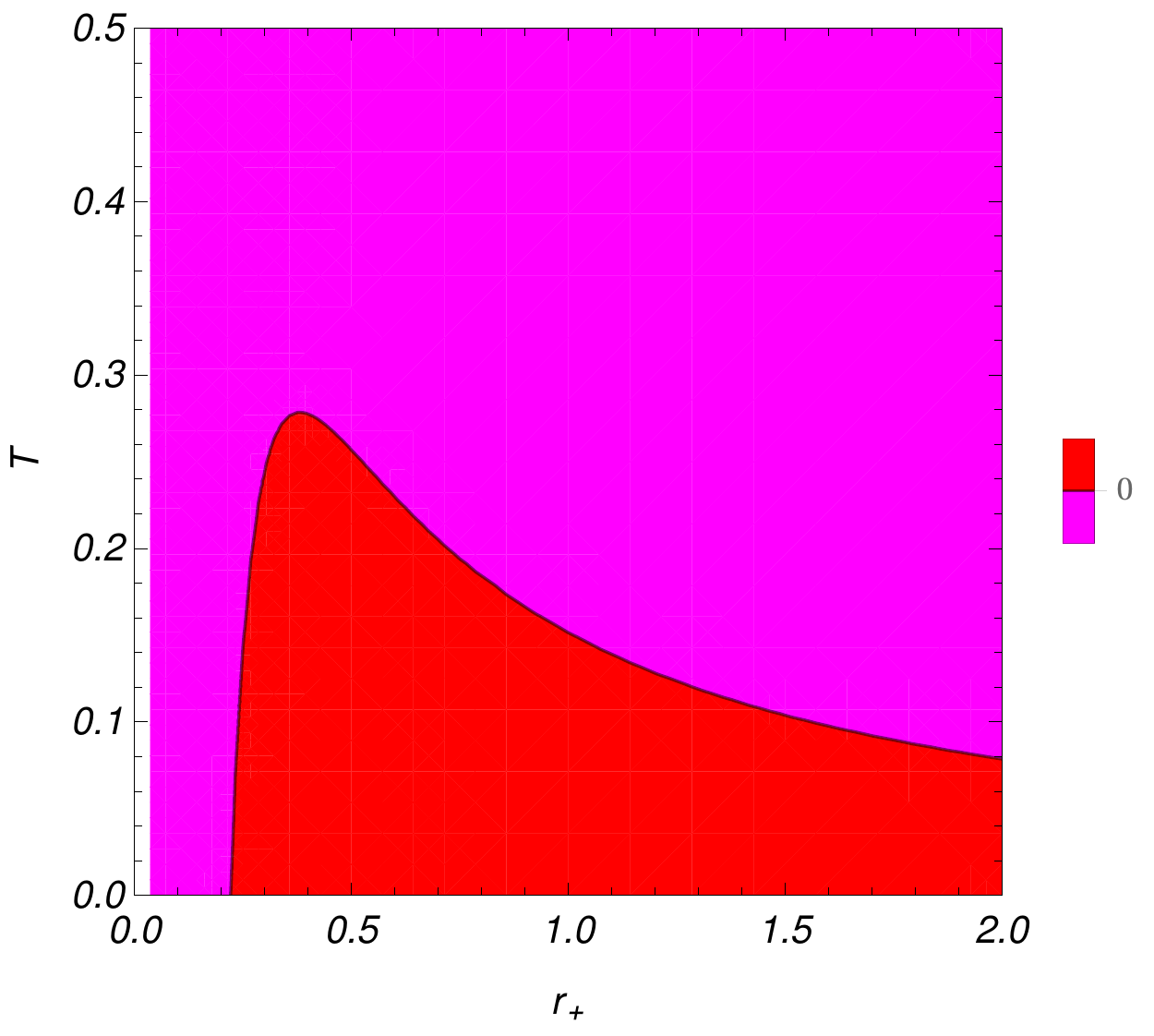}}  
\subfloat[$q_{M}=T=0.1$ and $g(\varepsilon)=1.1$.]{
        \includegraphics[width=0.25\textwidth]{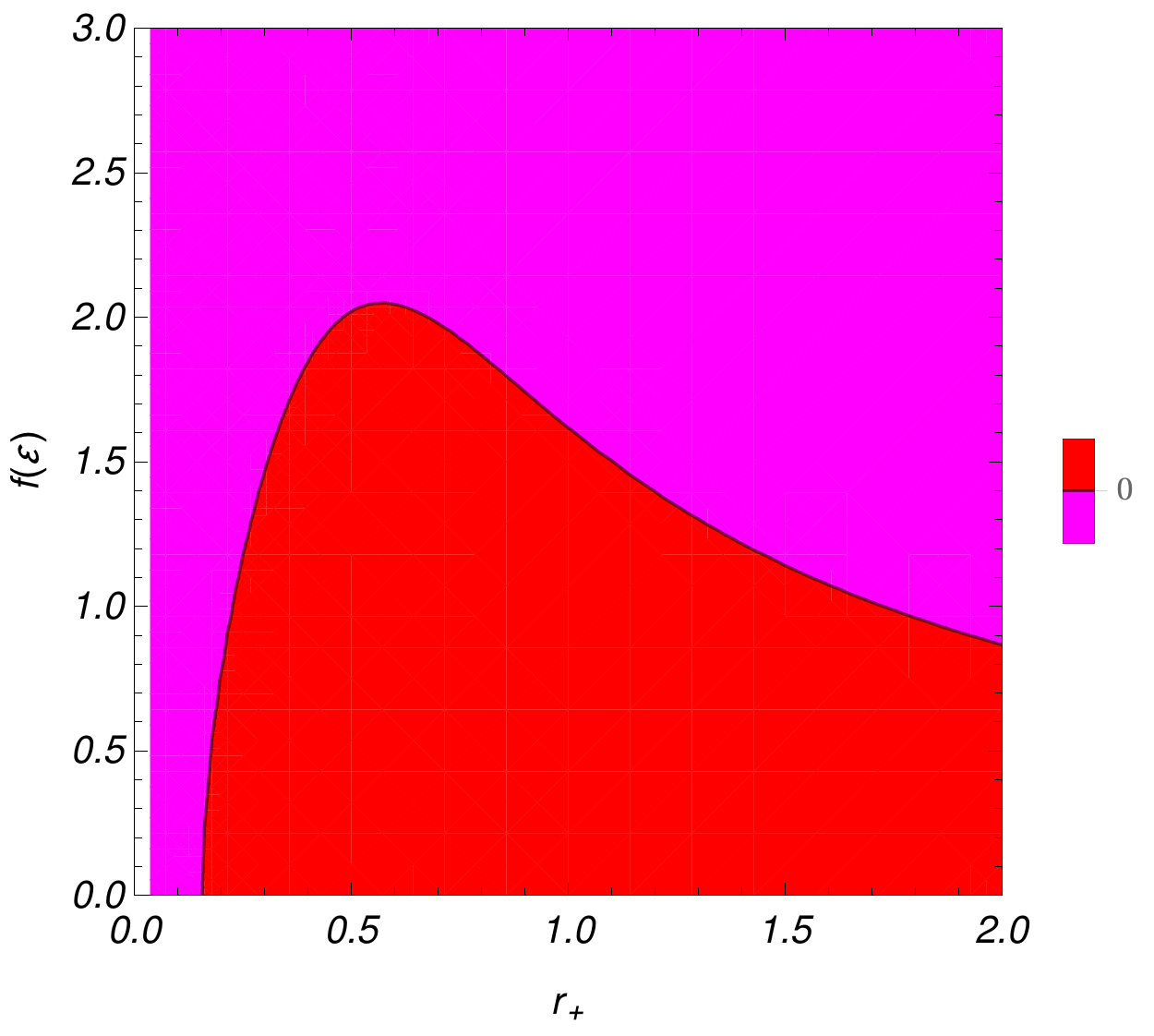}}  
\subfloat[$q_{M}=T=0.1$ and $f(\varepsilon)=1.1$.]{
        \includegraphics[width=0.25\textwidth]{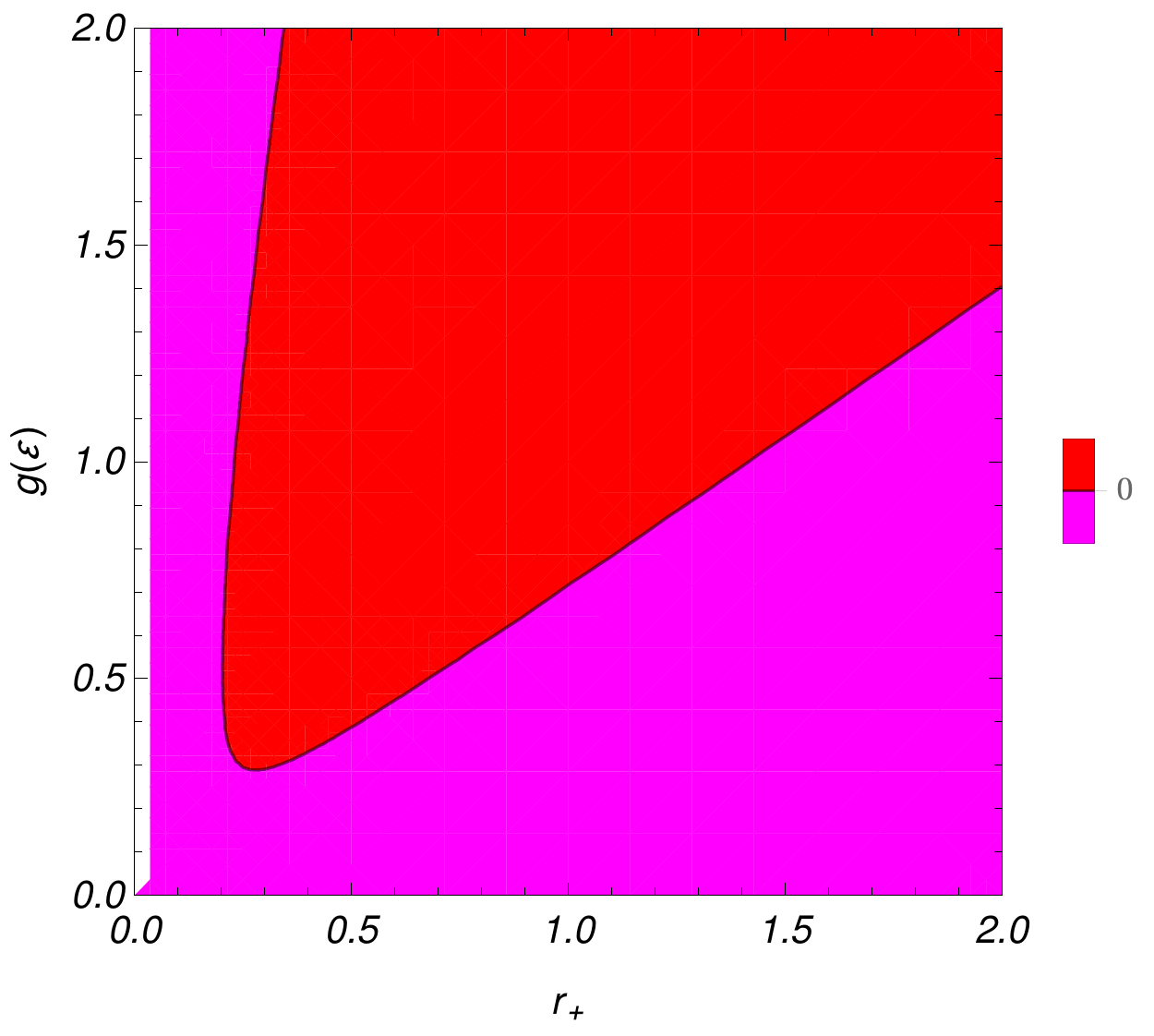}}  
\subfloat[$T=0.1$ and $g(\varepsilon)=f(\varepsilon)=1.1$.]{
        \includegraphics[width=0.25\textwidth]{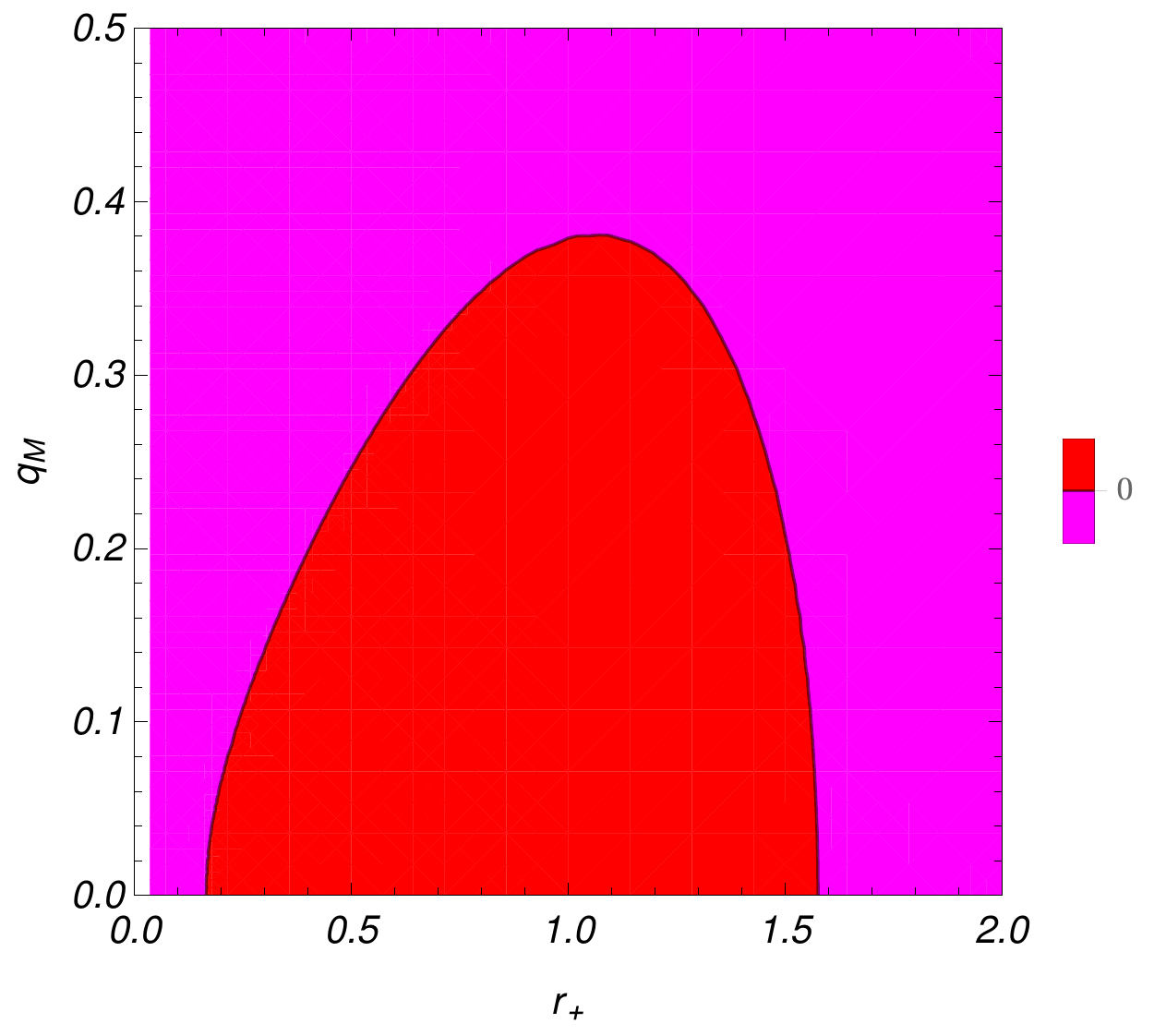}} \newline
\caption{Variation of the $(\frac{dP}{dr_{+}})_{q_{M},q_{E},T}$ as a
function of different parameters for $q_{E}=0.1$ and $k=1$.}
\label{Fig6}
\end{figure}

For hyperbolic and flat horizons, this expression is negative, therefore,
one can conclude that black holes for these horizons also admit the
mentioned principle. For the spherical black holes, the situation is
different. In this case, such expression could become positive valued
violating the principle. To express such possibility, we have plotted
diagrams in Fig. \ref{Fig6}. 

Accordingly, one can point to the existence of specific values for
temperature ($T_{S}$), magnetic charge ($q_{M_{S}}$) and $f(\varepsilon)$ ($%
f(\varepsilon)_{S}$) where for values smaller than them, pressure will
acquire a region of negativity. For $g(\varepsilon)$, in contrast, there is
special energy function, $g(\varepsilon)_{S}$, where for values larger than
that, pressure will have a region of negativity. That being said, one can
conclude that the region of negativity for pressure is a decreasing function
of temperature, magnetic charge and $f(\varepsilon)$ while it is an
increasing function of $g(\varepsilon)$. It should be noted that places
where signature of the $(\frac{dP}{dr_{+}})_{q_{M},q_{E},T}$ changes are
where pressure meets an extremum. Therefore, pressure can have up to two
extrema. Considering the high energy and asymptotic behaviors of the
pressure, one can state that the smaller extremum is a minimum and larger
one is a maximum. The region between such extrema is not physical, so
there is a phase transition taking place over this region. In section \ref%
{van der Waals like Behavior}, we will address the type of the phase
transition that was observed here and investigate the pressure in more
details.

\subsubsection{Gibbs free energy} \label{Gibbs free energy} 

The Gibbs free energy is a potential that can be
used to determine the equilibrium point at constant pressure and temperature
in the processes that a thermodynamical system goes through. At this
equilibrium point, the Gibbs free energy is minimized. Considering that we
are working in extended phase space, one can use the following relation to
obtain the Gibbs free energy for these black holes 
\begin{equation}
G=H-TS=M-TS=\frac{3g^2(\varepsilon) \left(3f^2(\varepsilon)
q_{E}^2+3g^2(\varepsilon) q_{M}^2+k r_{+}^2\right)-8 \pi P r_{+}^4}{48 \pi
f(\varepsilon) g^3(\varepsilon) r_{+}}.  \label{Gibbs}
\end{equation}

Contrary to temperature \eqref{temperature} and mass \eqref{mass}, Gibbs
free energy is a decreasing function of the pressure. In addition, it is an
increasing function of the electric and magnetic charges, and the effects of topological term depends on the horizon. For spherical and flat
horizons, the topological term positively affect the Gibbs free energy, while the
opposite stands for black holes with hyperbolic horizon.

The high energy limit of the Gibbs free energy is given by 
\begin{equation*}
\lim_{r_{+}\longrightarrow 0 }G=\frac{3(g^2(\varepsilon) q_{M}^2+
f^2(\varepsilon) q_{E}^2)}{16 \pi f(\varepsilon) g(\varepsilon) r_{+}}+\frac{%
k r_{+}}{16\pi f(\varepsilon) g(\varepsilon)} +O(r_{+}),
\end{equation*}
while the asymptotic behavior is 
\begin{equation*}
\lim_{r_{+}\longrightarrow \infty }G=-\frac{ P r_{+}^3}{6 f(\varepsilon)
g^3(\varepsilon)}+\frac{k r_{+}}{16\pi f(\varepsilon) g(\varepsilon)}+O(%
\frac{1}{r_{+}}).
\end{equation*}

The high energy limit of the Gibbs free energy is positive valued while the
asymptotic behavior is negative valued. Considering this, one can conclude
that at least one root exists for Gibbs free energy. The medium black holes
are governed by topological term. It is a matter of calculation to show that
Gibbs free energy could have up to two roots given by 
\begin{equation*}
\left. r_{+}\right\vert _{G=0}=\sqrt{\frac{3 g^2(\varepsilon) k}{16 \pi P }%
\pm \frac{3 g(\varepsilon) \sqrt{32 \pi P \left(f^2(\varepsilon)
q_{E}^2+g^2(\varepsilon) q_{M}^2\right)+g^2(\varepsilon) k^2}}{16 \pi P}}.
\end{equation*}

Considering the application of Gibbs free energy to determine the
equilibrium point, the first order derivation of it with respect to horizon
radius (volume) is obtained by 
\begin{equation*}
(\frac{dG}{dr_{+}})_{q_{M},q_{E},T,P}=-\frac{g^2(\varepsilon)
\left(3f^2(\varepsilon) q_{E}^2+3g^2(\varepsilon) q_{M}^2-k r_{+}^2\right)+8
\pi P r_{+}^4}{16 \pi f(\varepsilon) g^3(\varepsilon) r_{+}^2}.
\end{equation*}

The pressure, magnetic and electric charge terms enjoy the same sign whereas
the topological term has an opposite sign. Considering this, for flat and
hyperbolic horizons, the Gibbs free energy will be without any extremum,
whereas the spherical black holes could enjoy at least one extremum in their
Gibbs free energy. Considering this, one can extract extrema points in the
following form 
\begin{equation}
\left. r_{+}\right\vert _{(\frac{dG}{dr_{+}})=0}=\sqrt{\frac{%
g^2(\varepsilon) k}{16 \pi P }\pm \frac{ g(\varepsilon) \sqrt{%
g^2(\varepsilon) k^2-96 \pi P \left(f^2(\varepsilon)
q_{E}^2+g^2(\varepsilon) q_{M}^2\right)}}{16 \pi P}}.  \label{rootDG}
\end{equation}

\begin{figure}[!htb]
\centering
\subfloat[$q_{M}=0.1$ and $g(\varepsilon)=f(\varepsilon)=1.1$.]{
        \includegraphics[width=0.25\textwidth]{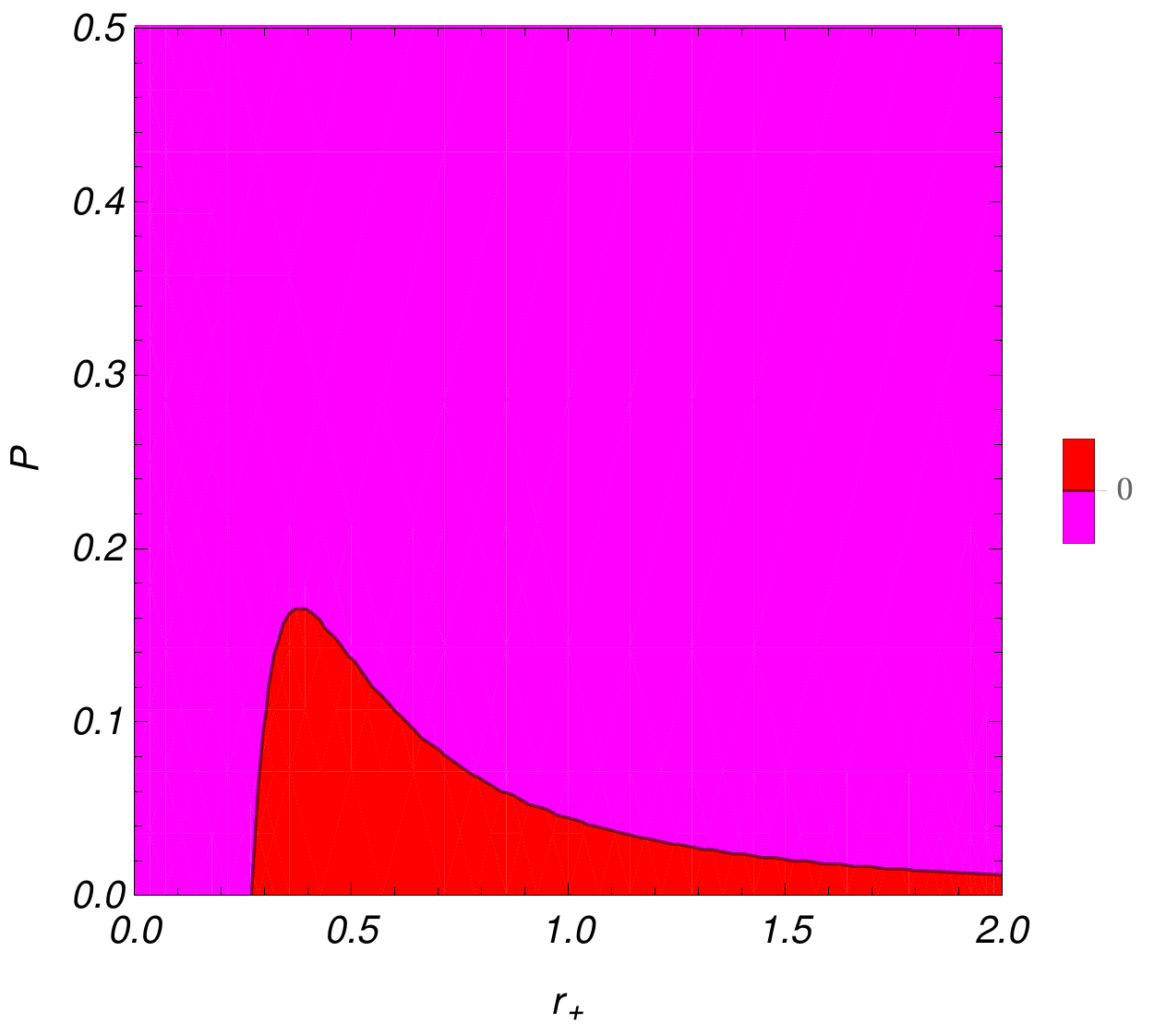}}  
\subfloat[$q_{M}=P=0.1$ and $g(\varepsilon)=1.1$.]{
        \includegraphics[width=0.25\textwidth]{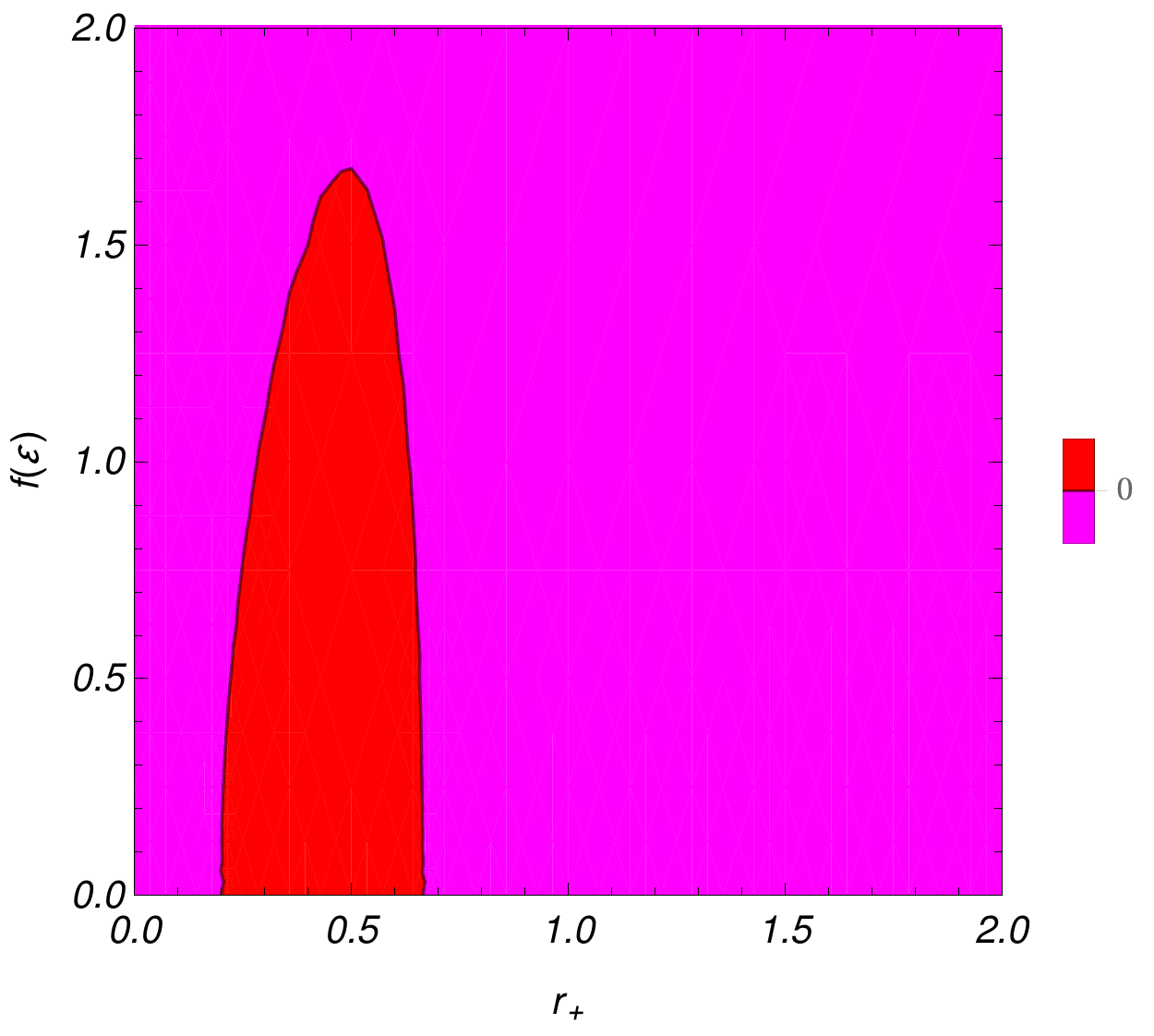}}  
\subfloat[$q_{M}=P=0.1$ and $f(\varepsilon)=1.1$.]{
        \includegraphics[width=0.25\textwidth]{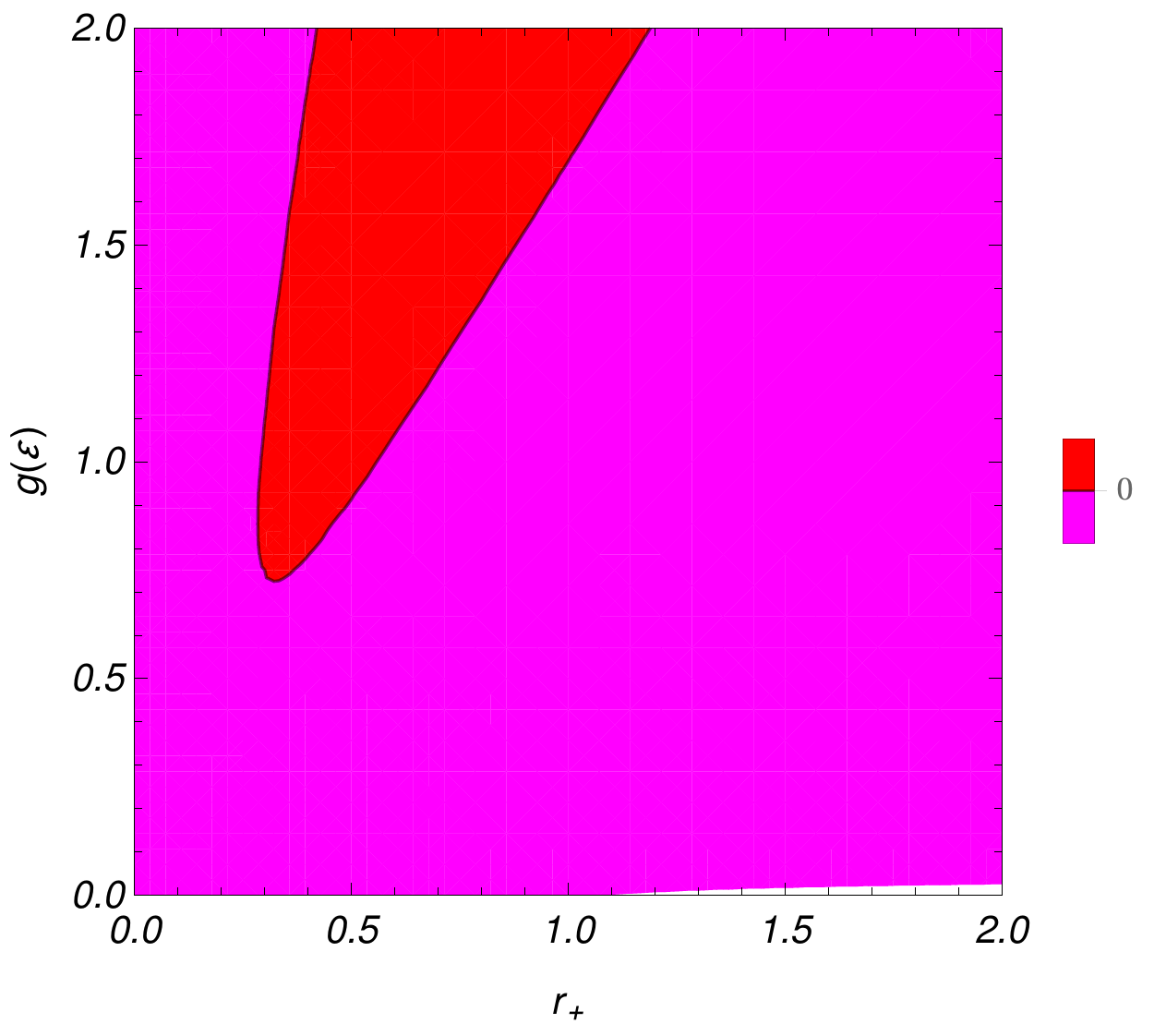}}  
\subfloat[$P=0.1$ and $g(\varepsilon)=f(\varepsilon)=1.1$.]{
        \includegraphics[width=0.25\textwidth]{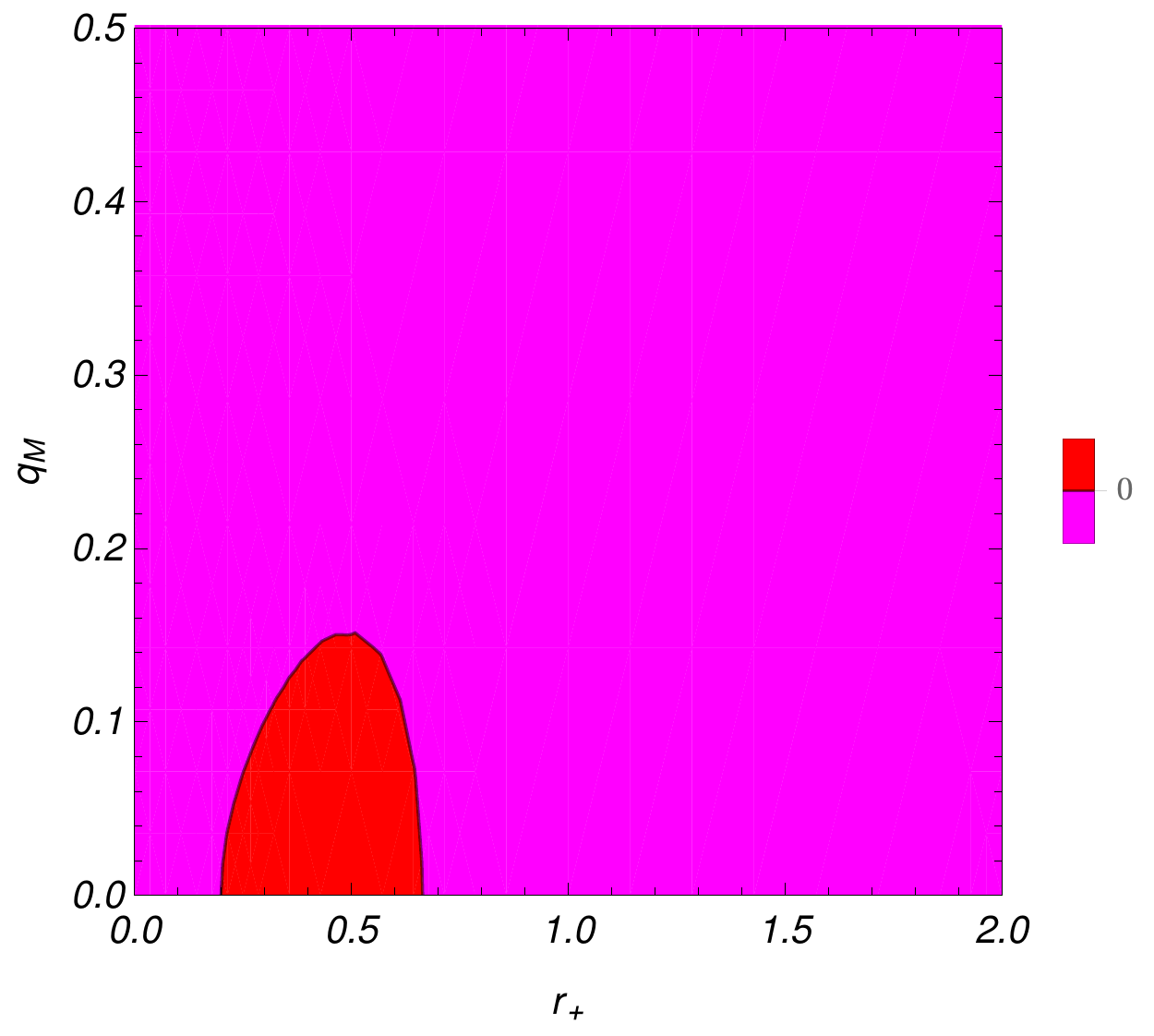}} \newline
\caption{Variation of the $(\frac{dG}{dr_{+}})_{q_{M},q_{E},T,P}$ as a
function of different parameters for $q_{E}=0.1$ and $k=1$.}
\label{Fig7}
\end{figure}

Evidently, the Gibbs free energy for spherical black holes ($k=1$) could
enjoy up to two extrema. Considering the high energy limit and asymptotic
behavior of the Gibbs free energy, if there are two extrema for it, the
first extremum will be a minimum while the second one is a maximum. In order
to elaborate this fact and investigate the effects of different parameters on
these extrema, we have plotted Fig. \ref{Fig7}.

The plotted diagrams for $(\frac{dG}{dr_{+}})_{q_{M},q_{E},T,P}$ are similar
to those that were obtained for $(\frac{dP}{dr_{+}})_{q_{M},q_{E},T}$
indicating that the effects of the rainbow functions and magnetic charge are same
as those that were reported before. The minimum (maximum) is an increasing
(decreasing) function of the pressure, magnetic charge and $f(\varepsilon)$
whereas, both the maximum and minimum are increasing functions of $%
g(\varepsilon)$.

The existence of equilibrium point indicates that there is at least two
distinguishable phases available for the thermodynamical system. In fact,
the equilibrium point is where two phases meet and phase transition could
take place between them. The Gibbs free energy should be a decreasing
function of the volume. Considering this, the region between two extrema,
where Gibbs free energy becomes an increasing function of the volume
(horizon radius) is not physically acceptable. Therefore, this region is not
accessible for the black holes and there is a phase transition over this
region for different phases available for the black holes. Once more, we point out
that rainbow functions have opposite effects on these two properties of the
Gibbs free energy. Later (in section \ref{van der Waals like Behavior}), we
will investigate the type of phase transition.

\subsubsection{Electric and magnetic properties} \label{Electric and magnetic properties} 

The total electric and magnetic charges of a system could be calculated by Gauss law. Using this law, one can show that the total electric and magnetic charges could be obtained as 
\begin{equation}
Q_{E}=\frac{f(\varepsilon) q_{E}}{4 \pi g(\varepsilon)},
\label{electric charge}
\end{equation}
\begin{equation}
Q_{M}=\frac{f(\varepsilon) q_{M}}{4 \pi g(\varepsilon)},
\label{magnetic charge}
\end{equation}
which show that due to the contribution of the gravity's rainbow, the total
electric and magnetic charges are modified. This is in contrast to what were
observed for Gauss-Bonnet \cite{GB}, dilatonic gravities \cite{Dilaton} and
Born-Infeld nonlinear electrodynamic generalization \cite{Dilaton}.

To obtain the electric and magnetic fields, we use the first law
of black hole thermodynamics. We employ \eqref{first
law} to obtain these quantities which results into 
\begin{equation}
\Phi_{E}=(\frac{dM}{dQ_{E}})_{P,S,Q_{M}}=\frac{q_{E}}{r_{+}},
\label{electric field}
\end{equation}
\begin{equation}
\Phi_{M}=(\frac{dM}{dQ_{M}})_{P,S,Q_{E}}=\frac{f^2(\varepsilon) q_{M}}{%
g^2(\varepsilon)}.  \label{magnetic field}
\end{equation}

The obtained electric field matches the one that was calculated before for
electrically charged black holes in the presence of gravity's rainbow \cite%
{GB,Dilaton}. It is independent of the rainbow functions. Interestingly and
contrary to electric field, the magnetic field is dependent of the rainbow functions and is affected by generalization of the rainbow gravity. But it
should be noted that the rainbow functions have opposite effects on this quantity
and if we choose the rainbow functions to be identical, the effects of
gravity's rainbow on this quantity will be removed.

\subsection{Stability of the solutions} \label{Stability of the solutions} 

In canonical ensemble, the thermal stability of the black hole is determined by the behavior of heat capacity. The signature of the heat capacity is a tool that is used for determining thermal
stability/instability of black holes. In previous sections, we
showed that these black holes enjoy more than one phase in their
thermodynamical state. Here, using the heat capacity, we determine the
number of the phases available for the black holes and their thermal
stability. It is worthwhile to mention that a black hole is thermally stable
(unstable) if its heat capacity is positive (negative). In addition, the
discontinuities in heat capacity could be interpreted as phase transition points.

The heat capacity for black holes under consideration is given by 
\begin{equation}
C= T\frac{(\frac{dS}{dr_{+}})_{q_{M},q_{E},P}}{(\frac{dT}{dr_{+}}%
)_{q_{M},q_{E},P}}=\frac{r_{+}^2(8 \pi P r_{+}^4-g^2(\varepsilon)
\left(f^2(\varepsilon) q_{E}^2+g^2(\varepsilon) q_{M}^2-k r_{+}^2\right))}{%
16 \pi g^2(\varepsilon) P r_{+}^4+2 g^4(\varepsilon) \left(3
f^2(\varepsilon) q_{E}^2+3 g^2(\varepsilon) q_{M}^2-k r_{+}^2\right)}.
\label{heat capacity}
\end{equation}

The first noticeable issue is that the sign of the topological, electric and
magnetic terms is different in numerator and denominator of heat capacity.
While in the numerator, mentioned terms are coupled with horizon radius
with higher power, in the denominator, they are coupled with higher power of $%
g(\varepsilon)$. The same coupling also takes place for the pressure term
where its sign is identical in both numerator and denominator of the
heat capacity. Remembering the role of heat capacity's sign and its
divergencies on determining the number of phases, we extract its roots in
the following form 
\begin{equation}
\left. r_{+}\right\vert _{C=0}=\sqrt{-\frac{g^2(\varepsilon) k}{16 \pi P }%
\pm \frac{g(\varepsilon) \sqrt{32 \pi P \left(f^2(\varepsilon)
q_{E}^2+g^2(\varepsilon) q_{M}^2\right)+g^2(\varepsilon) k^2}}{16 \pi P}}.
\label{rootC}
\end{equation}
while the divergencies are obtained by 
\begin{equation}
\left. r_{+}\right\vert _{C\longrightarrow \infty}=\sqrt{\frac{%
g^2(\varepsilon) k}{16 \pi P }\pm \frac{g(\varepsilon) \sqrt{%
g^2(\varepsilon) k^2-96 \pi P \left(f^2(\varepsilon)
q_{E}^2+g^2(\varepsilon) q_{M}^2\right)}}{16 \pi P}}.  \label{divC}
\end{equation}

The obtained roots for heat capacity \eqref{rootC} coincide with those
extracted from temperature \eqref{rootT}. Therefore, the heat capacity and
temperature share the same root. On the other hand, the divergencies of
heat capacity are identical to the extrema extracted for $(\frac{dG}{dr_{+}}%
)_{q_{M},q_{E},T,P}$. In previous sections, we established that the range
between roots of $(\frac{dG}{dr_{+}})_{q_{M},q_{E},T,P}$ is actually an
inaccessible region for the black holes. In addition, we pointed out that
such region and roots (which were equilibrium points) could only be observed
for black holes with spherical horizon. Therefore, here as well, one can
conclude that only for spherical black holes, existence of two divergencies
for these black holes could be observed.

\begin{figure}[!htb]
\centering
\subfloat[$q_{M}=0.1$, $k=1$ and $g(\varepsilon)=f(\varepsilon)=1.1$.]{
        \includegraphics[width=0.25\textwidth]{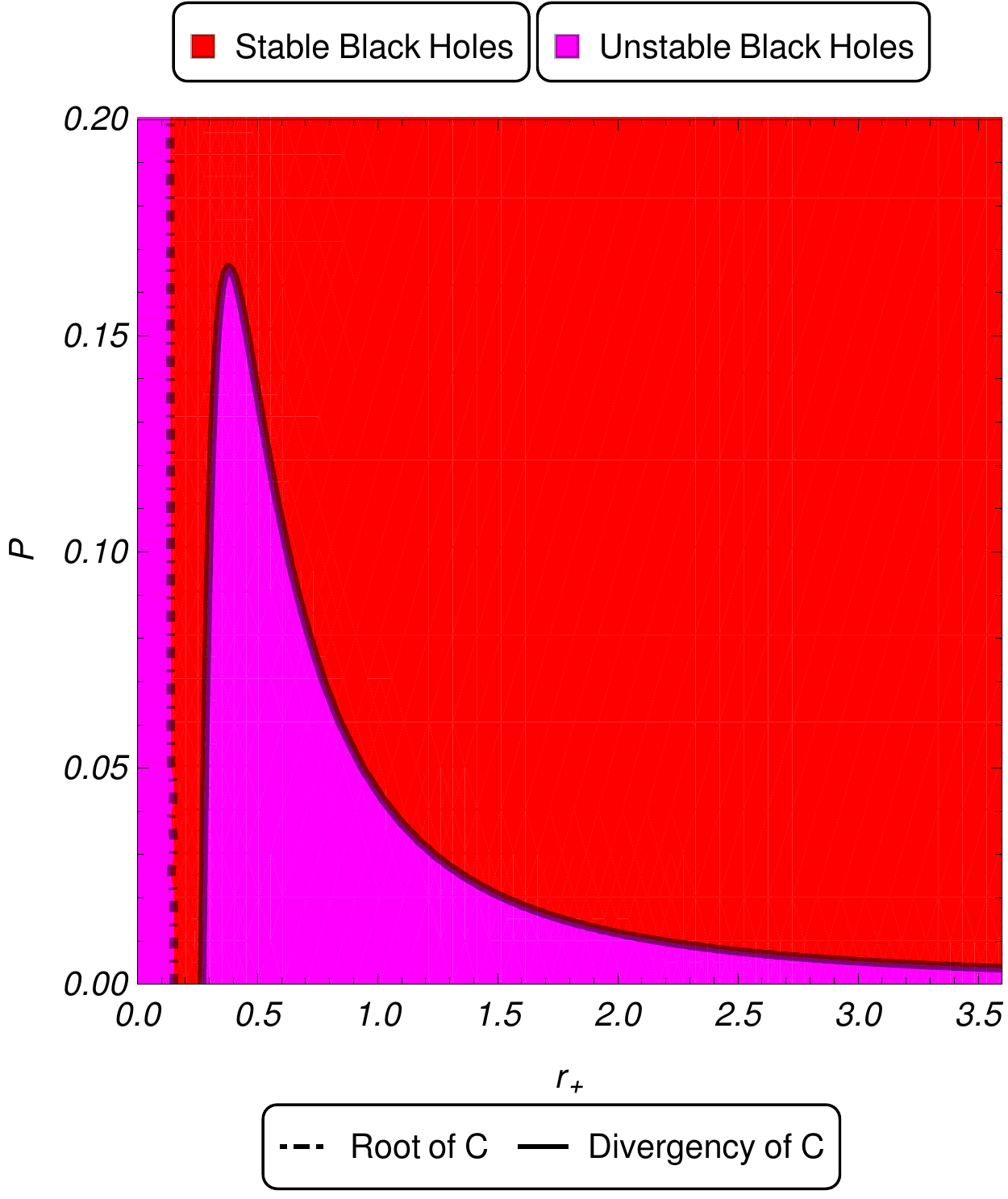}}  
\subfloat[$q_{M}=P=0.1$, $k=1$ and $g(\varepsilon)=1.1$.]{
        \includegraphics[width=0.25\textwidth]{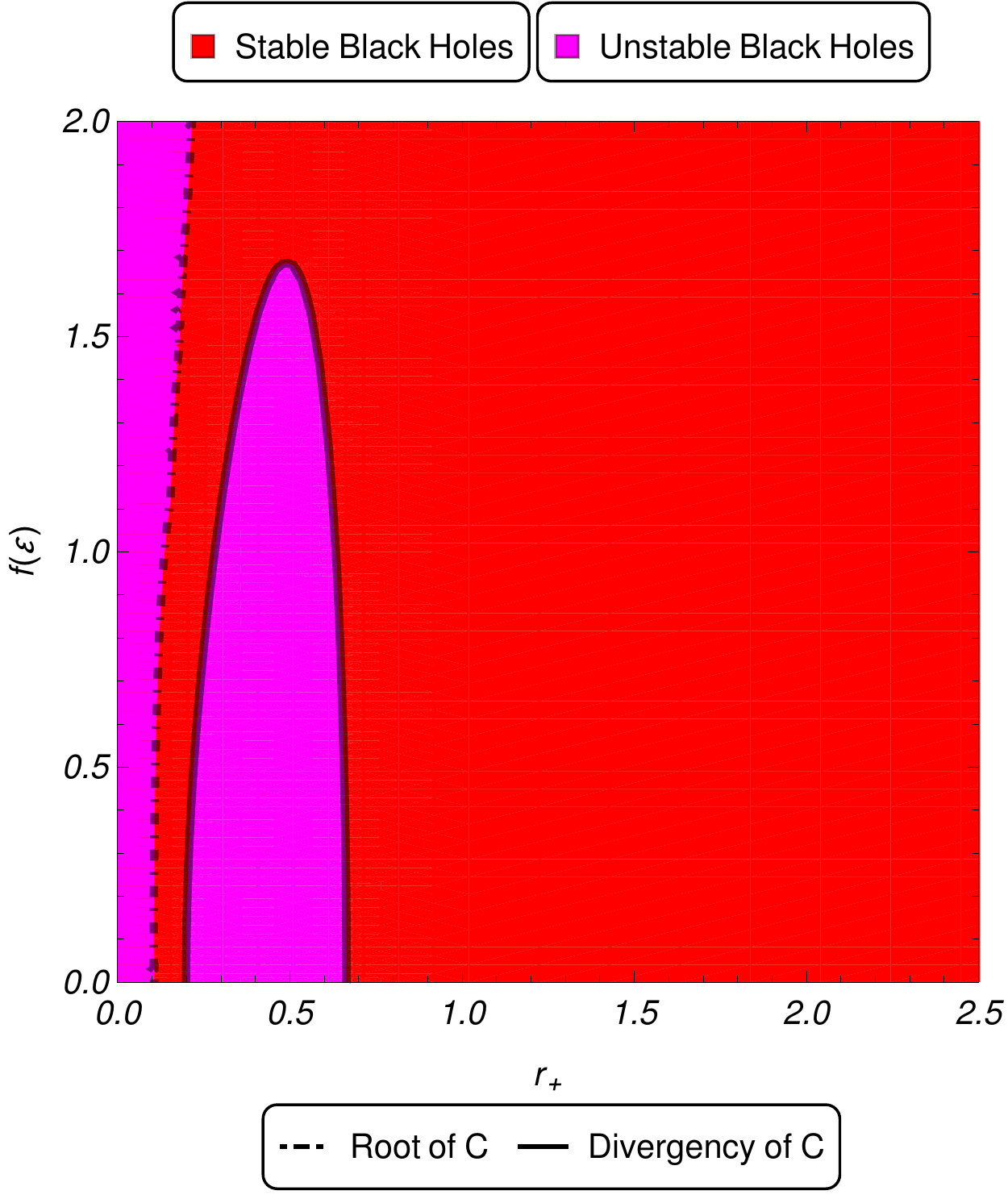}}  
\subfloat[$q_{M}=P=0.1$, $k=1$ and $f(\varepsilon)=1.1$.]{
        \includegraphics[width=0.25\textwidth]{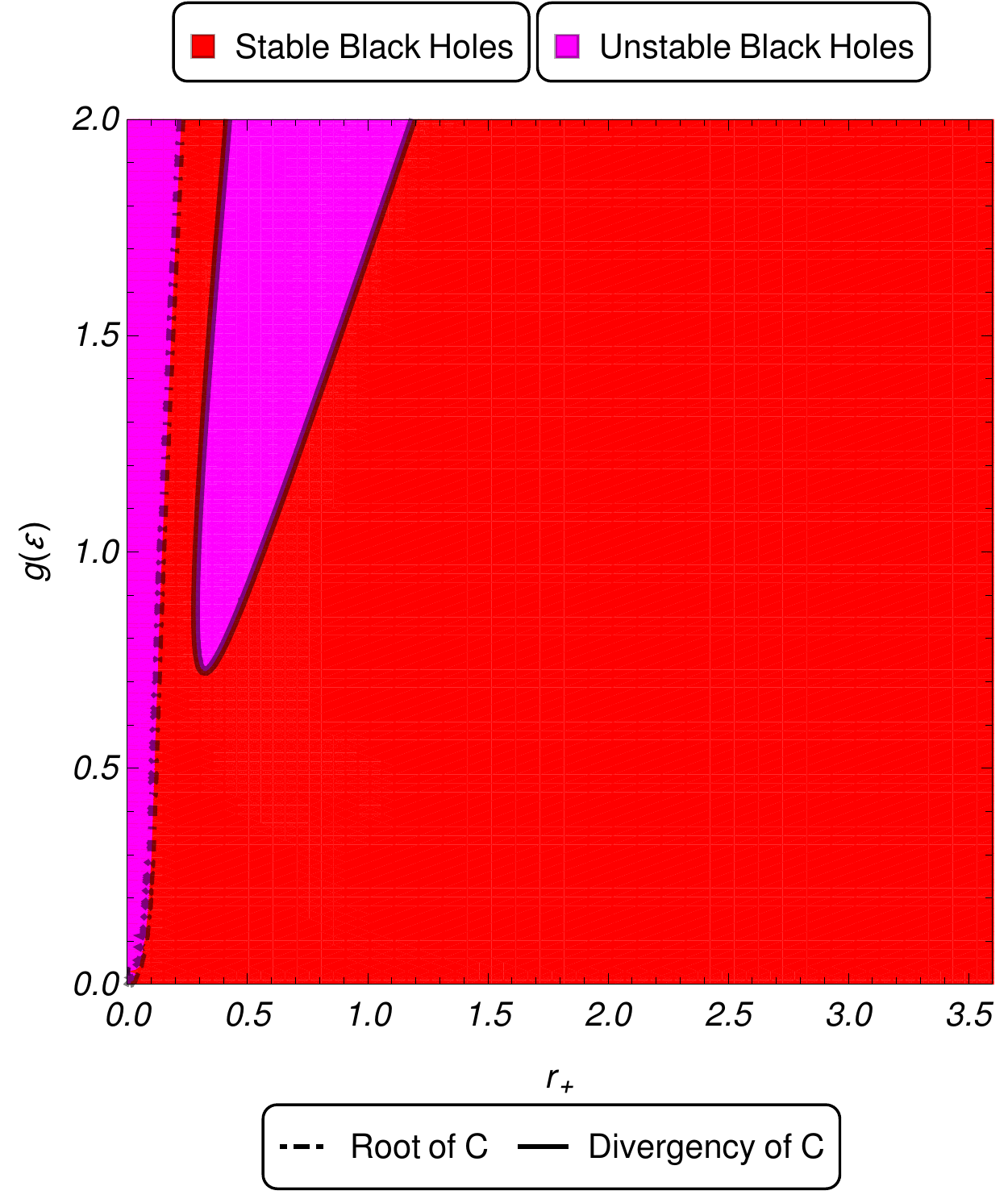}}\newline
\subfloat[$P=0.1$, $k=1$ and $g(\varepsilon)=f(\varepsilon)=1.1$.]{
        \includegraphics[width=0.25\textwidth]{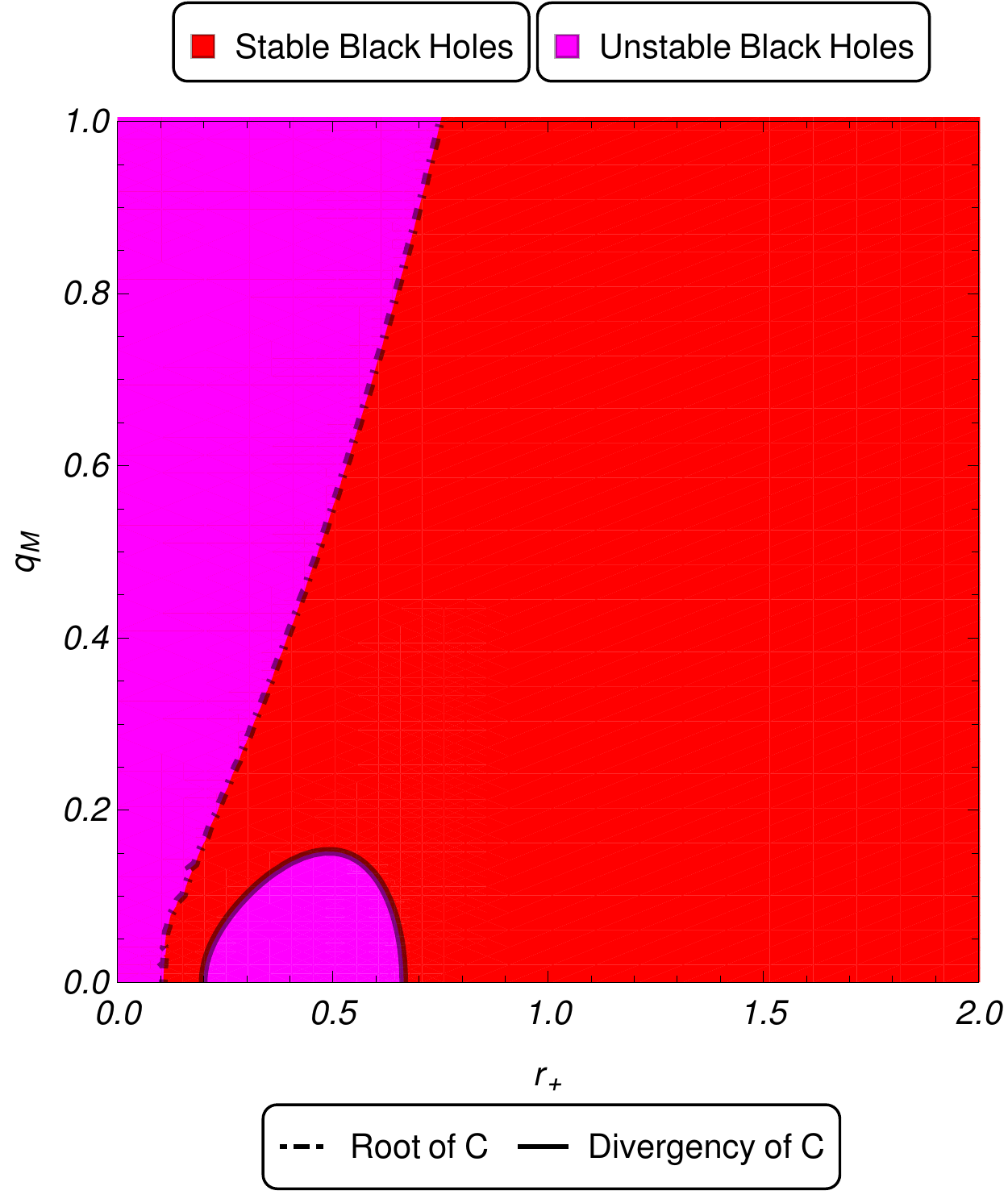}}  
\subfloat[$q_{M}=P=0.1$ and $g(\varepsilon)=f(\varepsilon)=1.1$.]{
        \includegraphics[width=0.25\textwidth]{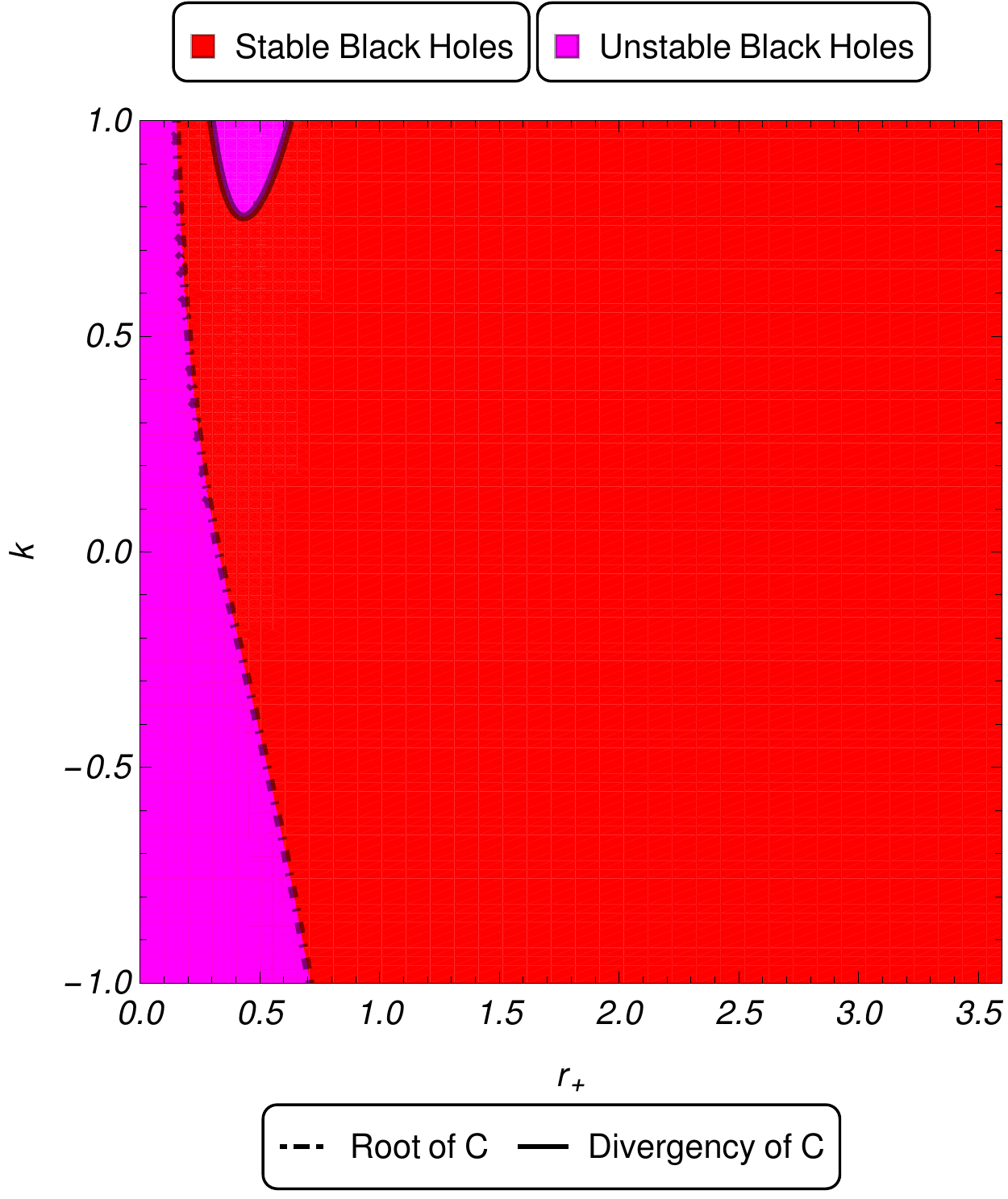}} \newline
\caption{Variation of the $C$ as a function of different parameters for $%
q_{E}=0.1$.}
\label{Fig8}
\end{figure}

The number of phases present in the thermodynamical state of the black
holes depends on number of roots and divergencies. In previous section,
we pointed out that these black holes could enjoy up to one root in their
temperature. Therefore, heat capacity has only one root. On the other hand, we pointed out that depending on values of
different parameters, spherical black holes could have up to two roots for $(%
\frac{dG}{dr_{+}})_{q_{M},q_{E},T,P}$ and black holes with flat and
hyperbolic horizons have no such roots. Therefore, one can also conclude
that heat capacity for spherical black holes could have up to two divergencies,
otherwise, there is no divergency for black holes. To make more clarification, we have plotted Fig. \ref{Fig8}
for variation of heat capacity as a function of different parameters.

As it was pointed out, the heat capacity has only up to one root. The place of this root is an increasing function of the rainbow
functions and magnetic charge, a decreasing function of the topological
factor and not significantly affected by pressure. The number of
divergencies in heat capacity is a decreasing function of the pressure,
magnetic charge and $f(\varepsilon)$ while it is an increasing function of $%
g(\varepsilon)$ and topological factor. In other words, the region between
two divergencies is a decreasing function of the pressure, magnetic charge
and $f(\varepsilon)$ and an increasing function $g(\varepsilon)$ and
topological factor.

Considering the effects of different parameters, the possible scenarios
regarding the number of phases for these black holes are the following
ones:

I) One root: there are two phases of small and large black holes. The
small one has negative temperature and therefore, not physical. Whereas,
the large black hole phase is thermally stable.

II) One root and one divergency: In this case, we have three phases of
small, medium and large black holes. The small black holes are not physical
due to negativity of the temperature. The medium and large phase are
separated by a divergence point. This divergency is actually the equilibrium
point that was discussed in section \ref{Gibbs free energy}. Therefore, this
is the point where two phases of medium and large black holes are in
equilibrium and going from one to the other could be achieved via a critical process.

III) One root and two divergencies: In this case, we have four
distinguishable phases for black holes; very small, small, medium and large
black holes. Since the temperature is negative in the very small black hole
phase, this phase is not a physical one. For small and large black hole
phases, heat capacity and temperature are positive, Gibbs free energy and
pressure have reasonable behaviors and mass is not negative. Therefore,
these two phases satisfy the thermodynamical principle. For the medium black
hole phase, although it has positive temperature and mass, the behaviors of
Gibbs free energy and pressure violate thermodynamically expected behaviors
and heat capacity is negative. Therefore, this phase is not physical and
accessible for the black holes. The small, medium and large black hole
phases are separated by two divergencies. This indicates that upon meeting
these divergencies, system go through a phase transition between them and
changes from small to larger black hole phases or vice versa.

Finally, we should point out that super magnetized and/or super pressurized
black holes admit only two phases in their thermodynamical structure where
they are separated by a root and only one of the phases is physical.
Therefore, there is no phase transition for black holes in this case. The
situation for rainbow functions is different. As we pointed out, the effects
of energy functions on the number of the phases and divergencies of heat
capacity (phase transitions) are opposite. While the number of phases and
phase transition points are decreasing functions of $f(\varepsilon)$, they
are increasing functions of $g(\varepsilon)$. The $f(\varepsilon)$
represents the effects of temporal coordinate while $g(\varepsilon)$
represents the effects of spatial coordinate. Considering this, one can
conclude that effective behavior of the temporal coordinate is toward
omitting phase transitions and instabilities in the system and acquiring a
uniform stable state while the opposite is true for spatial coordinates.

To complete our discussions here, we investigate the high energy limit of
heat capacity given by 
\begin{equation*}
\lim_{r_{+}\longrightarrow 0 }C=-\frac{r_{+}^2}{6 g^2(\varepsilon)}+\frac{k
r_{+}^4}{9g^2(\varepsilon) (g^2(\varepsilon) q_{M}^2+ f^2(\varepsilon)
q_{E}^2)}+O(r_{+}),
\end{equation*}
and asymptotic behavior obtained as 
\begin{equation*}
\lim_{r_{+}\longrightarrow \infty }C=\frac{r_{+}^2}{2 g^2(\varepsilon)}+\frac{k}{%
8 \pi P}+O(\frac{1}{r_{+}}).
\end{equation*}

Interestingly, contrary to other thermodynamical quantities, the high energy
limit and asymptotic behavior of the heat capacity are only governed by
horizon radius and one of the energy functions, $g(\varepsilon)$. The effect
of topological structure becomes noticeable in the second dominant term of
both limits.

\subsection{van der Waals like Behavior} \label{van der Waals like Behavior} 

In the last section, we established the possibility of existence of thermodynamical phase transition for our solutions. We also pointed out the number of thermodynamical phases present
for different cases. In this section, we would like to address the type of
phase transition. To do so, we first extract critical values for different
thermodynamical quantities and plot the corresponding diagrams to
investigate the critical behavior of these thermodynamical quantities. To do
so, we use the equation of state \eqref{pressure} and the properties of
inflection point. Considering the relation between volume and horizon radius,
the properties of inflection point could be extracted from 
\begin{equation}
\left( \frac{\partial P}{\partial r_{+}}\right)_{T,q_{M},q_{E}} =\left( 
\frac{\partial ^{2}P}{\partial r_{+}^{2}}\right)_{T,q_{M},q_{E}} =0,
\label{infel}
\end{equation}
which results into the following equation for obtaining critical horizon radius
(volume) 
\begin{equation}
6 (f^2(\varepsilon) q_{E}^2 + g^2(\varepsilon) q_{M}^2)-k r_{+}^2=0.
\label{critical expression}
\end{equation}

Evidently, only for spherical black holes, one can obtain critical horizon
radius, hence critical behavior. It is a matter of calculation to show that
critical horizon radius, temperature, pressure and Gibbs free energy are
given by
\begin{eqnarray}
r_{c}&=&\sqrt{\frac{6 (f^2(\varepsilon) q_{E}^2 + g^2(\varepsilon) q_{M}^2)}{%
k}},  \label{critical horizon} \\
\notag \\
T_{c}&=&\frac{g^2(\varepsilon) k^{\frac{3}{2}}}{3 \pi f(\varepsilon) \sqrt{6
(f^2(\varepsilon) q_{E}^2 + g^2(\varepsilon) q_{M}^2)}},
\label{critical temperature} \\
\notag \\
P_{c}&=&\frac{g(\varepsilon) k^2}{96 \pi (f^2(\varepsilon) q_{E}^2 +
g^2(\varepsilon) q_{M}^2)},  \label{critical pressure} \\
\notag \\
G_{c}&=&\frac{\sqrt{k (f^2(\varepsilon) q_{E}^2 + g^2(\varepsilon) q_{M}^2)}%
}{2 \sqrt{6} \pi f(\varepsilon) 2(\varepsilon)},  \label{critical Gibbs}
\end{eqnarray}

\begin{figure}[!htb]
\centering
\subfloat[$k=1$ and $g(\varepsilon)=f(\varepsilon)=1.1$.]{
        \includegraphics[width=0.25\textwidth]{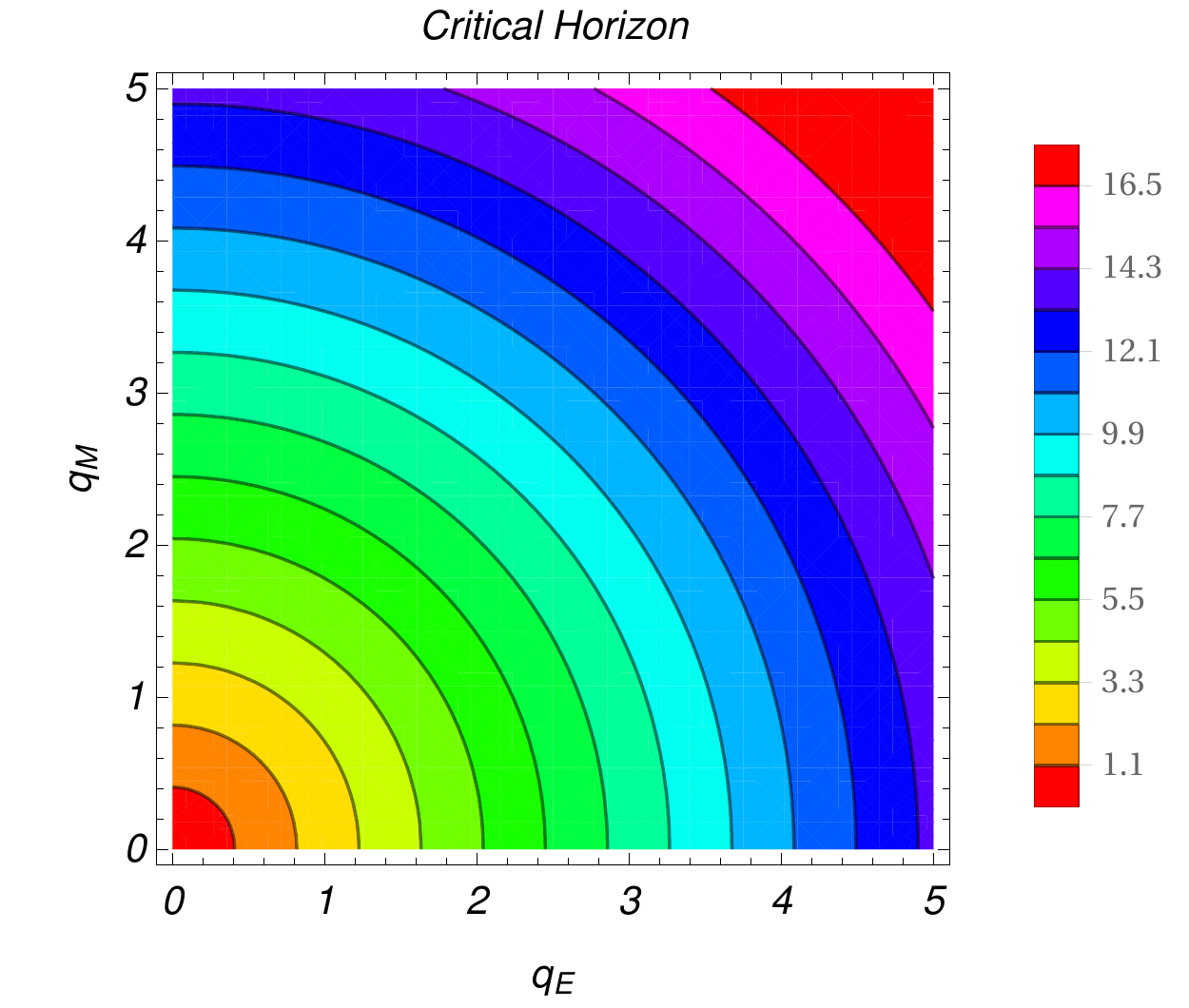}}  
\subfloat[$k=1$ and $q_{M}=q_{E}=0.1$.]{
        \includegraphics[width=0.25\textwidth]{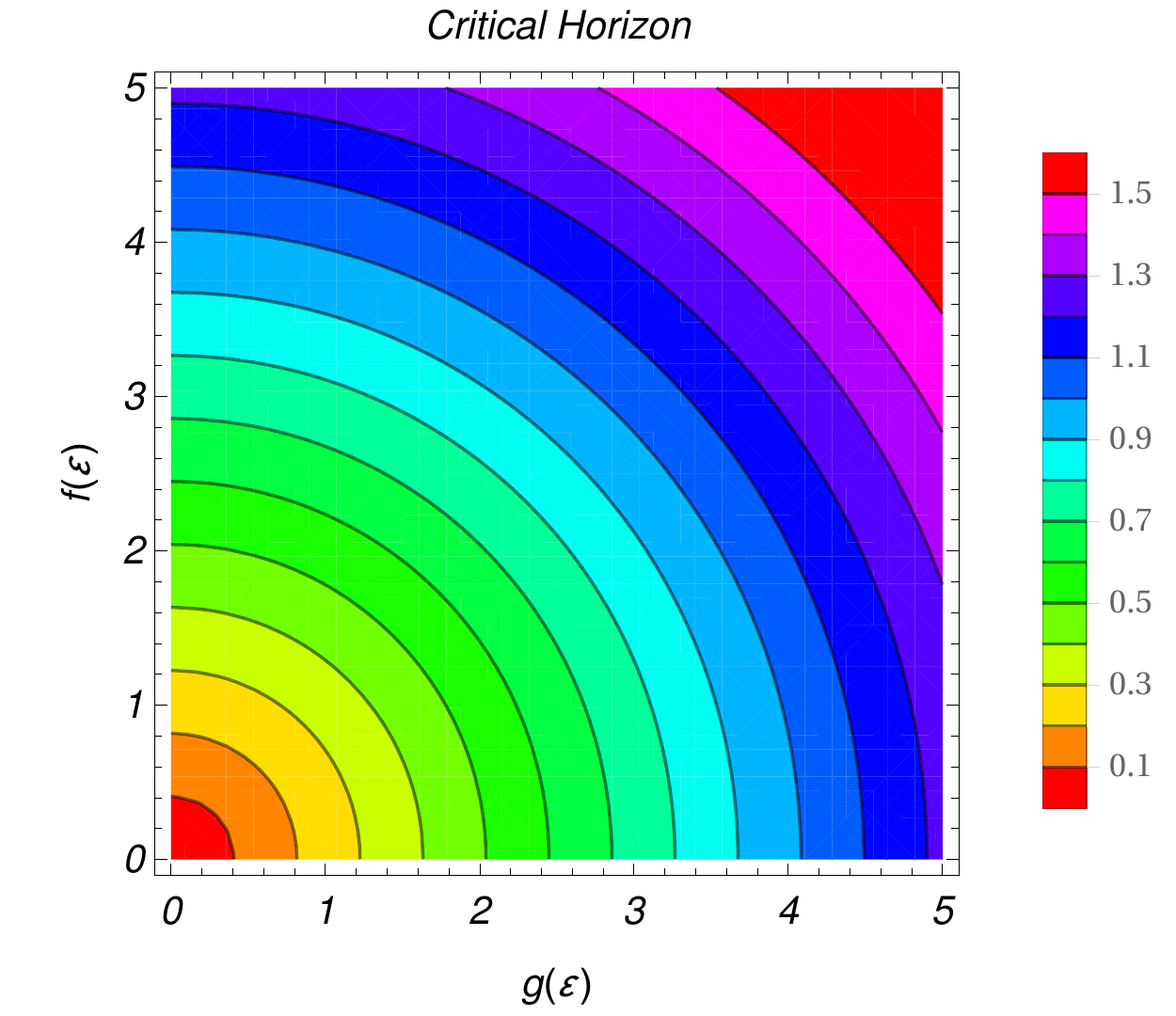}}  
\subfloat[$k=1$ and $g(\varepsilon)=f(\varepsilon)=1.1$.]{
        \includegraphics[width=0.25\textwidth]{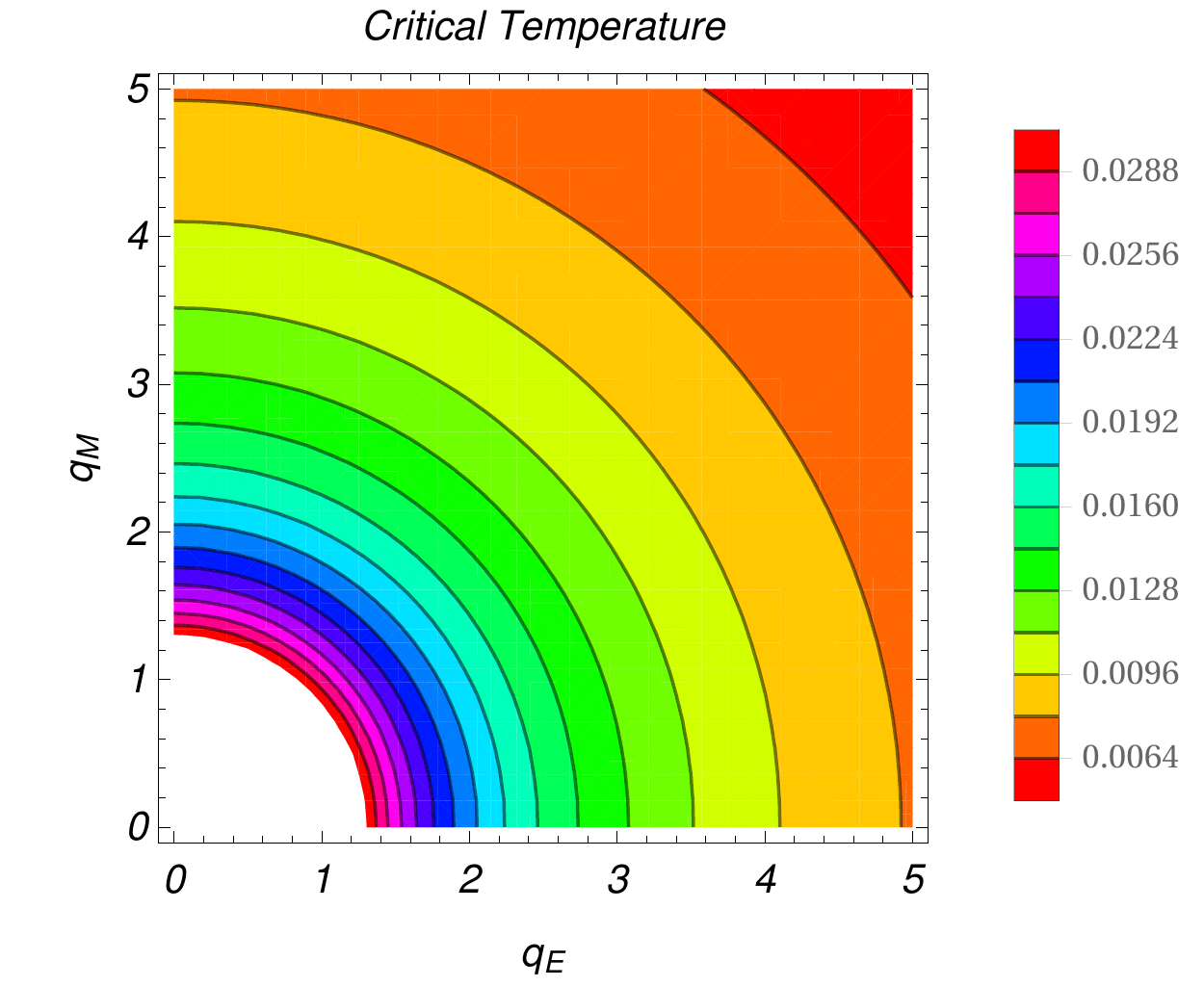}}  
\subfloat[$k=1$ and $q_{M}=q_{E}=0.1$.]{
        \includegraphics[width=0.25\textwidth]{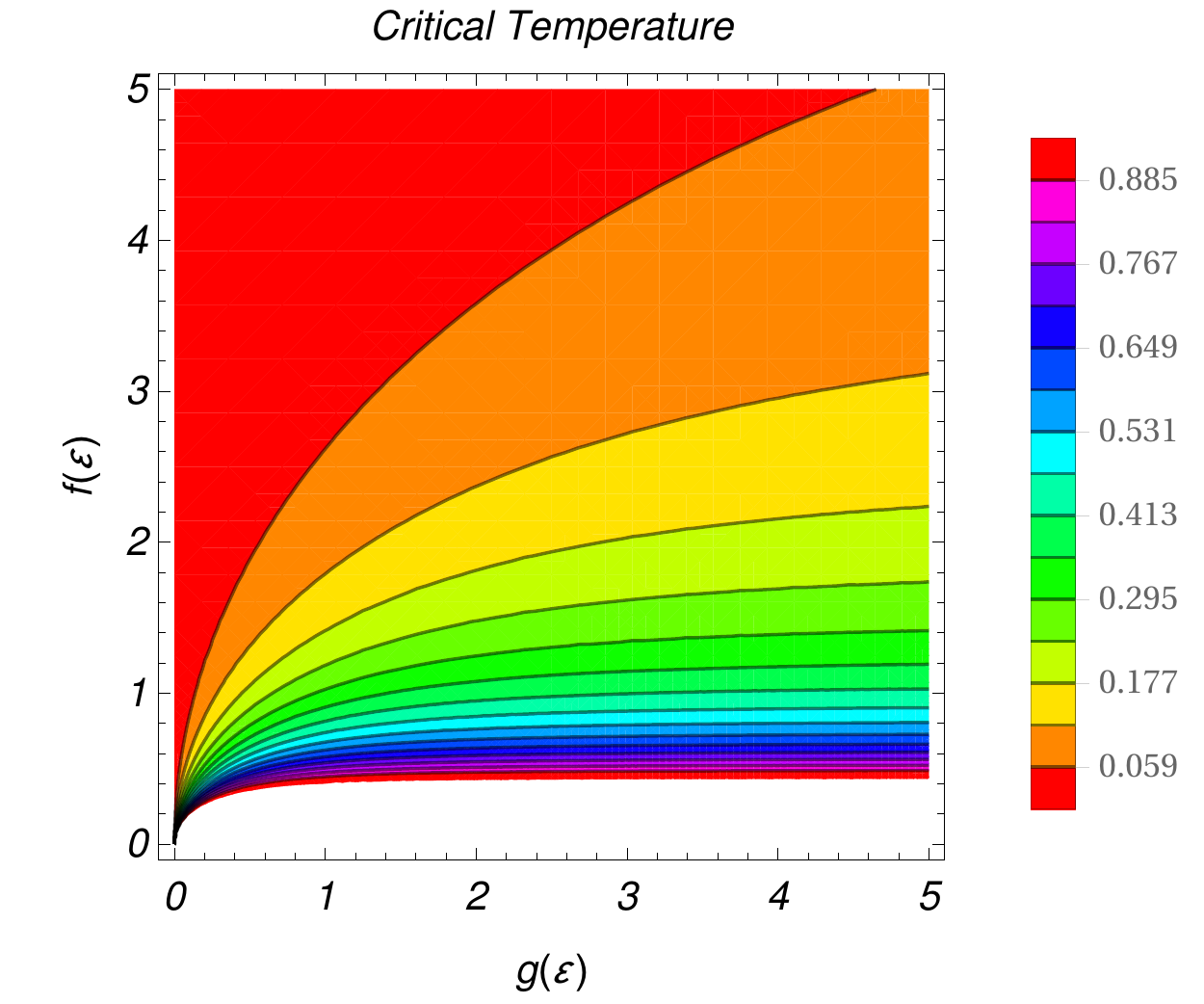}}\newline
\subfloat[$k=1$ and $g(\varepsilon)=f(\varepsilon)=1.1$.]{
        \includegraphics[width=0.25\textwidth]{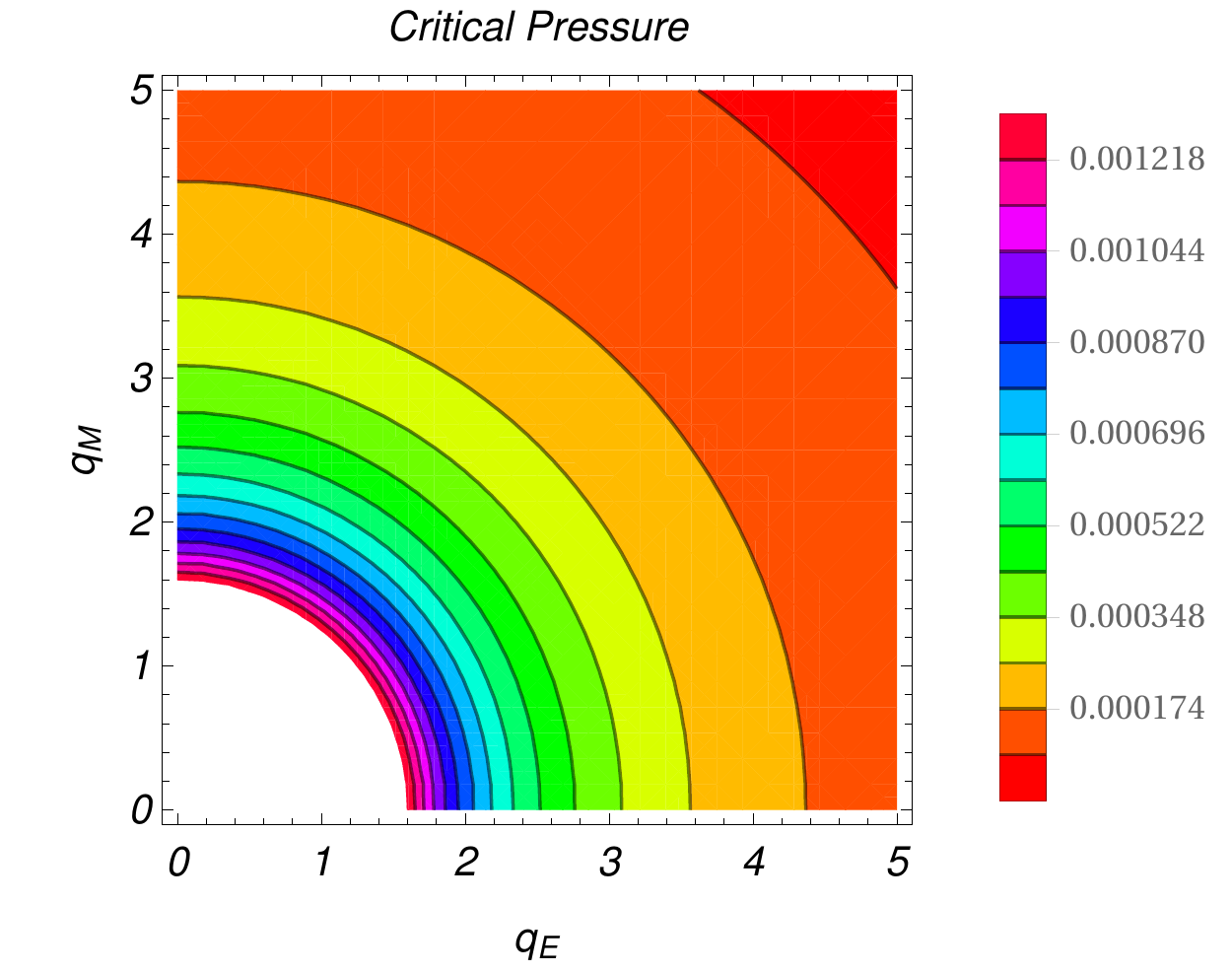}}  
\subfloat[$k=1$ and $q_{M}=q_{E}=0.1$.]{
        \includegraphics[width=0.25\textwidth]{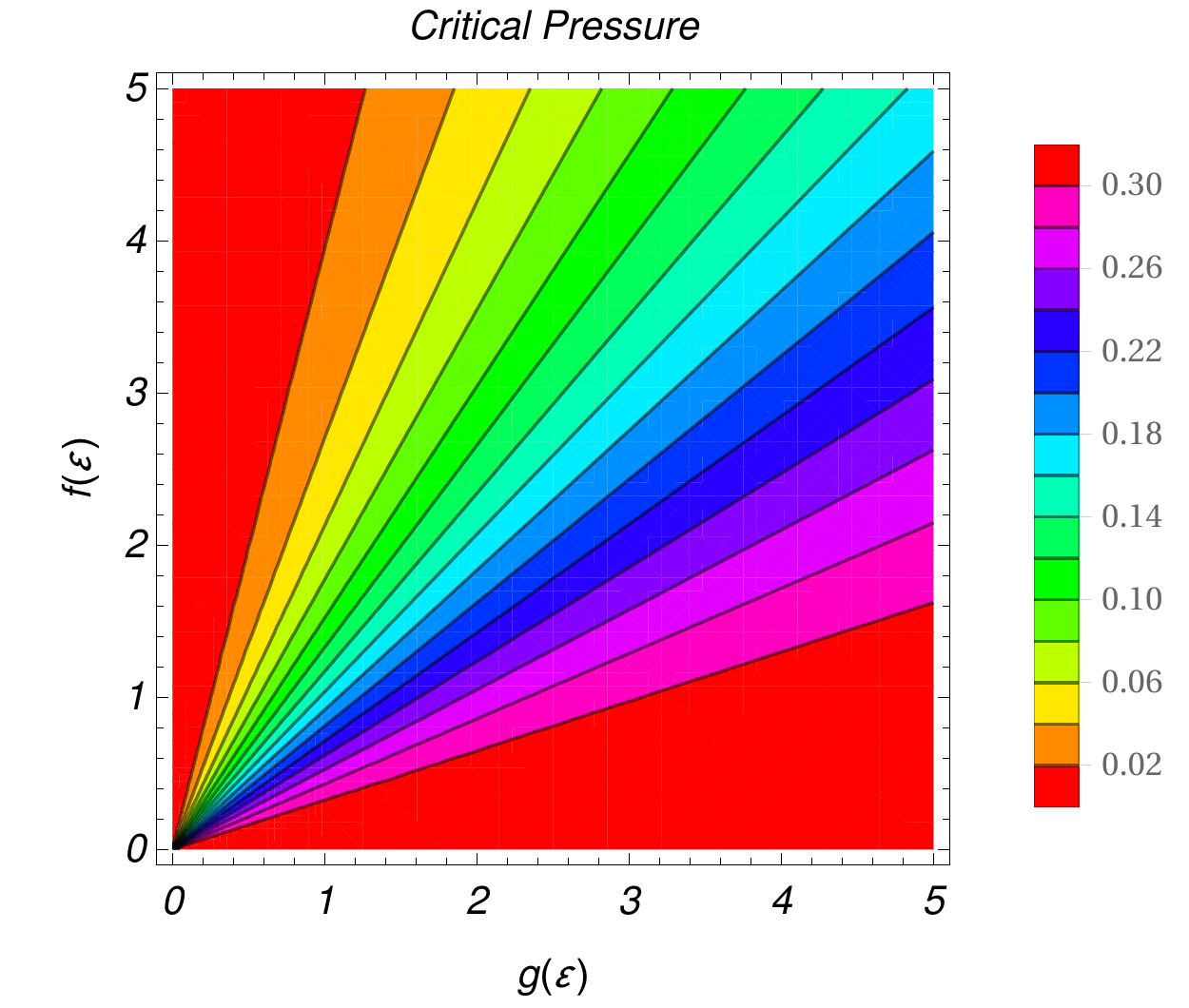}}  
\subfloat[$k=1$ and $g(\varepsilon)=f(\varepsilon)=1.1$.]{
        \includegraphics[width=0.25\textwidth]{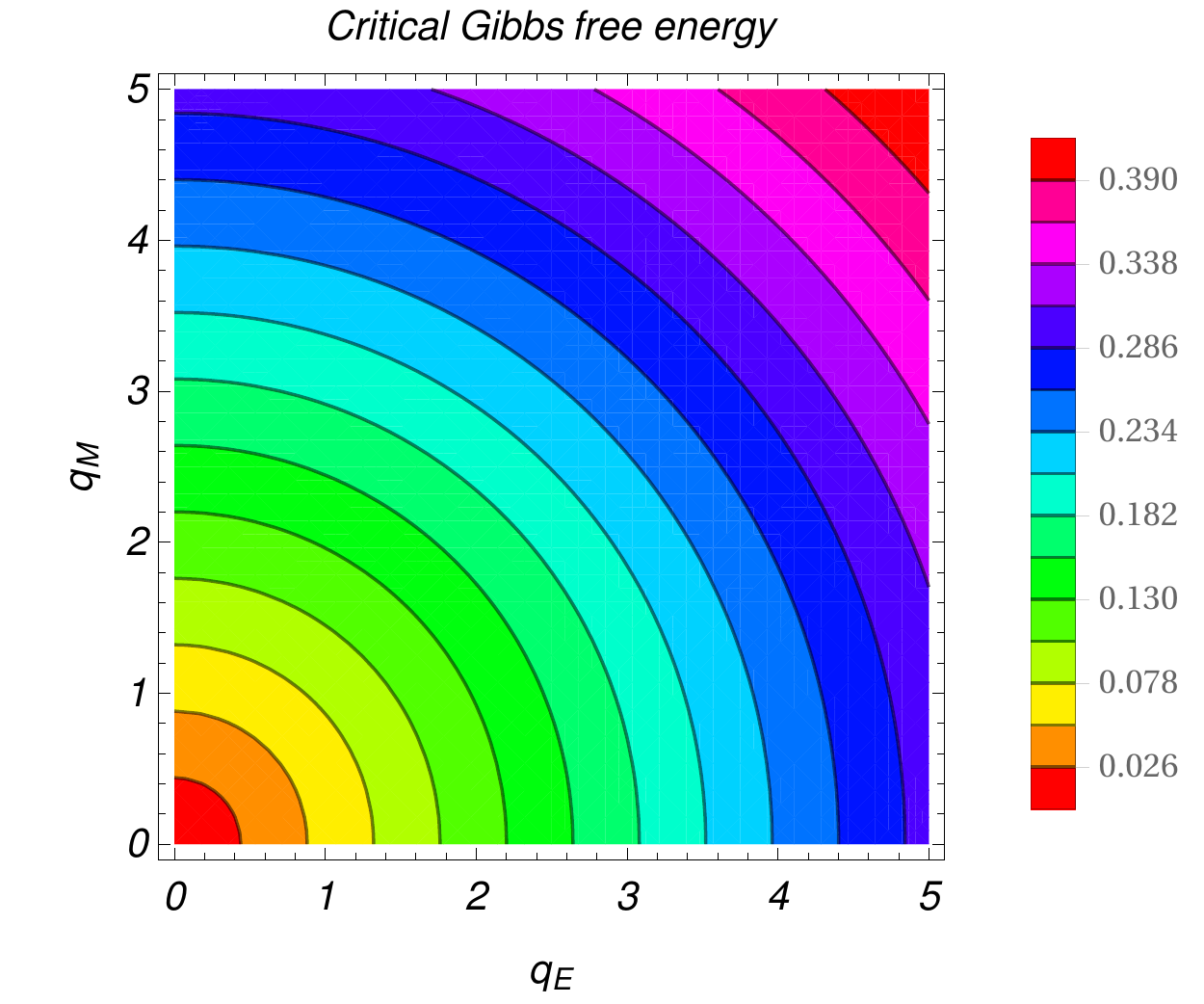}}  
\subfloat[$k=1$ and $q_{M}=q_{E}=0.1$.]{
        \includegraphics[width=0.25\textwidth]{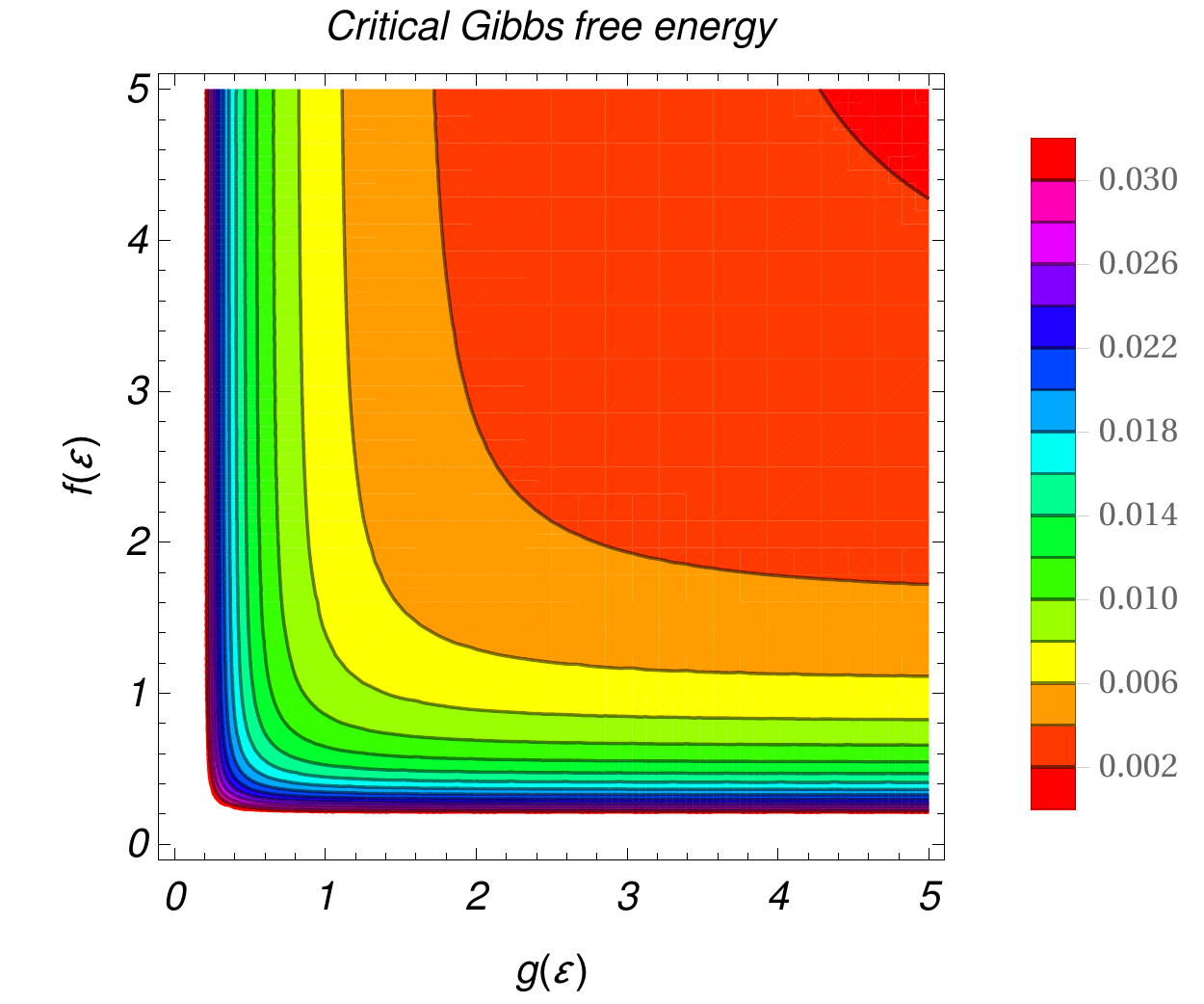}}\newline
\caption{Variation of the critical values as functions of black hole's
parameters.}
\label{Fig9}
\end{figure}

The obtained critical values also confirm that the critical behavior is only
possible for spherical black holes while it is absent for black holes with
hyperbolic and flat horizons. In order to study the effects of different
parameters on the thermodynamical quantities, we plot the necessary diagrams in
Fig. \ref{Fig9}.

\begin{figure*}[!htb]
\centering
\includegraphics[width=0.3\linewidth]{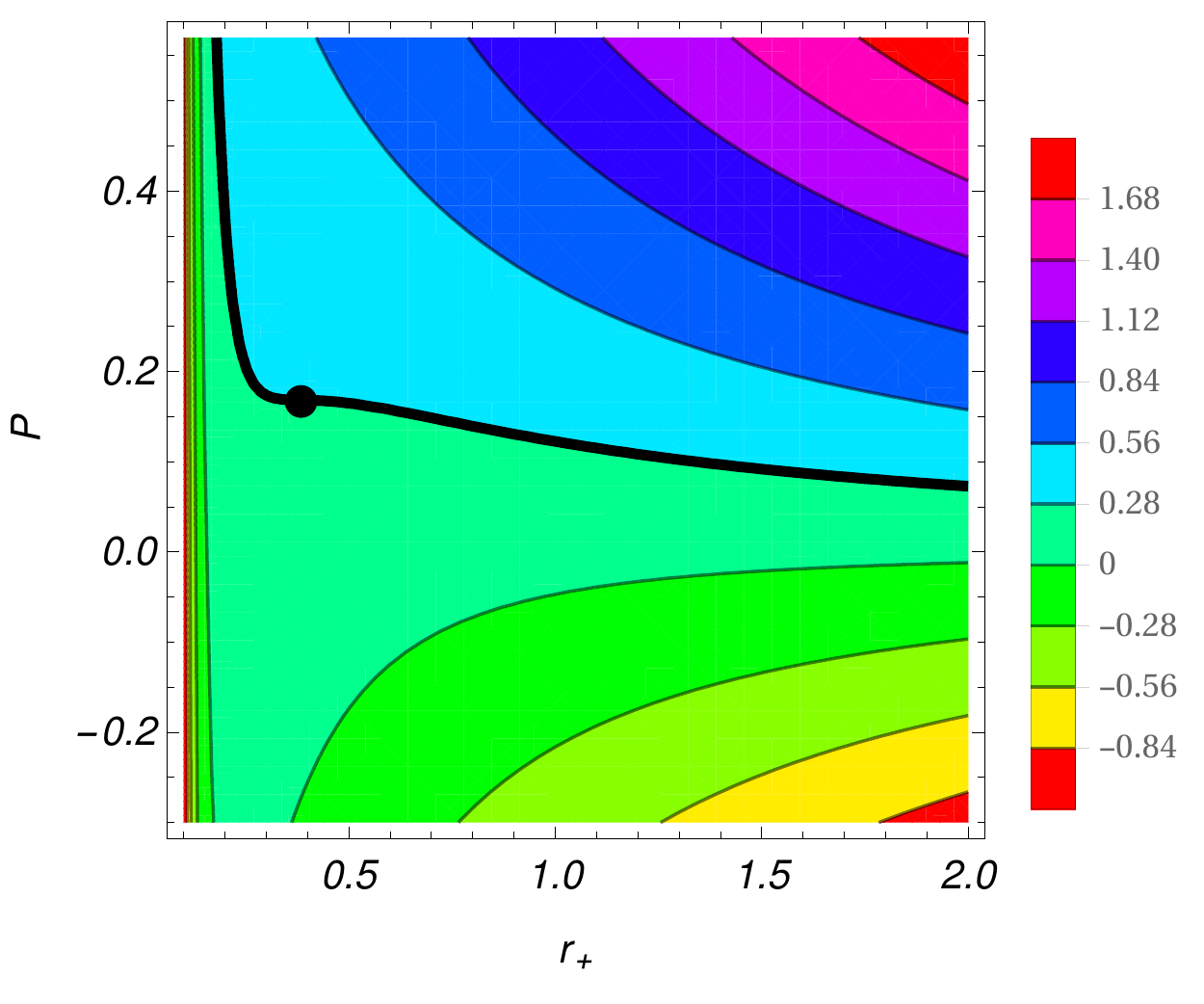}\hfil
\includegraphics[width=0.3\linewidth]{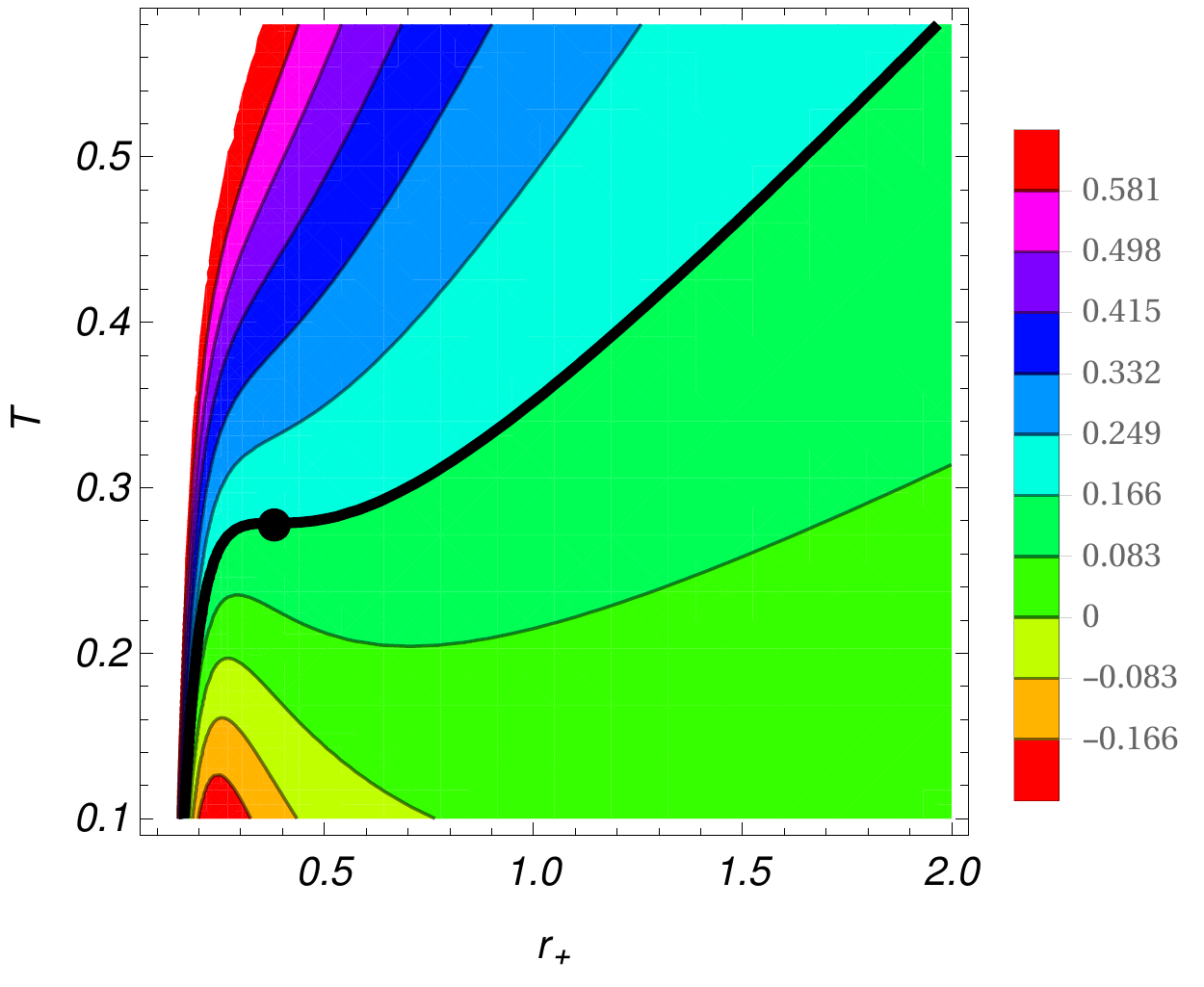}\hfil
\includegraphics[width=0.3\linewidth]{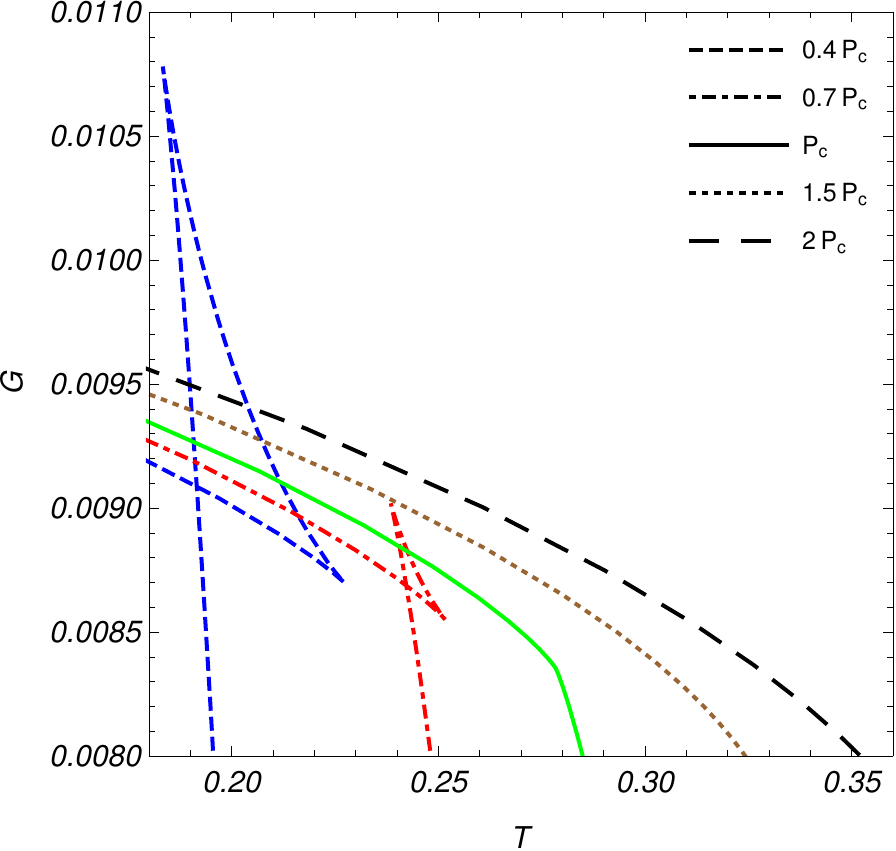}  
\caption{Van der Waals like phase diagrams for $q_{M}=q_{E}=0.1$, $f(\protect%
\varepsilon)=g(\protect\varepsilon)=1.1$ and $k=1$. The dot is the phase
transition point. \textbf{Left Panel:} Bold line corresponds to $T=T_{c}$. 
\textbf{Middle Panel:} Bold line corresponds to $P=P_{c}$. \textbf{Right
Panel:} Continuous line corresponds to $P=P_{c}$.}
\label{Fig10}
\end{figure*}

The critical horizon radius and Gibbs free energy are increasing functions
of the electric and magnetic charges while critical temperature and pressure
are decreasing functions of them. As for the rainbow functions, the
dependence of critical values on them is different from each case. 
The critical horizon radius is an increasing function of the both
energy functions, whereas the critical Gibbs free energy is a decreasing
function of both of them. Therefore, for these two critical values, the
effects of the rainbow functions are similar. Interestingly, at the critical
temperature and pressure, the effects of rainbow functions become opposite
to each other. While the critical temperature and pressure are increasing
functions of $g(\varepsilon)$, they are decreasing functions of $%
f(\varepsilon)$. The rate at which rainbow function are affecting critical
horizon radius and Gibbs free energy is similar. Whereas, such rate in
critical temperature and pressure is significantly different. This shows
that critical behavior of temperature and pressure are significantly
differently affected by one rainbow function compared to the other one.
Using the critical points, we plot the following diagrams to understand the type
of critical behavior (Fig. \ref{Fig10}).

For pressures smaller than the critical pressure, swallow tail and
subcritical isobars are formed in $G-T$ and $T-r_{+}$ diagrams. On the other
hand, for temperatures smaller than the critical temperature, there
are three different horizon radii (volumes) on the same isothermal diagrams for
the same pressure in $P-r_{+}$ diagram. The behaviors observed in the mentioned diagrams point
out that critical behavior of the black holes is similar to the one observed
for van der Waals of the liquid-gas system. This indicates that the phase
transition is first order. It could also be checked that the first order
derivation of Gibbs free energy with respect to thermodynamical volume is
finite at critical point. In conclusion, we find that the phase transition
that we observe for these black holes is first order and the thermodynamical
behavior of black holes around the critical point is van der Waals like.

\section{Conclusion}

In this paper, we investigated the properties of dyonic black holes in the
presence of gravity's rainbow. It was shown that due to the specific coupling of
the rainbow functions with temporal and other coordinates, one is able to
track the effects of different coordinates on the properties of the system
including thermodynamical quantities and behavior.

We found that existence of the black hole solutions was bounded by an upper
limit over rainbow functions and magnetic charge, whereas it was limited from below by the geometrical mass. Through our investigation of
thermodynamical quantities, it was confirmed that rainbow functions have
opposite effects (contrary to their effects on metric function). Such
behavior was found to be rooted in the special coupling of these functions with temporal and spatial coordinates. In fact, this coupling resulted into non-trivial and novel
effects on different thermodynamical quantities. For example, while the
electric potential was independent of rainbow function, the magnetic
potential was highly sensitive to variation of them. Subsequently, we observed
that the black holes would enjoy phase transition and multiple phases in
their structure. For super magnetized and pressurized cases, black holes
thermodynamically would have only one stable phase. On the contrary, for
small values of these parameters, the single-uniform-phase system would
be modified into a critically active one with phase transitions and upper
bounds imposed over its parameters. The effects of rainbow functions on such
properties and behaviors were complicated. While for super high values of one
of these rainbow functions, black holes would have single stable phase, in other cases, they would have critical behavior and separated stable phases. It
was also found that small black holes would have negative temperature and
hence, are non-physical. The type of phase transition reported for these
black holes was first order and critical behavior was van der Waals like.

\begin{acknowledgements}
We thank both Shiraz University and Shahid Beheshti University
Research Councils. This work has been supported financially partly
by the Research Institute for Astronomy and Astrophysics of
Maragha, Iran.
\end{acknowledgements}


\begin{thebibliography}{99}
 	
	
\bibitem{Magueijo1} J. Magueijo and L. Smolin, Class. Quantum Grav. \textbf{21}, 1725 (2004).


\bibitem{DSR1} G. Amelino-Camelia, Phys. Lett. B \textbf{510}, 255 (2001).

\bibitem{DSR2} J. Magueijo and L. Smolin, Phys.  Rev.  Lett. \textbf{88}, 190403 (2002).

\bibitem{DSR3} J. Magueijo and L. Smolin, Phys. Rev. D \textbf{67}, 044017 (2003).

 	
\bibitem{Amelino1} G. Amelino-Camelia, L. Freidel, J. Kowalski-Glikman, L. Smolin, Phys. Rev. D  \textbf{84}, 084010 (2011). 
	
 	
\bibitem{Magueijo2} J. Magueijo, Phys. Rev. Lett. \textbf{100}, 231302 (2008).

\bibitem{Amelino2}  G. Amelino-Camelia, M. Arzano, G. Gubitosi, and J.
Magueijo, Phys. Rev. D \textbf{87}, 123532 (2013).
	

\bibitem{Assanioussi}  M. Assanioussi, A. Dapor and J. Lewandowski,  Phys. Lett. B \textbf{751}, 302 (2015).
	
	

\bibitem{Mattingly} D. Mattingly, Living Rev. Rel. \textbf{8}, 5 (2005).
  
\bibitem{Lobo} I. P. Lobo, N. Loret and F. Nettel, Eur. Phys. J. C \textbf{77}, 451 (2017).
  
\bibitem{Nilsson} N. A. Nilsson and M. P. Dabrowski, Phys. Dark Univ. \textbf{18}, 115 (2017).
	
	
\bibitem{Chatrabhuti} A. Chatrabhuti, V. Yingcharoenrat and P. Channuie, Phys. Rev. D \textbf{93}, 043515 (2016).
	

\bibitem{Barrow} J. D. Barrow and J. Magueijo, Phys. Rev. D \textbf{88}, 103525 (2013).

\bibitem{Garattini1} R. Garattini and M. Sakellariadou, Phys. Rev. D \textbf{90}, 043521 (2014).
	
	
\bibitem{Awad} A. Awad, A. F. Ali and B. Majumder, JCAP \textbf{10}, 052 (2013).

\bibitem{Santos} G. Santos, G. Gubitosi and G. Amelino-Camelia, JCAP \textbf{08}, 005 (2015).

\bibitem{HMEP} S. H. Hendi, M. Momennia, B. Eslam Panah and S. Panahiyan, Phys. Dark Universe \textbf{16}, 26 (2017).
 	

\bibitem{Ali1} A. F. Ali, Phys. Rev. D \textbf{89}, 104040 (2014).

\bibitem{Ali2} A. F. Ali, M. Faizal and M. M. Khalil, Nucl. Phys. B \textbf{894}, 341 (2015).
 	

\bibitem{Ali3} A. F. Ali, M. Faizal and B. Majumder, Europhys. Lett. \textbf{109}, 20001 (2015).

\bibitem{Gim1} Y. Gim and W. Kim, JCAP \textbf{05}, 002 (2015).
 	

\bibitem{Galan} P. Galan and G. A. M. Marugan, Phys. Rev. D \textbf{70}, 124003 (2004). 

\bibitem{Kim} Y. W. Kim, S. K. Kim and Y. J. Park, [arXiv:1709.07755].

\bibitem{Gim2} Y. Gim, H, Um and W. Kim, JCAP \textbf{02}, 060 (2018).
 	

\bibitem{Garattini2} R. Garattini and E. N. Saridakis, Eur. Phys. J. C \textbf{75}, 343 (2015).


\bibitem{Garattini3} R. Garattini, [arXiv:1712.09728].


\bibitem{HBEP} S. H. Hendi, G. H. Bordbar, B. Eslam Panah and S. Panahiyan, JCAP \textbf{09}, 013 (2016).


\bibitem{Momeni} D. Momeni, S. Upadhyay, Y. Myrzakulov and R. Myrzakulov, Astrophys. Space Sci. \textbf{362}, 148 (2017).

\bibitem{Bakke} K. Bakke and H. Mota, [arXiv:1802.08711].


\bibitem{WD} H. L. Liu and G. L. Lu, [arXiv:1805.00333].
 	

\bibitem{Heydarzade} Y. Heydarzade, P. Rudra, F. Darabi, A. F. Ali, M. Faizal,  Phys. Lett. B \textbf{774}, 46 (2017).


\bibitem{He} M. He, P. Li, Z. Wang, J. C. Ding and J. B. Deng, Gen. Rel. Grav. \textbf{50}, 22 (2018).

\bibitem{Hendi1} S. H. Hendi and M. Faizal, Phys. Rev. D \textbf{92}, 044027 (2015). 

\bibitem{Hendi9} S. H. Hendi, B. Eslam Panah and S. Panahiyan, Prog. Theor. Exp. Phys. \textbf{2016},  103A02 (2016).

\bibitem{Hendi4} S. H. Hendi, B. Eslam Panah and S. Panahiyan, Phys. Lett. B \textbf{769}, 191 (2017).

\bibitem{Hendi6} S. H. Hendi, S. Panahiyan, S. Upadhyay, B. Eslam Panah, Phys. Rev. D \textbf{95}, 084036 (2017).

\bibitem{Bezerra} V. B. Bezerra, H. R. Christiansen, M. S. Cunha and C. R. Muniz, Phys. Rev. D \textbf{96}, 024018 (2017).


\bibitem{Leiva} C. Leiva, J. Saavedra and J. Villanueva, Mod. Phys. Lett. A \textbf{24}, 1443 (2009).
\bibitem{Ali4}  A. F. Ali, M. Faizal and B. Majumder, Europhys. Lett. \textbf{109}, 20001 (2015). 


\bibitem{Galan1} P. Galan and G. A. M. Marugan, Phys. Rev. D \textbf{74}, 044035 (2006). 

\bibitem{Ling1}  Y. Ling, X. Li, H. Zhang, Mod. Phys. Lett. A \textbf{22}, 2749 (2007).

\bibitem{Hendi2} S. H. Hendi, S. Panahiyan, B. Eslam Panah, M. Faizal and M. Momennia, Phys. Rev. D \textbf{94}, 024028 (2016).

\bibitem{Hendi3} S. H. Hendi, S. Panahiyan, B. Eslam Panah and M. Momennia, Eur. Phys. J. C \textbf{76}, 150 (2016).

\bibitem{Kim1} Y. W. Kim, S. K. Kim and Y. J. Park, Eur. Phys. J. C \textbf{76}, 557 (2016).

\bibitem{Hendi5} S. H. Hendi, S. Panahiyan, B. Eslam Panah, M. Faizal and M. Momennia, Phys. Rev. D \textbf{94}, 024028 (2016).

\bibitem{Alsaleh1} S. Alsaleh, Int. J. Mod. Phys. A \textbf{32}, 175007 (2017).

\bibitem{Alsaleh2} S. Alsaleh, Eur. Phys. J. Plus \textbf{132}, 181 (2017). 

\bibitem{Feng1} Z. W. Feng and S. Z. Yang, Phys. Lett. B \textbf{772}, 737 (2017).

\bibitem{Hendi7} S. H. Hendi, B. Eslam Panah, S. Panahiyan, M. Momenna, Eur. Phys. J. C \textbf{77}, 647 (2017).

\bibitem{Feng2} Z. W. Feng, [arXiv:1710.04496].

\bibitem{Dehghani1}  M. Dehghani, Phys. Lett. B \textbf{777}, 351 (2018).

\bibitem{Dehghani2}  M. Dehghani, Phys. Lett. B \textbf{781}, 553 (2018).

\bibitem{Hendi8} S. H. Hendi, H. Behnamifard and B. Bahrami-Asl, Prog. Theor. Exp. Phys. \textbf{2018}, 033E03 (2018).

\bibitem{PL} P. Li, M. He, J. C. Ding, X. R. Hu and J. B. Deng, [arXiv:1806.00361].


\bibitem{DS1} G. J. Cheng, R. R. Hsu and W. F. Lin, J. Math. Phys. \textbf{35}, 4839 (1994).

\bibitem{DS2} D. A. Lowe and A. Strominger, Phys. Rev. Lett. \textbf{73}, 1468 (1994).

\bibitem{DS3} H. Lu, Y. Pang and C.N. Pope, JHEP \textbf{11}, 033 (2013).

\bibitem{DS4} S. Li, H. Lu and H. Wei, JHEP \textbf{07}, 004 (2016).

\bibitem{DS5} M. Cardenas, O. Fuentealba and J. Matulich, JHEP \textbf{05}, 001 (2016).

\bibitem{DS6} P. Meessen, T. Ortin and P. F. Ramirez, JHEP \textbf{10}, 066 (2017).

\bibitem{DS7} K. A. Bronnikov, Grav. Cosmol. \textbf{23}, 343 (2017).

\bibitem{DS8} E. A. Davydov, [arXiv:1711.04198].

\bibitem{DS9} M. Bravo-Gaete and M. Hassaine, Phys. Rev. D \textbf{97}, 024020 (2018).


\bibitem{Caldarelli} M. M. Caldarelli, O. J. C. Dias and D. Klemm, JHEP \textbf{03}, 025 (2009).


\bibitem{Hartnoll} S. A. Hartnoll and P. Kovtun, Phys. Rev. D \textbf{76}, 066001 (2007).


\bibitem{Albash} T. Albash and C. V. Johnson, JHEP \textbf{09}, 121 (2008).


\bibitem{Goldstein} K. Goldstein, N. Iizuka, S. Kachru, S. Prakash, S. P.
Trivedi and A. Westphal, JHEP \textbf{10}, 027 (2010).


\bibitem{HD1} C. M. Chen, Y. M. Huang, J. R. Sun, M. F. Wu and S. J.
Zou, Phys. Rev. D \textbf{82}, 066003 (2010).

\bibitem{HD2} R. G. Cai and R. Q. Yang, Phys. Rev. D \textbf{90}, 081901 (2014).

\bibitem{Dutta} S. Dutta, A. Jainyand and R. Soniz, JHEP \textbf{12}, 060 (2013).

\bibitem{HRP} S. H. Hendi, N. Riazi and S. Panahiyan, Ann. Phys. (Berlin) \textbf{530}, 1700211 (2018).


\bibitem{GB} S. H. Hendi, S. Panahiyan, B. Eslam Panah, M. Faizal, and M.
Momennia, Phys. Rev. D \textbf{94}, 024028 (2016).


\bibitem{Dilaton} S. H. Hendi, B. Eslam Panah, S. Panahiyan and M. Momennia,
Eur. Phys. J. C \textbf{77}, 647 (2017).


\bibitem{Kubiznak} D. Kubiznak and R. B. Mann, JHEP \textbf{07}, 033 (2012).



\bibitem{HawkingT} S. W. Hawking, Commun. Math. Phys. \textbf{43}, 199
(1975).


\bibitem{Beckenstein} J. D. Beckenstein, Phys. Rev. D \textbf{7}, 2333
(1973).

\bibitem{Hawking} S. W. Hawking, Nature \textbf{248}, 30 (1974).

\end{thebibliography}
\end{document}